\documentclass[a4paper,useAMS,usenatbib]{mn2e}

\usepackage[total={17.8cm,24.0cm},centering]{geometry}
\usepackage[flushleft]{threeparttable}
\usepackage{graphicx}
\usepackage{float}
\usepackage{times}
\usepackage{color}
\usepackage{subcaption}
\usepackage{url}
\usepackage{amsmath}
\usepackage{accents}
\usepackage{xtab,afterpage}
\usepackage{lipsum}  
 \usepackage{pdflscape}

\def\afe{[$\alpha$/Fe]\:}

\def\mcyl{$m_{\rm{cyl}}$}

\newcommand{\sauron}{{\texttt {SAURON}}}

\title[Bars and the stellar populations of bulges]
{The imprints of bars on the vertical stellar population gradients of galactic bulges }

\author[Molaeinezhad et al.]
{A. Molaeinezhad$^1$\thanks{Email: molaei@ipm.ir},
	J. Falc\'on-Barroso$^{2,3}$,
	I. Mart\'inez-Valpuesta$^{2,3}$,
	H.G. Khosroshahi$^{1}$,
	\newauthor
	A. Vazdekis$^{2,3}$,
	F. La Barbera$^{6}$,
	R.F. Peletier$^{5}$,
	M. Balcells$^{2,4}$ \\
	$^{1}$School of Astronomy, Institute for Research in Fundamental Sciences (IPM), PO Box 19395-5746 Tehran, Iran \\
	$^{2}$Instituto de Astrof\'isica de Canarias, E-38200, La Laguna, Spain\\
	$^{3}$Depto. Astrof\'isica, Universidad de La Laguna (ULL), E-38206 La Laguna, Tenerife, Spain \\
	$^{4}$Isaac Newton Group of Telescopes, Apartado 321, 38700 Santa Cruz de La Palma, Canary Islands, Spain \\
	$^{5}$Kapteyn Astronomical Institute, University of Groningen, Postbus 800, 9700 AV Groningen, the Netherlands \\
    $^{6}$INAF – Osservatorio Astronomico di Capodimonte, I-80131 Napoli, Italy}

\begin{document}
 \date{Accepted 2017 January 09 . Received 2017 January 09 ; in original form 2016 July 26}

\maketitle
\label{firstpage}

\begin{abstract}
This is the second paper of a series aimed to study the stellar kinematics and 
population properties of bulges in highly-inclined barred galaxies. In this 
work, we carry out a detailed analysis of the stellar age, metallicity and [Mg/Fe] 
of 28 highly-inclined ($i > 65^{o}$) disc 
galaxies, from S0 to S(B)c, observed with the SAURON integral-field 
spectrograph. The sample is divided into two clean samples of barred and 
unbarred galaxies, on the basis of the correlation between the stellar velocity 
and h$_3$ profiles, as well as the level of cylindrical rotation within the 
bulge region. We find that while the mean stellar age, metallicity and [Mg/Fe] in 
the bulges of barred and unbarred galaxies are not statistically distinct, 
the [Mg/Fe] gradients along the minor axis (away from the disc) of barred galaxies 
are significantly different than those without bars. For barred galaxies, stars 
that are vertically further away from the midplane are in general more 
[Mg/Fe]--enhanced and thus the vertical gradients in [Mg/Fe] for barred galaxies 
are mostly positive, while for unbarred bulges the [Mg/Fe] profiles are typically 
negative or flat. This result, together with the old populations observed in the 
barred sample, indicates that bars are long-lasting structures, and therefore 
are not easily destroyed. The marked [Mg/Fe] differences with the bulges of 
unbarred galaxies indicate that different formation/evolution scenarios are 
required to explain their build-up, and emphasizes the role of bars in 
redistributing stellar material in the bulge dominated regions. 
\end{abstract}


\section{Introduction}
\label{sec:intro}
In the era of IFU spectroscopy, the spatially resolved spectroscopic 
information, along with modern stellar population models, have provided new 
insights into our understanding of the role of bars on the formation and 
evolution of disc galaxies. Barred galaxies represent a considerable fraction of 
the entire disc galaxy population 
\citep[e.g.][]{eskr2000,knapen2000,whyt2002,grosb2004,marin2007}. They are considered as key drivers of the internal secular evolution in disc galaxies by re- distribution of the angular momentum, triggering of star formation and the morphological transformation of galaxies in general, as demonstrated observationally 
\citep[e.g.][]{korm2013} and theoretically \citep[e.g.][]{athan2013}.

Most stellar population studies of barred systems focus on face-on disc galaxies 
and the stellar population gradients along the bars major axis \cite[see][and 
references therein]{sanc2016}. \citet{moor2006} analysed long-slit spectra of 38 
face-on spirals of type S0--Sbc and reported a trend towards younger ages for the 
bulges of barred galaxies. They concluded that a positive age gradient is a 
direct indicator of the existence of a bar. \citet{perez2011} studied the age 
and metallicity gradients in the bulge region of a sample of 20 early-type 
barred galaxies. Using the long-slit spectra along the bar major axis, they 
found that at a given velocity dispersion ($\sigma$), the bulges of barred 
galaxies seem to be more metal rich than the bulges of unbarred galaxies. They 
measured mostly negative metallicity gradients for the bulges in their sample, 
suggesting a strong link between the metallicity of the bulge and the presence 
of a bar. They also indicated that bulge populations distributions and gradient measurements 
could be biased against the existence of the bulge substructures, such as a nuclear 
ring or an inner disc. Similar results have been obtained by \citet{coel2011},  
that claim the current star formation is enhanced in the centres of barred 
galaxies, and consequently the measured mean stellar age of bulges is strongly 
affected by the bar.

A detailed and unbiased study of the extended stellar population in the bulges 
of barred and unbarred galaxies requires high inclinations to ensure minimal 
contamination by the stellar disc. Nonetheless, the number of works, dedicated 
to study the stellar populations in highly-inclined barred galaxies is still 
very low and is limited to a few nearby cases \citep[e.g.][]{jabl20072,will2011,perez2011}. Understanding the variations of 
stellar population parameters of bulges at different heights from the disc 
provides stronger constraints on the formation and evolution scenarios of bulges. 
More specifically, comparing the vertical gradients of the stellar 
population parameters in bulges with and without bars can provide very important 
new information to expand our understanding of bars nurture and its role in 
formation and evolution of disc galaxies.

Using long-slit spectroscopy, \cite{jabl20072} studied the vertical gradients of 
stellar population properties in the bulges of 32 edge-on galaxies, spanning a wide range of Hubble types, from S0 to Sc.
They found that the central parts of the bulges are mostly younger and more metal rich than their outer regions and 
the outer parts of the bulges have higher [$\alpha$/Fe] than the inner parts. 
They found no meaningful differences between the stellar populations of nearly 
edge-on barred and unbarred galaxies, although they were not in the position
to precisely distinguish bulges with and without bars. It has been widely accepted that 
boxy/peanut-shaped bulges (hereafter BP bulges), found in about half of edge-on 
galaxies, are bars viewed from the side 
\citep[e.g.][]{kuij1995,merr1999,bure1999,atha2005}. \citet{will2011} studied 
the stellar kinematics and populations in five edge-on galaxies, dominated by BP 
bulges, using long-slit spectroscopy. They measured the vertical gradients in 
the bulges and claimed that BP bulges in their sample do not form a homogeneous 
class of objects. Later, \citet{will2012}, analysed the central and the major 
axis profiles in the stellar populations of 28 highly inclined S0--Sb disc 
galaxies, presented in \citet{bure1999} and \citet{chun2004}, mostly with BP bulges. They found that at a given $\sigma$, gradients 
in BP bulges are shallower than those in elliptical galaxies, but that the 
stellar populations at the very centres of disc galaxies with or without BP 
bulges do not differ from those of early-type galaxies.\looseness-2  

As we stated earlier, only a limited number of studies have been dedicated to investigate 
the stellar population properties of bulges in highly-inclined barred galaxies, 
and on top of that, the results are mostly inconsistent \cite[see][and the references therein]{sanc2016}. This might be, at least 
in part, due to the lack of a reliable and robust method to distinguish barred 
and unbarred galaxies in highly-inclined systems. The presence of a bar is not generally apparent if the bar is oriented close to or exactly parallel to the line-of-sight, even in edge-on galaxies, \citep[see e.g.][]{bure2004,atha2016,mola2016}. Therefore, previous results might be highly biased \citep[see][]{pele2007}.

In the first paper of this series \citep[][hereafter Paper I]{mola2016} we 
studied the connection between bulge morphology and kinematics in 12 mid to 
highly-inclined disc galaxies, observed with the \sauron\ integral-field 
spectrograph. We suggested that the strong positive correlation between the 
stellar velocity ($V$) and the h$_3$ Gauss--Hermite moment appears to be the 
most reliable indicator for the presence of bars among all other bar 
diagnostics, even in cases with end-on orientation of the bar \citep[see][for 
more details]{bure2005,atha2016}. This is followed by introducing a method to 
quantify cylindrical rotation that is robust against inner bulge substructures 
such as inner discs, which are likely more common in barred galaxies. Our 
results also confirmed high levels of cylindrical rotation in barred galaxies 
with boxy/peanut bulges. In the current work, we study the stellar population 
properties and in particular, the vertical gradients along the minor axis in the 
bulges of 28 highly-inclined ($i > 65^{o}$) disc galaxies, observed with the \sauron\ 
integral-field spectrograph. Using the kinematic criteria introduced in Paper I, 
we are able to secure an unbiased classification of barred and unbarred galaxies 
in our sample, and compare the stellar population properties and vertical 
gradients in each class of bulges. 

The paper is organised as follows. In Section~\ref{sec:sample} we describe the 
sample and give a summary of optical properties and kinematics of the galaxies 
in our sample. Section~\ref{sec:population} is dedicated to the line-strength 
measurements and present the results of our stellar population analysis. In 
Section~\ref{sec:discuss} we briefly discuss the possible scenarios to interpret 
these results. Finally, we summarize our findings and draw our conclusions in 
Section~\ref{sec:conclusion}.

\begin{table*}
\caption{Properties of our sample of galaxies}
\label{tab:sample}
\begin{center}
\begin{tabular}{cccccccccccccc}
\hline
Galaxy  & Sample &  PA               & V$_{\rm{hel}}$     & $M_{K}$   &    T--type  & incl. &  $\sigma_{\rm{0}}$ &    Dust  &  \mcyl\          &  Bar   & $z_{\rm{disc}}$ & $x_{\rm{B}}$   & $z_{\rm{B}}$ \\
~       & ~      &  (deg)            & (km\,s$^{-1}$)     & (mag)     &    ~       & (deg) &  (km\,s$^{-1}$)     &    ~     &  ~               &  ~     & (arcsec)       & (arcsec)       & (arcsec) \\
(1)     & (2)    &  (3)              & (4)                & (5)       &    (6)     &  (7)  &  (8)                &   (9)    &  (10)            &  (11)  & (12)           & (13)           & (14) \\
\hline
NGC3098 & S2     &  \phantom{0}88.5  &    1397            & $-$22.72  &    $-$1.5  &  90   &    126.2  &    N     &    0.54$\pm$0.20 &    N?  & 1.5    & 10            &    \phantom{0}9 \\
NGC4026 & S2     &  177.5            &    \phantom{0}985  & $-$23.03  &    $-$1.8  &  84   &    158.1  &    N     &    0.54$\pm$0.13 &    Y   & 3.0    & 12            &    10 \\
NGC4036 & S2     &  261.2            &    1385            & $-$24.40  &    $-$2.6  &  75   &    181.9  &    F     &    0.26$\pm$0.24 &    N   & 2.0    & 12            &    \phantom{0}9 \\
NGC4179 & S2     &  142.8            &    \phantom{0}13   & $-$23.18  &    $-$1.9  &  86   &    167.5  &    N     &    0.59$\pm$0.12 &    Y   & 2.0    & 13            &    10 \\
NGC4251 & S2     &  \phantom{0}99.0  &    1066            & $-$23.68  &    $-$1.9  &  80   &    128.8  &    N     &    0.73$\pm$0.08 &    Y   & 2.0    & 12            &    \phantom{0}9 \\
NGC4270 & S2     &  109.8            &    2331            & $-$23.69  &    $-$2.0  &  80   &    139.6  &    N     &    0.58$\pm$0.12 &    Y   & 1.5    & 13            &    10 \\
NGC4346 & S2     &  \phantom{0}98.8  &    \phantom{0}832  & $-$22.55  &    $-$2.0  &  77   &    127.0  &    N     &    0.63$\pm$0.11 &    Y   & 3.5    & 12            &    12 \\
NGC4425 & S2     &  \phantom{0}25.8  &    1908            & $-$22.09  &    $-$0.6  &  90   &    \phantom{0}82.8  &    N     &    0.53$\pm$0.15 &    Y   & 3.0    & 14            &    \phantom{0}8 \\
NGC4435 & S2     &  \phantom{0}10.0  &    \phantom{0}791  & $-$23.83  &    $-$2.1  &  68   &    152.8  &    D     &    0.51$\pm$0.14 &    Y?   & 3.0    & 11            &    \phantom{0}9 \\
NGC4461 & S2     &  \phantom{0}8.1   &    1924            & $-$23.08  &    $-$0.8  &  71   &    133.0  &    N     &    0.58$\pm$0.15 &    Y   & 3.5    & 13            &    \phantom{0}9 \\
NGC4474 & S2     &  \phantom{0}79.4  &    1611            & $-$22.28  &    $-$2.0  &  89   &    \phantom{0}87.9  &    N     &    0.55$\pm$0.17 &    Y?   & 2.0    & 10            &    \phantom{0}9 \\
NGC4521 & S2     &  166.3            &    2511            & $-$23.92  &    $-$0.1  &  90   &    185.8  &    N     &    0.71$\pm$0.08 &    Y   & 2.0    & 10            &    \phantom{0}8 \\
NGC4710 & S2     &  \phantom{0}27.4  &    1102            & $-$23.53  &    $-$0.9  &  88   &    104.7  &    D     &    0.54$\pm$0.20 &    Y   & 3.0    & 19            &    13 \\
NGC4762 & S2     &  \phantom{0}29.6  &    \phantom{0}986  & $-$24.48  &    $-$1.8  &  90   &    133.7  &    N     &    0.66$\pm$0.13 &    Y   & 1.0    & 10            &    \phantom{0}7 \\
NGC5103 & S2     &  140.6            &    1273            & $-$22.36  &    --      &  90   &    111.2  &    N     &    0.37$\pm$0.21 &    N   & 2.0    & \phantom{0}9  &    \phantom{0}8 \\
NGC5326 & S1     &  130.0            &    2520            & $-$23.77  &    --      &  65   &    144.9  &    N     &    0.35$\pm$0.20 &    N   & 2.5    & 10            &    \phantom{0}8 \\
NGC5353 & S2     &  140.4            &    2198            & $-$25.11  &    $-$2.1  &  80   &    281.2  &    D     &    0.57$\pm$0.11 &    Y   & 3.0    & 15            &    \phantom{0}9 \\
NGC5422 & S1     &  152.3            &    1838            & $-$23.69  &    $-$1.5  &  90   &    161.8  &    N     &    0.69$\pm$0.08 &    Y   & 2.5    & 16            &    10 \\
NGC5475 & S1     &  166.2            &    1671            & $-$22.88  &    --      &  79   &    115.0  &    N     &    0.25$\pm$0.25 &    N   & 2.5    & 12            &    \phantom{0}8 \\
NGC5574 & S2     &  \phantom{0}62.7  &    1589            & $-$22.30  &    $-$2.8  &  89   &    \phantom{0}81.9  &    N     &    0.51$\pm$0.23 &    Y   & 2.0    & 11            &    \phantom{0}9 \\
NGC5611 & S2     &  \phantom{0}64.6  &    1968            & $-$22.20  &    $-$1.9  &  74   &    137.4  &    N     &    0.40$\pm$0.24 &    N   & 1.5    & 10            &    \phantom{0}9 \\
NGC5689 & S1     &  \phantom{0}84.0  &    2160            & $-$24.00  &    --      &  81   &    157.4  &    D     &    0.64$\pm$0.09 &    Y   & 3.0    & 17            &    13 \\
NGC5707 & S1     &  \phantom{0}35.0  &    2212            & $-$23.22  &    --      &  80   &    131.8  &    N     &    0.25$\pm$0.20 &    N   & 1.5    & 11            &    \phantom{0}8 \\
NGC5746 & S1     &  170.0            &    1727            & $-$24.99  &    --      &  81   &    202.8  &    D     &    0.61$\pm$0.09 &    Y   & 3.5    & 22            &    15 \\
NGC5838 & S1     &  \phantom{0}40.1  &    1341            & $-$24.13  &    $-$2.6  &  72   &    246.0  &    N     &    0.47$\pm$0.17 &    Y   & 2.0    & 11            &    \phantom{0}9 \\
NGC5854 & S2     &  \phantom{0}54.8  &    1663            & $-$23.30  &    $-$1.1  &  74   &    104.7  &    N     &    0.38$\pm$0.26 &    Y   & 3.5    & 13            &    10 \\
NGC5864 & S2     &  \phantom{0}65.6  &    1874            & $-$23.62  &    $-$1.7  &  74   &    110.7  &    N     &    0.59$\pm$0.12 &    Y   & 1.5    & 13            &    10 \\
NGC6010 & S1     &  102.9            &    2022            & $-$23.53  &    --      &  90   &    159.2  &    D     &    0.45$\pm$0.19 &    Y   & 3.5    & 11            &    10 \\
\hline
\end{tabular}
\end{center}
\begin{flushleft}
\small NOTES: 
(1) Galaxy name.
(2) Sample -- S1: Molaeinezhad et al. (2016) sample. S2: Extracted from the ATLAS3D survey.
(3) Position angle (N--E) of the dust free minor axis. 
(4) Heliocentric velocity, taken from \citet{capp2011}.For galaxies not in the ATLAS3D sample, values are taken from NED.
(5) Total galaxy absolute magnitude in $K$-band, taken from \citet{pele1997} and \citet{capp2011}, for samples S1 and S2, respectively.
(6) Morphological T--type from HyperLeda \citep{patu2003}.
(7) Inclinations derived from the best-fitting mass-follow-light JAM modelling \citep{capp2013}, or if not available, derived from disc ellipticity in $R$-band from \cite{pele1997}, corrected for finite disc thickness.
(8) Central velocity dispersion, defined as the maximum $\sigma$ within the bulge analysis window, from our data.
(9) Dust features -- D: dusty disc, F: dusty filament, N: none \citep[Taken from][]{capp2011}.
(10) Level of cylindrical rotation (\mcyl) within the bulge analysis window and its uncertainty. Values of \mcyl\ are generally between 1 (pure cylindrical rotation) and 0 (no sign of cylindrical rotation).
(11) Detection of a bar, based on the kinematic analysis as in Paper I. The Question mark (?) indicates cases where bar detection is uncertain.
(12) Vertical extent of the region, close to the disc plane which is most likely disrupted by contamination of dust and/or central components.
(13) and (14) Analysis window of the bulge along the major and minor axis respectively. See \S\ref{sec:sample} for more details.
\end{flushleft}
\end{table*}

\section{Sample \& IFU observations}
\label{sec:sample}

\subsection{Sample Selection \& Properties}

Our sample consists of 28 highly-inclined ($i > 65^{o}$) galaxies with high S/N IFU 
observations, that can be used to study the stellar population gradients along the minor axis. 
The sample is drawn from the complete, magnitude limited sample of 260 nearby 
galaxies in the ATLAS3D survey \citep{capp2011}, complemented with the sample of 
highly inclined galaxies introduced in Paper I. For the ATLAS3D survey, we 
focused on disc galaxies, and only kept those galaxies whose inclination, as measured via the 
best fitting mass-follow-light JAM models \citep{capp2012}, are above 
65$^{\circ}$. This criterion allows us to access the clean vertical extent of the 
bulge, less disrupted by contamination of dust, disc and/or central components. 
We visually inspected the sample to discard galaxies with irregular morphology 
or low quality IFU data (with poor spatial sampling in the regions of interest). 
In addition we discarded galaxies where the emission line \textit{cleaning} of 
the spectra was particularly problematic in the short wavelength range provided 
by the \sauron\ spectrograph. Our final sample comprises 20 galaxies from the 
ATLAS3D sample (hereafter S2), along with 8 galaxies from our original sample of 
Paper I (hereafter S1). The basic photometric properties of these 28 galaxies 
are presented in Table \ref{tab:sample}. \looseness-2

\subsection{Observations}

The \sauron\ observations of both samples S1 and S2 are described in detail in 
Paper I and \citet{capp2011}, respectively. Briefly, the spectroscopic 
observations of the S1 sample were carried out between October 1999 and 2011 
with the \sauron\ integral-field spectrograph \citep{baco2001} attached to the 
4.2-m William Herschel Telescope (WHT) of the Observatorio del Roque de los 
Muchachos at La Palma, Spain. We used the low spatial resolution mode of 
\sauron\, which gives a $33\arcsec\times41\arcsec$ field-of-view (FoV), with a 
spatial sampling of $0\farcs94\times0\farcs94$. This setup produces 1431 spectra 
per pointing over the \sauron\ FoV. Additionally, a dedicated set of $146$ 
lenses provides simultaneous sky spectra $1\farcm9$ away from the main field. The 
spectral resolution delivered by the instrument is $\sim$4.2~\AA\ (FWHM) and 
covers the narrow spectral range 4800-5380~\AA. 
The IFU observing strategy and 
the data reduction of the S2 as described in \citet{capp2011} is similar to the 
S1 sample, which guarantees the homogeneity of data for our sample. The stellar kinematics 
data, published here for S1 is the same as was presented in Paper I. 
Kinematics maps of the S2 sample were presented in details in \citet{kraj2011}. 
In order to ensure meaningful and accurate stellar population parameters all the 
data were Voronoi binned, following \citet{2003MNRAS.342..345C}, to a target 
SNR=60. We describe our procedures to extract the population parameters in 
\S\ref{sec:population}.\looseness-2

\subsection{Sample morphological and kinematic classification}

\subsubsection{Morphological analysis}

In our study of the stellar population properties of bulges, the choice of 
highly-inclined galaxies ensures minimal contamination by the stellar disc. 
However, at these inclinations the bar component is not easily distinguishable if it is not inclined enough or has not vertically thickened enough to be clearly recognized above the disc plane, and/or beyond a possible large classical bulge. It is worth noting that 
throughout this study, the ‘bulge’ of a galaxy is referred to the excess of 
light that is above the main exponential disc. 
\\
A major concern in such studies is the disc contamination, that could severely 
bias the bulge morphology analysis. For this purpose, as we did in Paper I, we 
have used Galfit \citep{peng2002} to find the best exponential disc model and 
subtract it from the original image. These subtracted images have been used to 
delimit the region of interest in our galaxies: the 'photometric bulge'. The 
extent of the bulge in the radial ($x$) and vertical ($z$) directions has been 
evaluated by integrating the residual images, along the minor (or major) axis, 
and determining the region in which more than 90 per cent of light comes from the bulge.
 We have made use of $i-$band images from the Sloan Digital Sky 
survey DR10 \citep{ahn2014} for this purpose. In addition, we have produced 
unsharp masked (median filtering) images to highlight the high frequency 
features of the images and reveal possible structures with no radial symmetries. 
These are presented in Appendix~\ref{app:unsharp}, together with the bulge 
region obtained from our Galfit analysis. The radial and vertical extent of our 
bulges (in arcsec) are listed in Table \ref{tab:sample}.

\subsubsection{Kinematic classification}

\citet{bure2005} suggested, using N-body simulations, a number of major-axis 
stellar kinematic features that can be used as bar diagnostics in highly 
inclined systems: (1) a ‘double-hump’ rotation curve; (2) a broad central 
velocity dispersion peak with a plateau (and possibly a secondary maximum) at 
moderate radii and (3) an h$_3$ profile correlated with velocity over the 
projected bar length. All these diagnostics have been successfully expanded 
beyond the disc plane by \citet{lann2015}, who studied the imprints of bars on 
the 2D line-of-sight kinematics of simulated disc galaxies. In Paper I, we 
tested this approach on 12 mid- to highly inclined galaxies, all of them 
included in the S1 sample, and confirmed that the correlation between the 
stellar line of sight velocity ($V$) and the h$_3$ Gauss--Hermite parameter is a 
very reliable bar diagnostic tool. The success of this method is the ability to 
unveil the hidden bar, when the bar is not clearly visible in photometric data, 
due to its orientation or strength. We have taken advantage of this approach 
to detect the presence of bars in our sample of galaxies. The diagnostic plots 
showing the degree of correlation can be found in Appendix~\ref{app:unsharp}.

\begin{figure}
\centering
\includegraphics[width=0.5\textwidth]{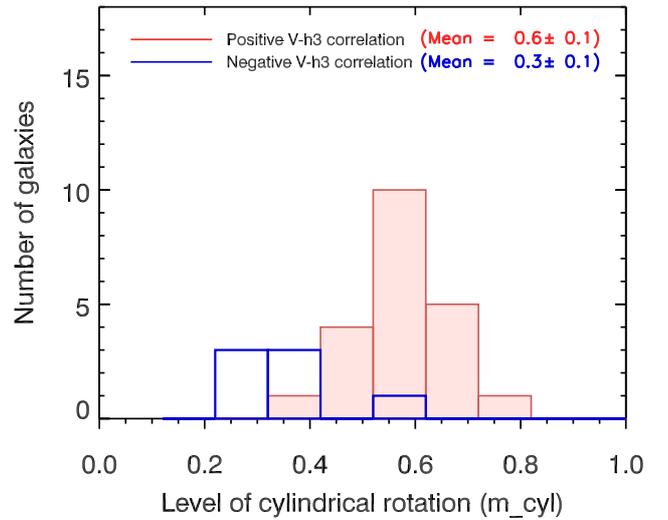}
\caption{Distribution of the level of cylindrical rotation (\mcyl) for the 
galaxies with positive $V$--h$_3$ correlation (red histogram) and those 
with mostly negative or null correlation (blue histogram).}
\label{fig:cyl}
\end{figure}

In addition, a typical kinematic feature of most barred galaxies with BP bulges 
is cylindrical rotation \citep[e.g.][]{korm1982,bure1999,falc2006,mola2016}. In 
such systems, the mean stellar rotation speed shows little difference at 
different height above the disc plane. In Paper I, we introduced a method to 
quantify the level of cylindrical rotation in bulges. This quantity (\mcyl) 
shows the importance of this property in bulges and is generally a value between 
+1 which indicates pure cylindrical rotation and 0 where there is no sign of 
cylindrical rotation within the bulge analysis window. Figure~\ref{fig:cyl} 
shows the distribution of \mcyl\ for the galaxies in our sample, separated into 
two categories of galaxies with positive $V$--h$_3$ correlations and those with 
negative or null $V$--h$_3$ correlations. As expected, the galaxies with 
relatively high values of \mcyl\ are mostly those showing positive correlation 
between $V$ and h$_3$, while those with lower \mcyl\ values do not show a 
significant correlation. The \mcyl\ values and related uncertainties are 
presented in Table \ref{tab:sample}. 

As our goal is to have two clean classes of barred and unbarred systems, we use 
the combination of both bar diagnostic tools, the correlation $V$--h$_3$ and the 
level of cylindrical rotation, to define the galaxies belonging to each class. 
In this study, "barred" galaxies refer to those systems that show remarkably 
positive correlation between $V$ and h$_3$ in the areas above the mid-plane 
within the bulge analysis window, while those galaxies with non-positive 
$V$--h$_3$ and \mcyl$<$0.55 are marked as "unbarred" system. Not surprisingly, 
all galaxies with BP bulges in our sample belong to the "barred" class.

\begin{figure*}
	\captionsetup[subfigure]{labelformat=empty}
	\begin{subfigure}{0.37\textwidth}
		\includegraphics[width=1.0\textwidth]{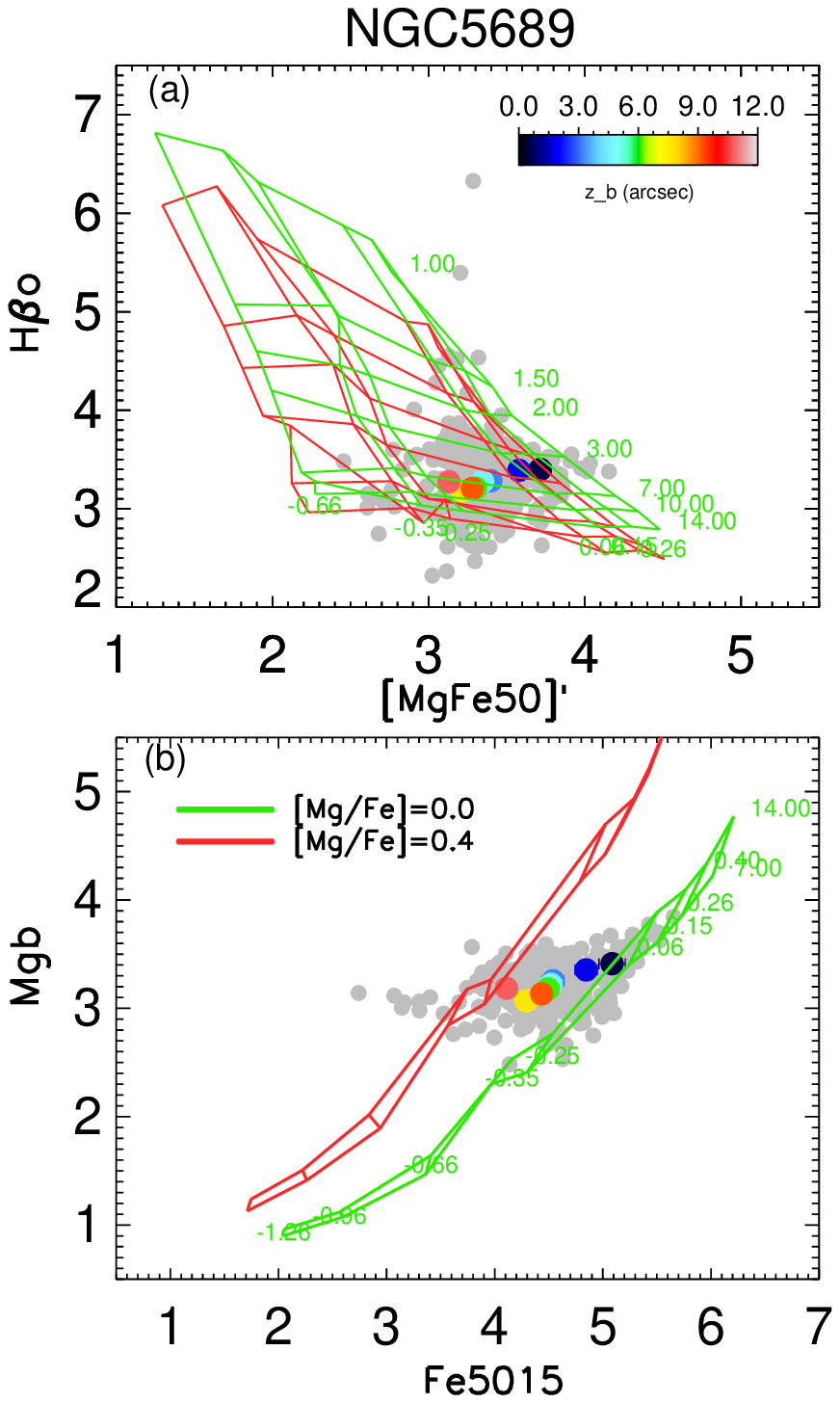}
	\end{subfigure}%
	\begin{subfigure}{0.6\textwidth}
		\centering
		\includegraphics[width=1.0\textwidth]{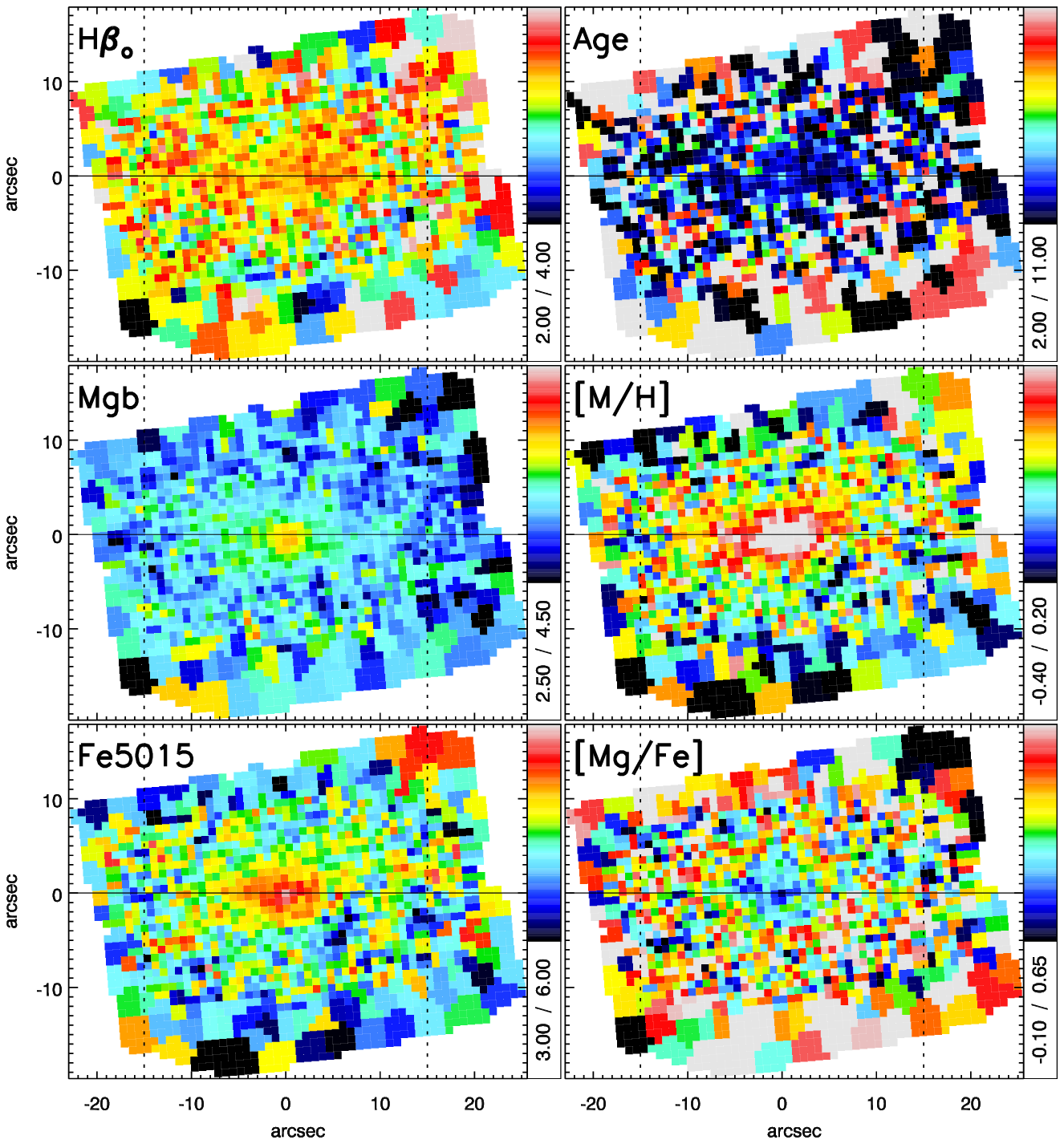}
	\end{subfigure}
	\caption{Left column, panel (a): The index-index diagram of H$\beta_{o}$ 
		versus [MgFe50]' for all Voronoi bins (grey filled circles) within the bulge 
		analysis window of NGC\,5689, a representative barred galaxy with BP bulge in 
		our sample. Left column, panel (b): The Mg$b$ versus Fe5015 diagram for this 
		galaxy. Overlaid are the MILES SSP models for different ages and 
		metallicities. The green and red grids correspond to the scaled-solar and 
		$\alpha$--enhanced SSP models, respectively. Coloured symbols indicate average 
		Voronoi bin values at various heights from the major axis. Middle column: The absorption 
		line-strength maps of H$\beta_{o}$, Mg$b$ and 
		Fe5015, respectively (from top to bottom). Right Column: computed SSP-equivalent maps of age, metallicity and [Mg/Fe], respectively.}
	\label{fig:5689}
\end{figure*}

\section{Stellar population quantities}
\label{sec:population}
Besides stellar kinematics, we have also measured line-strength 
indices of the S1 and S2 galaxies in the recently defined Line Index System 
(LIS) LIS-8.4~\AA\ \citep[][hereafter VAZ10]{vazd2010}. This method has the 
advantage of circumventing the use of the so-called Lick/IDS fitting functions for 
the model predictions, which requires the determination of often uncertain 
offsets to account for differences in the flux calibration between models and 
observations. 

The wavelength range provided by \sauron\ (4800--5380~\AA) constrains the set of spectral 
indices that can be used to estimate stellar population parameters, i.e. 
age, metallicity, and [Mg/Fe].  
We have used the three available indices that are measured across the
full field for all galaxies, namely the H$\beta_{\rm{o}}$ age indicator 
(i.e. the optimized H$\beta$ index of \citet{cervantes09}), as well as 
the metallicity- and [Mg/Fe]--sensitive indices Fe5015 and Mg$b$ \citep{trag1998}. 
Notice that the [Mg/Fe] abundance ratio is of particular interest for the present study, as 
it measures -- for a fixed IMF --  
the different timescale of the ejection of magnesium and iron into the interstellar medium.
Thus, the [Mg/Fe] is believed to be a "chemical clock" to estimate star-formation 
timescales for the bulk of a stellar population \citep[e.g.][]{thom2011,vazd2015, mart2016}. \looseness-2

Our analysis relies on $\alpha$--MILES stellar population models, from \citet[][hereafter V15]{vazd2015}. The models, 
constructed from MILES  stellar spectra, have been computed, in a self-consistent manner, at both solar scale and for 
\afe$=+0.4$,  over a range of metallicities, with the aid of theoretical stellar spectra from 
\citet{coel2005,coel2007}, and using BaSTi (either scaled-solar or $\alpha$--enhanced) isochrones from \citet{piet2004, piet2006}. Notice that throughout the present work, we adopt the notation [Mg/Fe], rather than [$\alpha$/Fe], as our estimate of abundance ratios relies entirely on Mg- and Fe-sensitive features, i.e. Mg$b$5177 and Fe5015.

The $\alpha$--MILES models cover the (optical) MILES spectral range ($\lambda\lambda \sim 3500$--$7400$~\AA), in the metallicity and age range of $\rm -2.27 \le [Z/H] \le +0.4$ and $\rm  0.03 \le t \le 14$~Gyr, respectively. 
For both scaled-solar and $\alpha$--enhanced SSP models, we have produced an interpolated grid of model 
line-strengths with a  step of $\sim 0.06$~Gyr in age, and $\sim 0.017$~dex in metallicity. The scaled-solar and 
$\alpha$--enhanced grids are then interpolated/linearly-extrapolated to cover the [Mg/Fe] range from $-0.1$ to $+0.8$~dex, 
with a step of $0.009$~dex in [Mg/Fe]. Notice that the extrapolation does not affect at all our results,
as the integrated, average profiles of [Mg/Fe] for barred and unbarred galaxies exceeds only slightly (by $\sim 0.05$~dex;
see Fig.~\ref{fig:afe_profile}) the value of [Mg/Fe]$=+0.4$ for which V15 $\alpha$--enhanced models have been computed.

For each spectrum, we derive the SSP-equivalent age, metallicity, and [Mg/Fe],
by minimizing the following equation:
\begin{eqnarray}
\chi^2 & = & \rm \sum_i \frac{ (  EW_{obs,i} - EW_{mod,i} )^2 }{ \sigma_{obs,i}^{2}},
\end{eqnarray}
where the index $\rm i$ runs over the available indices (H$\beta_{\rm{o}}$, Fe5015, and Mg$b$), 
$\rm EW_{obs,i}$ and $\rm EW_{mod,i}$ are observed and model line-strengths, while  
$\rm \sigma_{obs,i}$ denote uncertainties on observed line-strengths.
The minimization is performed over the grid of SSP model predictions, with varying age,
metallicity, and [Mg/Fe] (see above). Uncertainty on best-fit parameters are estimated from
$N=1000$ bootstrap iterations, where the fitting is repeated after
shifting observed line strengths according to their uncertainties.

Figure~\ref{fig:5689} shows the line-strength indices and resulting stellar 
population parameters for NGC\,5689, a representative barred galaxy with BP 
bulge in our sample. In the left column, panel (a) shows the H$\beta_{o}$ 
versus [MgFe50]' for all Voronoi bins (grey filled circles) within the bulge analysis 
window of NGC5689. Coloured symbols indicate  representative values of Voronoi 
bins at various heights from the major axis (see \S\ref{vertical} for more details). 
Overlaid are the MILES SSP models for different ages and metallicities. Panel (b) presents the Mg$b$ versus Fe5015 
diagram for this galaxy. The green and red grids correspond to the scaled-solar 
and [Mg/Fe]--enhanced SSP models, respectively.  
The middle column of the figure shows the H$\beta_{\rm{o}}$, Fe5015, 
and Mg$b$ index maps, while the right column presents the age, metallicity and 
[Mg/Fe] maps obtained with the procedure outlined above.

\subsection{Comparison of the mean stellar population parameters of bulges of barred and unbarred galaxies}

The majority of stellar population studies in bulges focus on the integrated 
properties inside a certain aperture, near the centre. These  studies could 
therefore strongly depend on the presence (or not) of different subcomponents 
\cite[e.g. a young stellar central disc, nuclear cluster; see][]{pele2007}. This 
effect is particularly important in highly-inclined galaxies, where the main 
disc contamination is a problematic issue. To avoid these difficulties, 
we limit our analysis of the bulge stellar populations to a window well beyond the disc plane.
 More specifically, the regions of 
 interest in our bulges are limited to the clean (e.g. dust free) side of the 
galaxies, between $-x_{\rm{B}}$ and $+x_{\rm{B}}$ along the major axis and 
$z_{\rm{disc}}$ to $z_{\rm{B}}$ along the minor axis of galaxies. The 
$z_{\rm{disc}}$ indicates the vertical extent of the region, close to disc plane that might be contaminated by dust and/or central components (evaluated by visual inspection of the velocity maps). 

\begin{figure}
\centering
\includegraphics[width=0.45\textwidth]{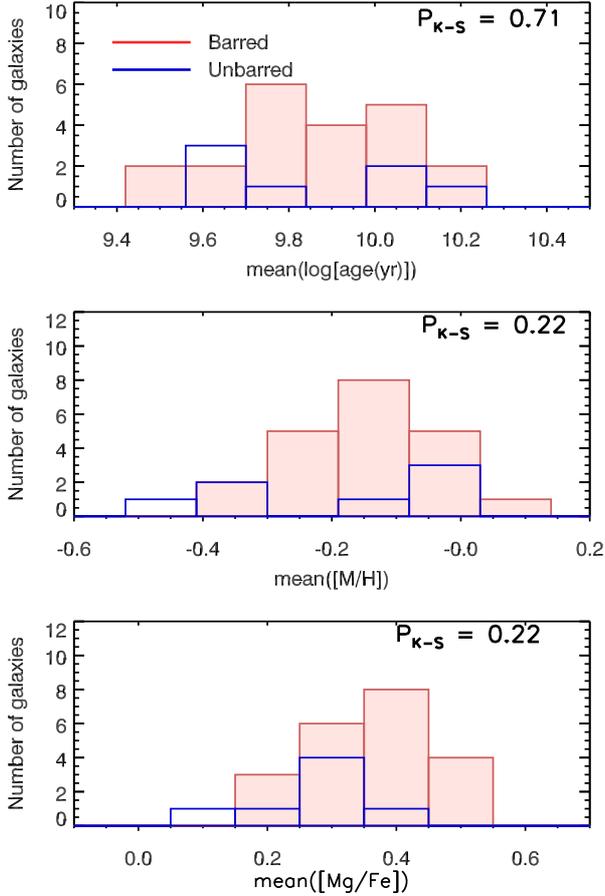}
\caption{Distribution of mean SSP-equivalent age, metallicity and [Mg/Fe] within 
the bulge analysis window of barred (red histogram) and unbarred (blue 
histogram) galaxies in our sample. $P_{K-S}$ indicates the probability that the two 
distributions are drawn from the same population.}
\label{fig:mean_ssp}
\end{figure}

In order to investigate stellar populations of bars and the influence of bars on 
bulges of nearby galaxies, we compare the key stellar population parameters 
within the bulge analysis windows of our sample. Figure~\ref{fig:mean_ssp} shows 
the distributions of the mean SSP-equivalent age, metallicity and [Mg/Fe] for our 
two classes of bulges. A Kolmogorov--Smirnov test (hereafter K--S) indicates 
that the difference between the two populations of bulges is not significant. 
This is in agreement with results of \citet{jabl20072} who found no significant 
difference between the stellar populations of early type (S0--S(B)b) edge-on 
barred and unbarred galaxies. 

\begin{figure}
\centering
		\includegraphics[width=0.4\textwidth]{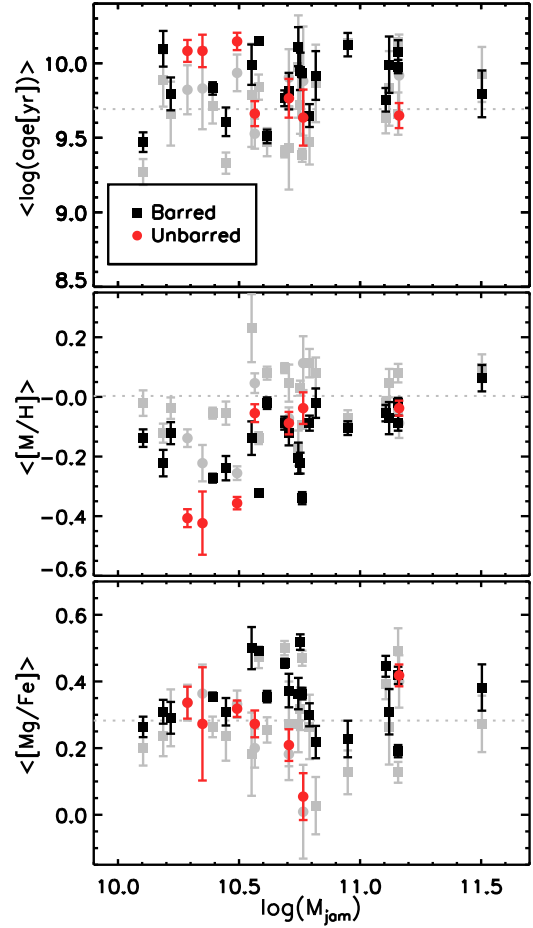}
\caption{Distribution of mean SSP equivalent age, metallicity and [Mg/Fe] versus 
the dynamical mass (M$_{\rm{JAM}}$) estimates, within one effective radius 
($R_{\rm{eff}}$) of both classes of barred and unbarred galaxies in our sample. 
Overploted as grey filled circles show stellar populations of galaxies, measured 
in the central regions of the galaxies, within a circular aperture with a radius 
of $z_{\rm{disc}}$. Horizontal dotted line represents the mean stellar 
populations of the central apertures.}
\label{fig:mean_sigma_ssp}
\end{figure}

A key aspect to constrain bulge formation scenarios is the correlation of the stellar population properties of bulges, with the total mass of their parent galaxies.
Figure~\ref{fig:mean_sigma_ssp} shows the 
distribution of the mean SSP-equivalent age, metallicity and [Mg/Fe] of the bulges 
(within the bulge analysis window) versus the dynamical mass (M$_{\rm{JAM}}$) 
estimates from \citep{capp2013}, within one 
effective radius ($R_{\rm{eff}}$) of galaxies. The central stellar populations 
(within a circular aperture with a radius of $z_{\rm{disc}}$) are also shown 
(grey filled circles). As this figure shows, no significant correlations of the 
stellar population parameters with the dynamical mass are found, except for 
metallicity, where more massive galaxies harbour on average more metal rich 
bulges. In good agreement with the results by \citet{jabl20072}, most of bulges in our sample, independent of the presence of the bar, are more metal rich, younger and less [Mg/Fe]--enhanced in the central regions. \citet{jabl20072} argued that the outer regions of bulges, which are generally older and more metal poor than central regions consist of the first stars to form, supporting those scenarios in which star formation proceeds from the outer parts inwards. Here, we provide a different explanation and interpret the observed stellar population properties in the central regions of bulges as due to higher (and possibly unsought) contamination level of main disc components in stellar population patterns of bulges.

\subsection{Vertical gradients}
\label{vertical}

Exploring the possible influence of bar on the formation and evolution of bulges is 
not possible without understanding the variations of stellar population 
parameters with radius. Here, we have taken advantage of the spatially resolved 
stellar population maps to study the vertical gradients of the SSP-equivalent age, 
metallicity and [Mg/Fe] in our sample. 

For this purpose we have defined, on the clean (i.e. dust-free) side of our 
galaxies,  a set of pseudo-slits (from $-x_{\rm{B}}$ to $+x_{\rm{B}}$), parallel 
to the major axis of the bulge, at various heights from the major axis. The slits cover the 
whole vertical extent of the bulge from $z=0$ to $z=z_{\rm{B}}$. The outer slits 
(with respect to the major axis) are wider to compensate for the small number of 
contributed data points (Voronoi bins), so that the number of bins within all 
slits is almost the same. The vertical profiles are constructed by finding the 
mean value of stellar population parameter in question in each slit at different 
heights from the disc plane, normalised to $z_{\rm{B}}$. The related errors 
are calculated as the standard error of the means in each pseudo--slit. In order 
to evaluate the vertical gradients in age, metallicity and [Mg/Fe] in our bulges, 
we fit an error-weighted straight line to the individual profiles excluding the 
regions close to disc plane ($z_{\rm{disc}}$), which are most likely affected 
by contamination of dust and/or central components. The slope of the lines 
(vertical gradients) and the related errors for all galaxies in our sample 
are listed in Table~\ref{tab:result}.

\begin{figure*}
	\centering
	\includegraphics[width=0.95\textwidth]{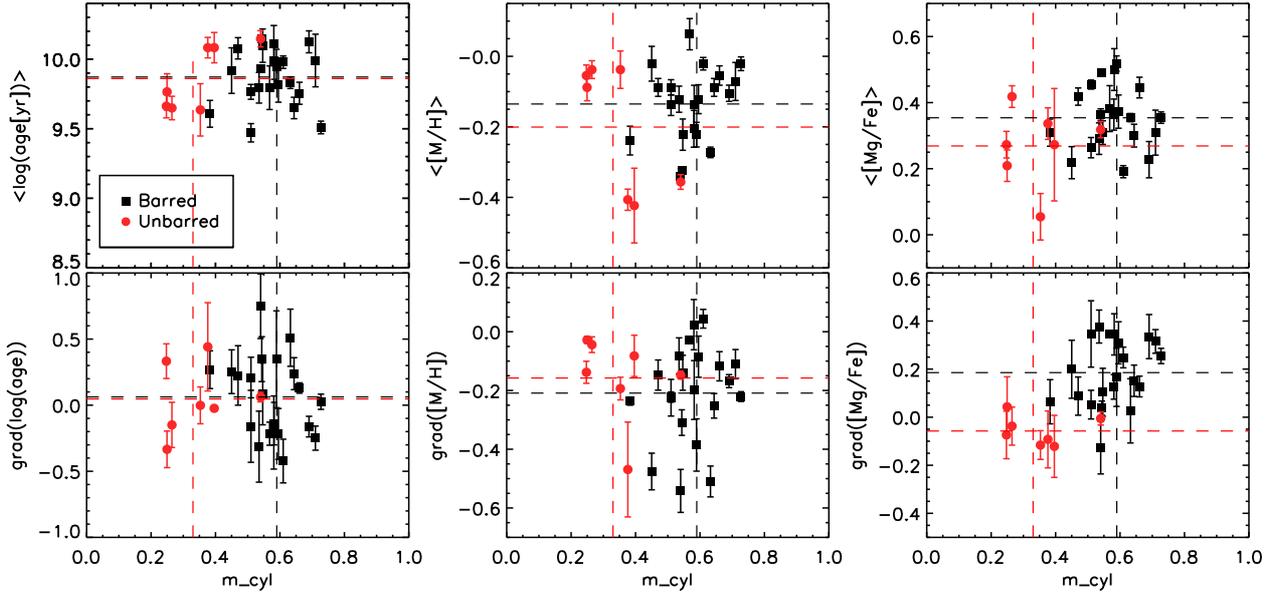}
	\caption{Mean and vertical gradients of the stellar age, metallicity and [Mg/Fe] of bulges as a function of the level of cylindrical rotation ($m_{cyl}$) for all galaxies in our sample. In each panel, the vertical dashed lines indicate the mean level of cylindrical rotation for barred and unbarred bulges, respectively, while the horizontal dashed lines show the averaged stellar population properties (upper panels) and vertical gradients (lower panels), respectively.} 
	\label{fig:cyl_slp_mean}
\end{figure*}

Figure \ref{fig:cyl_slp_mean} demonstrates the means and the vertical gradients of the SSP-age, metallicity and [Mg/Fe] of bulges as a function of level of cylindrical rotation ($m_{cyl}$) for all galaxies in our sample. These results reveal a trend towards steeper positive vertical gradients in [Mg/Fe] for bulges with higher level of cylindrical rotation, which are mostly barred systems (lower right panel). However, as noted earlier, cylindrical rotation cannot be considered a reliable kinematic property to identify the full population of barred galaxies, as this feature strongly depends on bar orientation and galaxy inclination.

\begin{figure*}
\captionsetup[subfigure]{labelformat=empty}
\centering
\begin{subfigure}{0.765\textwidth}
\centering
\includegraphics[width=1.0\textwidth]{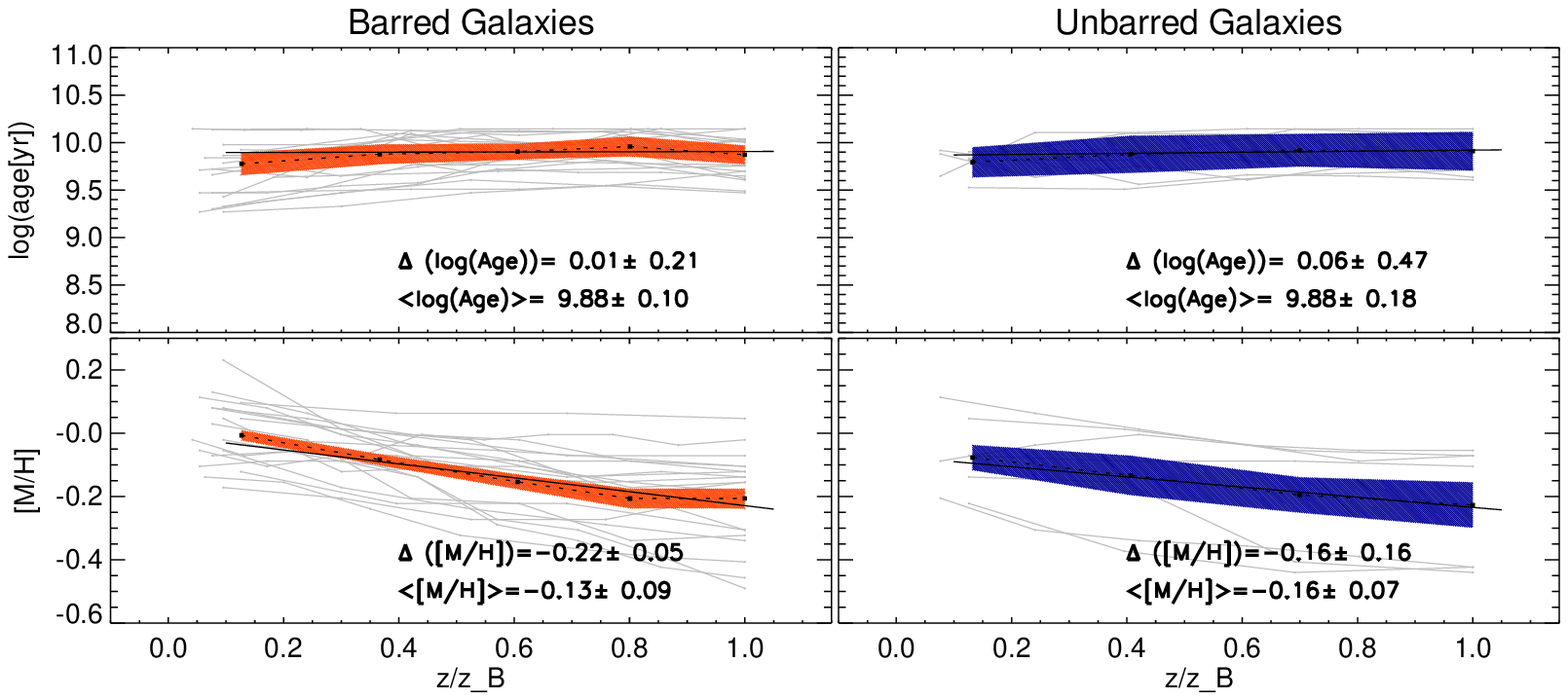}
\end{subfigure}%
\begin{subfigure}{0.235\textwidth}
\centering
\includegraphics[width=1.0\textwidth]{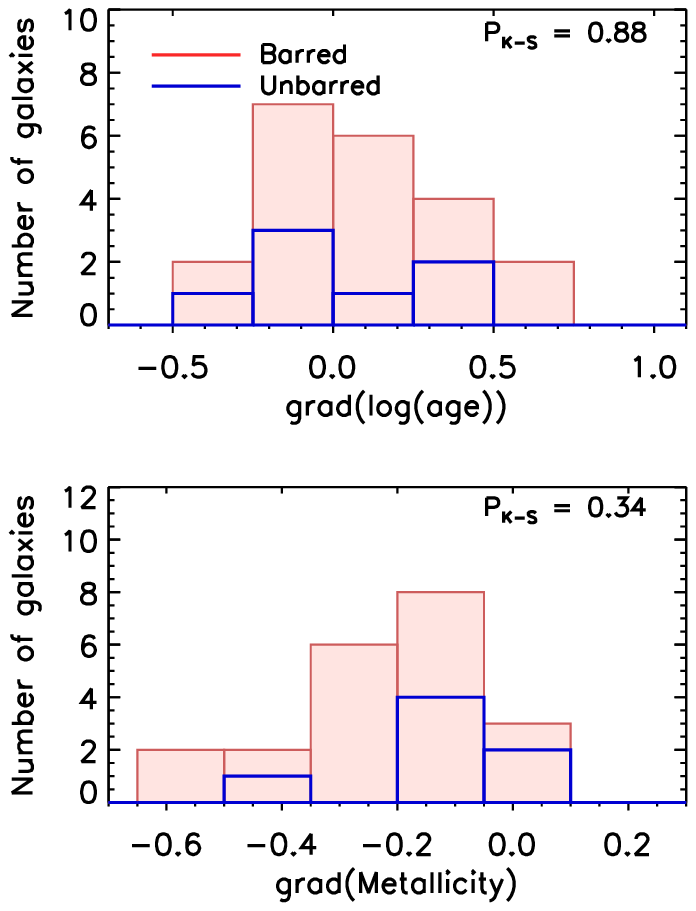}
\end{subfigure}
\centering
\caption{Left panels: The vertical profiles of the SSP equivalent metallicity and age for the 2 classes of bulges in our sample. The grey lines correspond to the individual profiles for each galaxy. For each class of bulges, the individual data points are binned so that each non-overlapping bin contains the same number of data points. Hatched regions indicate the error-weighted averaged profiles and the corresponding uncertainty, following the maximum-likelihood method, in each bin. Lines represent the best linear fits to 
the data sets, while data points belong to the regions, close to the disc plane 
are not included in the fitting process. Right panels: Distribution of the 
vertical gradients in age and metallicity for both classes of bulges in 
our sample. The gradients have been calculated in dex/$z_{\rm{B}}$, where the $z_{\rm{B}}$ is the vertical extent of the bulge/bar in arcsec.}
\label{fig:slp_all}
\end{figure*}

In the left panels of Fig.~\ref{fig:slp_all}, we present the vertical profiles of the SSP equivalent metallicity and age for the 2 classes of bulges in our sample, together with the error-weighted averaged profiles. To derive the averaged profiles, the individual data points are binned so that each non-overlapping bin contains the same number of data points. The error weighted average and the corresponding error in each bin are determined using the maximum-likelihood method \citep[see][]{tayl1982}, so that data points with smaller uncertainties in each bin contributed more to the computed mean than those with larger error bars. The right panels show the distribution of 
average values in each class. The individual SSP-equivalent population profiles, 
along the minor axis for all bulges in our sample are presented in 
Appendix~\ref{app:single_profiles}. The figure shows that the mean age profiles 
for our sample are rather consistent, in mean values and slope, for both barred 
and unbarred galaxies. The metallicity gradient is negative in both cases too, 
although it appears to be somewhat flatter for unbarred systems. Mean values, 
though, are very similar for both types of bulges.

More interestingly, as shown in Fig.~\ref{fig:afe_profile}, the distribution 
and gradients of [Mg/Fe] are remarkably different for barred and unbarred galaxies.
[Mg/Fe] gradients are mostly positive in barred galaxies, while for unbarred ones they are 
mostly flat. A Kolmogorov-Smirnov (K-S) test gives a probability of $10^{-4}$ 
that vertical [Mg/Fe] gradients of the barred and unbarred bulges are drawn from 
the same distribution.\\ 

A major concern, in this kind of studies is that our gradient measurements could be 
severely biased against the physical scale height of the bulges. To test this, 
we have computed the correlation between the [Mg/Fe] gradients scaled to the 
physical size of the bulge along the minor axis, in kpc units and normalised to 
$z_{\rm{B}}$, in arcsec. Our results, not shown here, are very consistent in 
both cases without major differences, except for few cases with relatively small 
bulges. We have also investigated whether the gradients measured in our galaxies 
depend on the potential level of contamination introduced by the projection of 
main disc behind the bulges in those galaxies with lowest inclinations in our 
sample. We have confirmed that the average and slope values obtained for the 
full sample are consistent with those for the least inclined galaxies 
alone.

These results leave us with a very intriguing picture with very similar average 
stellar population values for barred and unbarred galaxies, consistent 
mean age profiles, slightly steeper metallicity gradients in barred galaxies, 
but most notably a drastic different in the vertical [Mg/Fe] gradient between them. In 
the next section we will describe the implications of these results and outline 
possible scenarios that might explain them. We will also compare them 
with those from the closest barred galaxy known: the Milky Way.

\section{Discussion}
\label{sec:discuss}

The results presented here pose a challenge to many theories about the formation 
and evolution of barred galaxies. A common approach to investigate their 
influence on the formation and evolution of bulges is to compare the stellar 
population properties of galaxies with and without bars. If bars follow a 
different evolutionary path, we would expect different abundance patterns in 
bulges of barred and unbarred galaxies.

Most bulges studies in the literature show similar and relatively broad 
distributions in mean SSP-equivalent ages and mostly no vertical gradients in 
age, independent of the bulge classification 
\citep[e.g.][]{falc2002,jabl20072,pele2007, more2008}. This is in agreement with earlier 
findings for early-type galaxies \citep[e.g.][]{mehl2003,sanc2006}. Furthermore, 
the flatness of the vertical age profiles in barred galaxies in our sample could 
in principle support those formation scenarios of bulges in which the bulge 
forms with very little influence from the disc (e.g. mergers or monolithic 
collapse), as expected in bulges of unbarred galaxies. It is worth noting that 
the fact that old stellar populations are observed in bulges of galaxies does not necessarily mean
 that bars are old structures, as bar might have formed recently 
out from old stars, coming from the disc  \citep{sanc2011}. The fact that barred galaxies contain 
old stellar populations is, however, in marked contrast to the internal secular evolution scenario 
expected in these systems, in which the barred bulges are produced through radial and vertical 
transport of disc material (which is typically young), due to instabilities and resonances \citep[see][]{IMV+Shlosman04,moor2006,korm2004}. 
Originally, it was thought that if bulges formed secularly, stars that have been 
scattered furthest from the disc are the oldest stars and therefore a positive 
age gradient is expected \citep{free2008}. Considering the stellar age pattern 
of bulges in our sample, and more specifically barred galaxies, one could 
na\"\i vely conclude that a pure secular evolution scenario via the bar buckling 
process, which is the widely acceptable mechanism to form bulges in barred 
galaxies, cannot explain the observed properties of bulges, because then the stars 
at large scale heights should be younger. 
The only solution we see that bars were formed long time ago, and were not destroyed later on. The disc itself was later rejuvenated with newly formed stars from infalling gas close to the midplane, as we are seeing in our own Milky Way. This leads to these bulges having old ages overall, except in the midplane, where in general younger stars are being seen.

\begin{figure}
	\centering
	\includegraphics[width=0.45\textwidth]{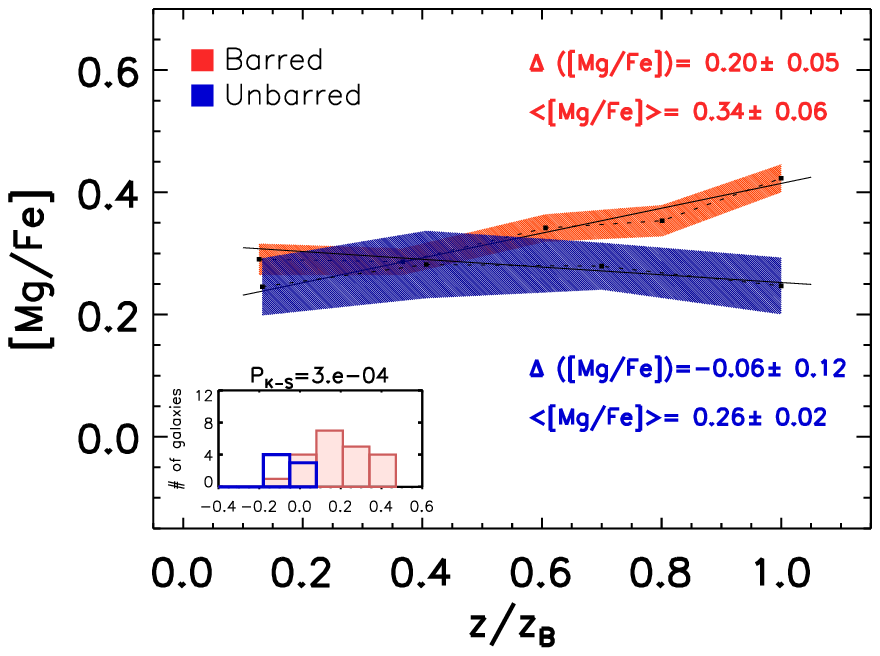}
	\caption{The integrated, error weighted average profiles of the variations of the [Mg/Fe], with increasing height (z) from the disc plane for both barred and unbarred galaxies in our sample. The inset histogram shows the distribution of the vertical gradients in [Mg/Fe] for both classes of bulges in our sample. Gradients have been measured in the range $z_{disc} < z < z_{B}$, normalized by the vertical extent of the bulge ($z_{B}$) in arcsec. $P_{KS}$ gives the probability that the two distributions are drawn from the same populations, as derived from a K-S test.} 
	\label{fig:afe_profile}
\end{figure}

Both sample of bulges show negative metallicity gradients along the minor axis, 
a behaviour that have been formerly interpreted as a direct evidence against the 
secular evolution scenario of bulges. However, more recent numerical models show 
that, the negative metallicity gradient in secularly formed bulges can be 
consistently reproduced, depending on the original disc radial gradients 
\citep{mart2013}, vertical gradients of the composition of two-component disc 
\citep{bekk2011} or both \citep{dima2014}. Other bulge formation scenarios, 
mergers and monolithic collapse, also predict such metallicity gradients 
\citep{egge1962,lars1974,arim1987}. Therefore, the negative metallicity 
gradients along the minor axis of bulges of barred and unbarred galaxies do 
not allow us to single out the bulge formation and evolutionary scenarios.

The most striking feature found in our data is that bulges of barred galaxies are in general more [Mg/Fe]--enhanced at higher latitudes than similar regions of bulges in unbarred systems. This result appears to reveal that both kinds of bulges have different origins. 
Bulges of unbarred galaxies are often denoted as 'classical' or 'elliptical-like' bulges based on the remarkable resemblance they have in many properties (from morphology to stellar populations) to elliptical galaxies. If, as suggested in the literature, they have similar formation and evolution channels, then bulges of unbarred galaxies have been formed through merger of monolithically collapsed clouds at high redshifts \citep{baug1996}, similar to low-luminosity ellipticals \citep[e.g.][]{moor2006}. In that situation the material that forms those bulges comes typically from low-mass systems where the [Mg/Fe] is typically low. While there is some level of enrichment in this process (that is why the typical [Mg/Fe] values are higher than in disc material), the final mixture of all that material is lower than the one they could have if they were made of primordial stars (those that formed quickly at very early epochs).


For barred galaxies, the high [Mg/Fe] abundance ratio found at high latitudes implies that stars at higher z in barred galaxies formed much faster compared to those in the inner regions. The reason why this is not clearly seen in the age gradients, where one may expect a more positive gradient in barred, relative to unbarred systems, is that beyond $\sim$6 Gyr it is more and more difficult to distinguish formation events on time-scales of the order of less than 1Gyr. Our age diagnostic, the H$\beta_{\rm{o}}$ index, will not give us that level of precision with the typical observational uncertainties we have in the H$\beta_{\rm{o}}$ equivalent width \citep[see][for details]{cervantes09}.

Since young populations are not observed at high z in barred galaxies, it is clear that the bar was made of stars that formed long time ago. The alternative is that the bar will selectively push old material vertically from a disc made of young and old stars (like the ones we observed in the nearby Universe), which is not easy to explain. The high [Mg/Fe] observed in those outer regions seem to imply that whatever process that formed these stars with such high [Mg/Fe] must have formed very quickly. We are thus saying that the bars we observed today (at least in our sample) were bars that formed long time ago. Numerical simulations of barred galaxies suggest they are made from disc instabilities and go through a particularly important buckling phase where a lot of material gets pumped-up to higher latitudes above the disc plane \citep[see][]{IMV+Shlosman04,atha2005,deba2006,MV+06,saha2012}. If true, our observations are setting an important constraint on the time that process occurred. The mean age values and especially the positive [Mg/Fe] gradients suggest this process took place very early on in the life of galaxies. These results go along the lines of recent works in the literature pointing at the early formation of bars \citep[see e.g.][]{seidel15,seidel16} and those analysing the evolution of the bar fraction in disc galaxies as a function of redshift \citep[see][]{shet2008,melv2014}.

The Milky Way (MW) is a particular case of a barred system where we can 
directly resolve the distribution of the different kinematic and chemical 
stellar populations. The bulge is known to consist of predominantly old stars 
($\sim$\,10\,Gyr) with a very small fraction (up to 3.5\%) of the stars younger 
than 5\,Gyr \citep{clar2008,clar2011}. The origin of these intermediate-age 
stars is still under debate and different mechanisms have been suggested to 
interpret the observational results \citep[see][for a detailed review of all 
arguments]{gonz2016}. The metallicity distribution function of starts in the 
Galactic bulge is broad, asymmetric \citep{rich1990,lbat1995,zocc2003,zocc2008} 
with a clear vertical gradient of $\sim$\,0.6\,dex/kpc decreasing towards high 
latitudes \citep[e.g.][]{zocc2008,ness2013}.
 The vertical metallicity gradient of the MW bulge is around -0.6 dex/kpc \citep{zocc2008}, -0.06 dex/deg=-0.43 dex/kpc \citep{gonz2013} and -0.45 dex/kpc \citep{ness2013}. In any case, the three measurements are higher than the average value we measured in bulges of barred galaxies ($\nabla$[M/H]=-0.3 dex/kpc). In principle, it could seem that the MW is somehow different to the rest of the galaxies. This difference could be partially due to different method we use to measure the vertical gradients. In case of MW, we are always using the minor axis, but in external galaxies we are averaging over the entire extent of the bulge (including those bulge regions away from the minor axis). In order to understand the difference we have used the simulations by \citet{mart2013} to compute the gradients along the minor axis and also averaging within longitude $-5^{\circ} < l < +5^{\circ}$. The vertical gradients decrease by 28\%, from -0.46 dex/kpc to -0.33 dex/kpc respectively. Moreover, it is worth noting that, when looking at bulges of external galaxies we are integrating foreground and background light, and in the MW we always have a magnitude limit avoiding many background stars. These two explanations can explain the differences with the MW.

 Finally, the $\alpha$-element 
abundances of the MW bulge show that stars with [Fe/H]$<-0.3$ are more 
$\alpha$-enhanced than metal-rich stars ([Fe/H]$>-0.3$) 
\citep{cunh2006,lecu2007,rich2007}. While the measurements in the MW come from 
individual stars, they seem to be consistent with those from unresolved stellar populations 
in our galaxy sample. In this respect, as shown for the stellar kinematics in Paper I, the MW is no different 
from the typical barred galaxy in the nearby Universe.

On the theoretical side, there has been mounting evidence supporting the 
long-lasting nature of bars \citep[e.g.][]{atha2003,MV+06,deba2006,atha2013}. 
More recently Debattista et al. (2016, private communication) studied, in the 
context of the MW analysis, the influence of an evolving long lasting bar on the 
stellar populations in the bulge dominated region, offering a 
somewhat alternative scenario. By means of a pure N--body simulation, they show 
that the stellar populations of disc material that have different initial 
in-plane kinematics separate when a bar forms. The  population that has hotter 
radial kinematics also has hotter vertical kinematics and becomes the vertically 
thick and box-shaped  part of the bar, while the radially cooler populations is 
vertically thin. In this scenario, the declining metallicity profile towards 
higher latitudes and vertically rising abundance ratio in the Galactic bulge are 
interpreted as a result of this separation of initially nearly co-spatial 
populations. This is an interesting evolutionary scenario, but how the original 
superposition of differentiated discs could be formed is still unknown. While this scenario is also consistent with our observations of more distant galaxies, we are not yet in position to robustly disentangle the different chemo-kinematic components of our galaxies, as done in the MW. In fact, while we agree on the final set of observed properties in the galaxies we see today, we provide a different explanation to the way we get to those properties. Current attempts 
to do so are limited to very dramatic cases of large-scale counter-rotating disc 
components \citep[e.g.][]{coccato13,johnston13}. Meanwhile, more chemo-dynamical 
simulations are also necessary to confirm this issue and explain possible 
segregation processes, induced by bars, to produce the vertical gradients in 
stellar population of secularly formed bulges.

\section{Conclusions}
\label{sec:conclusion}

This paper investigates the imprints of bars on the stellar population 
properties of bulges and more specifically, the vertical gradients of stellar 
populations. For this purpose, we have carried out a detailed analysis of the 
stellar age, metallicity and [Mg/Fe] of 28 highly-inclined ($i > 65^{o}$) disc galaxies 
($i > 65^{o}$) , from S0 to S(B)c, observed with the \sauron\ integral field 
spectrograph. The choice of high-inclination galaxies ensures minimal 
contamination by the stellar disc. Following the approach, applied in Paper I, 
the sample is divided into two clean samples of barred (n=21) and unbarred 
galaxies (n=7). Comparing the stellar population properties and vertical 
gradients of bulges in these two classes, we find that while : 

\begin{itemize}
 \item The distribution of the mean stellar age, metallicity and [Mg/Fe] in the 
 bulges of barred and unbarred galaxies are not statistically distinct.
 \item Galaxies in our sample span a wide range of vertical gradients in SSP-equivalent 
 stellar ages and there is no significant difference between barred and unbarred galaxies.
 \item Both classes of bulges present negative metallicity and the gradients are 
not statistically distinct for barred and unbarred galaxies. A similar pattern 
has been reported for the BP bulges of MW.
\end{itemize}

We see a strong difference in the vertical gradient in [Mg/Fe] in barred and unbarred galaxies.
The vertical gradients in [Mg/Fe] for barred galaxies are mostly positive, 
while for unbarred galaxies the profile is almost flat. In other words, 
bulges of barred galaxies are in general more [Mg/Fe]--enhanced at higher 
latitudes than regions, close to disc plane and therefore demonstrate a positive 
[Mg/Fe] gradients along the minor axis.

Such high level of [Mg/Fe] ratio in barred bulges together with the relatively old 
population of bar material, as old as unbarred bulges in our sample, suggests 
that bars are long-lasting structures made of old material. The origin of such 
[Mg/Fe]--enhanced stars is not yet clear, but a simple interpretation 
would be that the buckling phase leading to the formation of the bar took place 
long time ago ($>$\,10\,Gyr). More detailed chemo-dynamical studies are still 
needed to shed more light on this issue. What it seems clear is that the 
positive vertical gradient in [Mg/Fe] is a characteristic feature of barred bulges 
and therefore different evolution mechanisms are required to interpret the 
stellar population differences observed between bulges of barred and unbarred galaxies.


\section*{Acknowledgments}
A.M. wishes to thank the School of Astronomy, IPM and the Iranian National 
Observatory (INO) for providing support while working on this paper. A.M. also 
acknowledges the Isaac Newton Group of Telescopes (ING), the Instituto de 
Astrof\'isica de Canarias (IAC) and the Kapteyn Astronomical Institute for hospitality and support while this paper was 
in progress. The authors acknowledge support from the Spanish Ministry of 
Economy and Competitiveness (MINECO) through grants AYA2009-11137, 
AYA2016-77237-C3-1-P and AYA2014-58308-P. Funding for SDSS-III has been provided 
by the Alfred P. Sloan Foundation, the Participating Institutions, the National 
Science Foundation, and the U.S. Department of Energy Office of Science. The 
SDSS-III web site is http://www.sdss3.org/. 

\begin{landscape}
	\begin{table}
		\caption {Linear fit age, metallicity and [Mg/Fe] gradients to the bulge region and the mean values within the bulge region.}
		\label{tab:result}
		\begin{center}
			\begin{tabular}{lcrrrrrrrrr}
				\hline
				Galaxy  & Bar & $\langle$age$\rangle$ & $\langle$[M/H]$\rangle$ & $\langle$[Mg/Fe]$\rangle$ &    $\nabla$log(age)    &     $\nabla$[M/H]     &   $\nabla$[Mg/Fe] &    $\nabla$log(age)    &     $\nabla$[M/H]     &   $\nabla$[Mg/Fe]   \\
				~       & ~   &   (Gyr)  & (dex)                   &   (dex)                    &    (dex/$z_{\rm{B}}$)                &     (dex/$z_{\rm{B}}$)                &   (dex/$z_{\rm{B}}$)   &    (dex/kpc)                &     (dex/kpc)                &   (dex/kpc)             \\
				(1)     & (2) &  (3) & (4) & (5) & (6) & (7) & (8) & (9) & (10) & (11) \\
				\hline
										NGC3098 & unbarred & $14.00 \pm 0.61$ & $-0.36 \pm 0.04$ &  $0.32 \pm 0.01$ & $ 0.06 \pm 0.04$ & $-0.15 \pm 0.01$ & $-0.01 \pm 0.03$ &  $  0.08  \pm  0.04$  &   $ -0.19  \pm 0.02$  &   $ -0.01  \pm 0.04$  \\  
										NGC4026 & BP       & $14.00 \pm 3.04$ & $-0.32 \pm 0.10$ &  $0.49 \pm 0.02$ & $ 0.35 \pm 0.17$ & $-0.31 \pm 0.04$ & $ 0.04 \pm 0.03$ &  $  0.54  \pm  0.21$  &   $ -0.48  \pm 0.07$  &   $  0.06  \pm 0.05$  \\  
										NGC4036 & unbarred & $ 4.45 \pm 0.99$ & $-0.04 \pm 0.02$ &  $0.42 \pm 0.05$ & $-0.15 \pm 0.17$ & $-0.04 \pm 0.03$ & $-0.04 \pm 0.08$ &  $ -0.19  \pm  0.21$  &   $ -0.06  \pm 0.03$  &   $ -0.05  \pm 0.10$  \\  
										NGC4179 & BP       & $ 8.82 \pm 3.97$ & $-0.22 \pm 0.13$ &  $0.52 \pm 0.08$ & $ 0.35 \pm 0.37$ & $-0.39 \pm 0.09$ & $ 0.17 \pm 0.12$ &  $  0.49  \pm  0.47$  &   $ -0.54  \pm 0.13$  &   $  0.23  \pm 0.17$  \\  
										NGC4251 & BP       & $ 3.23 \pm 0.22$ & $-0.02 \pm 0.07$ &  $0.35 \pm 0.08$ & $ 0.03 \pm 0.06$ & $-0.22 \pm 0.02$ & $ 0.25 \pm 0.03$ &  $  0.04  \pm  0.09$  &   $ -0.38  \pm 0.03$  &   $  0.43  \pm 0.05$  \\  
										NGC4270 & Bar      & $12.91 \pm 3.03$ & $-0.21 \pm 0.04$ &  $0.36 \pm 0.05$ & $-0.21 \pm 0.28$ & $ 0.02 \pm 0.09$ & $ 0.13 \pm 0.05$ &  $ -0.20  \pm  0.27$  &   $  0.02  \pm 0.08$  &   $  0.12  \pm 0.05$  \\  
										NGC4346 & Bar      & $ 6.77 \pm 3.04$ & $-0.27 \pm 0.16$ &  $0.35 \pm 0.10$ & $ 0.51 \pm 0.22$ & $-0.51 \pm 0.05$ & $ 0.03 \pm 0.13$ &  $  0.78  \pm  0.26$  &   $ -0.78  \pm 0.08$  &   $  0.04  \pm 0.20$  \\  
										NGC4425 & BP       & $ 6.23 \pm 1.67$ & $-0.12 \pm 0.03$ &  $0.29 \pm 0.12$ & $-0.31 \pm 0.27$ & $-0.08 \pm 0.06$ & $ 0.38 \pm 0.07$ &  $ -0.50  \pm  0.39$  &   $ -0.13  \pm 0.10$  &   $  0.60  \pm 0.11$  \\  
										NGC4435 & Bar      & $ 5.82 \pm 1.51$ & $-0.09 \pm 0.06$ &  $0.45 \pm 0.10$ & $-0.16 \pm 0.27$ & $-0.22 \pm 0.06$ & $ 0.35 \pm 0.14$ &  $ -0.34  \pm  0.55$  &   $ -0.48  \pm 0.13$  &   $  0.75  \pm 0.30$  \\  
										NGC4461 & Bar      & $ 9.77 \pm 1.55$ & $-0.14 \pm 0.08$ &  $0.50 \pm 0.11$ & $-0.14 \pm 0.16$ & $-0.20 \pm 0.10$ & $ 0.34 \pm 0.09$ &  $ -0.16  \pm  0.18$  &   $ -0.22  \pm 0.11$  &   $  0.39  \pm 0.10$  \\  
										NGC4474 & Bar      & $12.50 \pm 2.37$ & $-0.22 \pm 0.05$ &  $0.31 \pm 0.05$ & $ 0.08 \pm 0.23$ & $-0.14 \pm 0.06$ & $ 0.10 \pm 0.10$ &  $  0.16  \pm  0.42$  &   $ -0.26  \pm 0.11$  &   $  0.19  \pm 0.19$  \\  
										NGC4521 & Bar      & $ 9.77 \pm 1.61$ & $-0.07 \pm 0.04$ &  $0.31 \pm 0.10$ & $-0.25 \pm 0.09$ & $-0.11 \pm 0.05$ & $ 0.32 \pm 0.05$ &  $ -0.30  \pm  0.10$  &   $ -0.13  \pm 0.06$  &   $  0.38  \pm 0.06$  \\  
										NGC4710 & BP       & $ 8.55 \pm 2.72$ & $-0.34 \pm 0.13$ &  $0.36 \pm 0.06$ & $ 0.75 \pm 0.24$ & $-0.54 \pm 0.07$ & $-0.13 \pm 0.11$ &  $  0.89  \pm  0.27$  &   $ -0.64  \pm 0.09$  &   $ -0.15  \pm 0.13$  \\  
										NGC4762 & Bar      & $ 5.68 \pm 0.48$ & $-0.05 \pm 0.04$ &  $0.45 \pm 0.04$ & $ 0.13 \pm 0.04$ & $-0.12 \pm 0.05$ & $ 0.13 \pm 0.05$ &  $  0.58  \pm  0.04$  &   $ -0.52  \pm 0.05$  &   $  0.37  \pm 0.05$  \\  
										NGC5103 & unbarred & $12.09 \pm 3.84$ & $-0.41 \pm 0.16$ &  $0.34 \pm 0.05$ & $ 0.44 \pm 0.33$ & $-0.47 \pm 0.16$ & $-0.09 \pm 0.12$ &  $  1.05  \pm  0.70$  &   $ -1.11  \pm 0.38$  &   $ -0.22  \pm 0.28$  \\  
										NGC5326 & unbarred & $ 4.32 \pm 0.80$ & $-0.04 \pm 0.06$ &  $0.05 \pm 0.05$ & $-0.00 \pm 0.14$ & $-0.19 \pm 0.04$ & $-0.12 \pm 0.06$ &  $ -0.00  \pm  0.11$  &   $ -0.13  \pm 0.03$  &   $ -0.08  \pm 0.04$  \\  
										NGC5353 & BP       & $ 6.23 \pm 1.08$ & $ 0.06 \pm 0.01$ &  $0.38 \pm 0.10$ & $-0.22 \pm 0.09$ & $-0.03 \pm 0.01$ & $ 0.35 \pm 0.01$ &  $ -0.29  \pm  0.10$  &   $ -0.04  \pm 0.02$  &   $  0.47  \pm 0.01$  \\  
										NGC5422 & BP       & $13.32 \pm 1.56$ & $-0.10 \pm 0.05$ &  $0.23 \pm 0.11$ & $-0.16 \pm 0.08$ & $-0.17 \pm 0.02$ & $ 0.34 \pm 0.09$ &  $ -0.16  \pm  0.08$  &   $ -0.17  \pm 0.02$  &   $  0.34  \pm 0.09$  \\  
										NGC5475 & unbarred & $ 4.59 \pm 0.98$ & $-0.05 \pm 0.04$ &  $0.27 \pm 0.05$ & $ 0.33 \pm 0.13$ & $-0.14 \pm 0.04$ & $-0.07 \pm 0.10$ &  $  0.59  \pm  0.17$  &   $ -0.25  \pm 0.07$  &   $ -0.13  \pm 0.18$  \\  
										NGC5574 & Bar      & $ 2.95 \pm 0.56$ & $-0.14 \pm 0.07$ &  $0.26 \pm 0.03$ & $ 0.21 \pm 0.15$ & $-0.22 \pm 0.02$ & $ 0.05 \pm 0.06$ &  $  0.30  \pm  0.20$  &   $ -0.32  \pm 0.03$  &   $  0.08  \pm 0.09$  \\  
										NGC5611 & unbarred & $12.09 \pm 0.24$ & $-0.42 \pm 0.03$ &  $0.27 \pm 0.04$ & $-0.02 \pm 0.01$ & $-0.08 \pm 0.07$ & $-0.12 \pm 0.13$ &  $ -0.04  \pm  0.02$  &   $ -0.12  \pm 0.10$  &   $ -0.18  \pm 0.15$  \\  
										NGC5689 & BP       & $ 4.45 \pm 0.85$ & $-0.09 \pm 0.08$ &  $0.30 \pm 0.07$ & $ 0.24 \pm 0.12$ & $-0.25 \pm 0.04$ & $ 0.15 \pm 0.07$ &  $  0.14  \pm  0.10$  &   $ -0.15  \pm 0.03$  &   $  0.09  \pm 0.04$  \\  
										NGC5707 & unbarred & $ 5.82 \pm 1.68$ & $-0.09 \pm 0.01$ &  $0.21 \pm 0.05$ & $-0.33 \pm 0.14$ & $-0.03 \pm 0.01$ & $ 0.04 \pm 0.12$ &  $ -0.44  \pm  0.16$  &   $ -0.04  \pm 0.02$  &   $  0.06  \pm 0.17$  \\  
										NGC5746 & BP       & $ 9.50 \pm 2.98$ & $-0.02 \pm 0.03$ &  $0.19 \pm 0.07$ & $-0.42 \pm 0.17$ & $ 0.04 \pm 0.03$ & $ 0.25 \pm 0.04$ &  $ -0.24  \pm  0.14$  &   $  0.03  \pm 0.02$  &   $  0.14  \pm 0.02$  \\  
										NGC5838 & Bar       & $11.95 \pm 2.84$ & $-0.09 \pm 0.05$ &  $0.42 \pm 0.05$ & $ 0.22 \pm 0.22$ & $-0.15 \pm 0.05$ & $ 0.09 \pm 0.08$ & $  0.29  \pm  0.27$  &   $ -0.19  \pm 0.07$  &   $  0.11  \pm 0.10$  \\  
										NGC5854 & BP       & $ 4.05 \pm 0.87$ & $-0.24 \pm 0.07$ &  $0.31 \pm 0.05$ & $ 0.27 \pm 0.14$ & $-0.24 \pm 0.01$ & $ 0.06 \pm 0.09$ &  $  0.36  \pm  0.17$  &   $ -0.31  \pm 0.02$  &   $  0.09  \pm 0.12$  \\  
										NGC5864 & BP       & $ 6.50 \pm 1.53$ & $-0.12 \pm 0.04$ &  $0.37 \pm 0.10$ & $-0.22 \pm 0.19$ & $-0.09 \pm 0.07$ & $ 0.31 \pm 0.09$ &  $ -0.26  \pm  0.22$  &   $ -0.10  \pm 0.08$  &   $  0.37  \pm 0.10$  \\  
										NGC6010 & Bar       & $ 8.27 \pm 1.88$ & $-0.02 \pm 0.15$ &  $0.22 \pm 0.09$ & $ 0.25 \pm 0.17$ & $-0.48 \pm 0.06$ & $ 0.20 \pm 0.12$ & $  0.29  \pm  0.18$  &   $ -0.54  \pm 0.07$  &   $  0.23  \pm 0.14$  \\
				\hline
			\end{tabular}
		\end{center}
		\begin{flushleft}
			\small NOTES: 
			(1) Galaxy name.
			(2) Bar presence based on our kinematic analysis as in Paper I. 
			(3), (4) and (5) Mean SSP-equivalent age, metallicity and [Mg/Fe] within the bulge analysis window, respectively.
            (6), (7) and (8) Vertical gradients in age, metallicity and [Mg/Fe]. Gradients calculated in the range
			$z_{disc} < z < z_{B}$, normalized by the vertical extent of the bulge ($z_{B}$) in arcsec. (9), (10) and (11) Vertical gradients in age, metallicity and [Mg/Fe], expressed in physical units (dex/kpc).
		\end{flushleft}
	\end{table}
\end{landscape}

\bibliographystyle{mn2e}
\bibliography{mybib}

\begin{thebibliography}{}

\bibitem[\protect\citeauthoryear{{Ahn}, {Alexandroff}, {Allende Prieto},
  {Anders}, {Anderson}, {Anderton}, {Andrews}, {Aubourg}, {Bailey}, {Bastien}
  \& et al.}{{Ahn} et~al.}{2014}]{ahn2014}
{Ahn} C.~P.,  {Alexandroff} R.,  {Allende Prieto} C.,  {Anders} F.,  {Anderson}
  S.~F.,  {Anderton} T.,  {Andrews} B.~H.,  {Aubourg} {\'E}.,  {Bailey} S.,
  {Bastien} F.~A.,    et al. 2014, \apjs, 211, 17

\bibitem[\protect\citeauthoryear{{Arimoto} \& {Yoshii}}{{Arimoto} \&
  {Yoshii}}{1987}]{arim1987}
{Arimoto} N.,  {Yoshii} Y.,  1987, \aap, 173, 23

\bibitem[\protect\citeauthoryear{{Athanassoula}}{{Athanassoula}}{2003}]{atha2003}
{Athanassoula} E.,  2003, \mnras, 341, 1179

\bibitem[\protect\citeauthoryear{{Athanassoula}}{{Athanassoula}}{2005}]{atha2005}
{Athanassoula} E.,  2005, \mnras, 358, 1477

\bibitem[\protect\citeauthoryear{{Athanassoula}}{{Athanassoula}}{2013}]{athan2013}
{Athanassoula} E.,  2013, {Bars and secular evolution in disk galaxies:
  Theoretical input}.
p.~305

\bibitem[\protect\citeauthoryear{{Athanassoula}}{{Athanassoula}}{2016}]{atha2016}
{Athanassoula} E.,  2016, Galactic Bulges, 418, 391

\bibitem[\protect\citeauthoryear{{Athanassoula}, {Machado} \&
  {Rodionov}}{{Athanassoula} et~al.}{2013}]{atha2013}
{Athanassoula} E.,  {Machado} R.~E.~G.,    {Rodionov} S.~A.,  2013, \mnras,
  429, 1949

\bibitem[\protect\citeauthoryear{{Bacon}, {Copin}, {Monnet}, {Miller},
  {Allington-Smith}, {Bureau}, {Carollo}, {Davies}, {Emsellem}, {Kuntschner},
  {Peletier}, {Verolme} \& {de Zeeuw}}{{Bacon} et~al.}{2001}]{baco2001}
{Bacon} R.,  {Copin} Y.,  {Monnet} G.,  {Miller} B.~W.,  {Allington-Smith}
  J.~R.,  {Bureau} M.,  {Carollo} C.~M.,  {Davies} R.~L.,  {Emsellem} E.,
  {Kuntschner} H.,  {Peletier} R.~F.,  {Verolme} E.~K.,    {de Zeeuw} P.~T.,
  2001, \mnras, 326, 23

\bibitem[\protect\citeauthoryear{{Baugh}, {Cole} \& {Frenk}}{{Baugh}
  et~al.}{1996}]{baug1996}
{Baugh} C.~M.,  {Cole} S.,    {Frenk} C.~S.,  1996, \mnras, 283, 1361

\bibitem[\protect\citeauthoryear{{Bekki} \& {Tsujimoto}}{{Bekki} \&
  {Tsujimoto}}{2011}]{bekk2011}
{Bekki} K.,  {Tsujimoto} T.,  2011, \mnras, 416, L60

\bibitem[\protect\citeauthoryear{{Bureau} \& {Athanassoula}}{{Bureau} \&
  {Athanassoula}}{2005}]{bure2005}
{Bureau} M.,  {Athanassoula} E.,  2005, \apj, 626, 159

\bibitem[\protect\citeauthoryear{{Bureau}, {Athanassoula}, {Chung} \&
  {Aronica}}{{Bureau} et~al.}{2004}]{bure2004}
{Bureau} M.,  {Athanassoula} E.,  {Chung} A.,    {Aronica} G.,  2004, in
  {Block} D.~L.,  {Puerari} I.,  {Freeman} K.~C.,  {Groess} R.,   {Block}
  E.~K.,  eds, Penetrating Bars Through Masks of Cosmic Dust Vol.~319 of
  Astrophysics and Space Science Library, {Bar-Driven Evolution and 2D
  Spectroscopy of Bulges}.
p.~139

\bibitem[\protect\citeauthoryear{{Bureau} \& {Freeman}}{{Bureau} \&
  {Freeman}}{1999}]{bure1999}
{Bureau} M.,  {Freeman} K.~C.,  1999, \aj, 118, 126

\bibitem[\protect\citeauthoryear{{Cappellari} \& {Copin}}{{Cappellari} \&
  {Copin}}{2003}]{2003MNRAS.342..345C}
{Cappellari} M.,  {Copin} Y.,  2003, \mnras, 342, 345

\bibitem[\protect\citeauthoryear{{Cappellari}, {Emsellem}, {Krajnovi{\'c}},
  {McDermid}, {Scott}, {Verdoes Kleijn} \& et al.}{{Cappellari}
  et~al.}{2011}]{capp2011}
{Cappellari} M.,  {Emsellem} E.,  {Krajnovi{\'c}} D.,  {McDermid} R.~M.,
  {Scott} N.,  {Verdoes Kleijn} G.~A.,    et al. 2011, \mnras, 413, 813

\bibitem[\protect\citeauthoryear{{Cappellari}, {Scott}, {Alatalo}, {Blitz},
  {Bois}, {Bournaud}, {Bureau} \& {Crocker}}{{Cappellari}
  et~al.}{2013}]{capp2013}
{Cappellari} M.,  {Scott} N.,  {Alatalo} K.,  {Blitz} L.,  {Bois} M.,
  {Bournaud} F.,  {Bureau} M.,    {Crocker} A.~F.,  2013, \mnras, 432, 1709

\bibitem[\protect\citeauthoryear{{Cappellari}, {Scott}, {Alatalo}, {Blitz},
  {Bois}, {Bournaud}, {Bureau}, {Crocker}, {Davies} \& {Davis}}{{Cappellari}
  et~al.}{2013}]{capp2012}
{Cappellari} M.,  {Scott} N.,  {Alatalo} K.,  {Blitz} L.,  {Bois} M.,
  {Bournaud} F.,  {Bureau} M.,  {Crocker} A.~F.,  {Davies} R.~L.,    {Davis}
  T.~A.,  2013, \mnras, 432, 1709

\bibitem[\protect\citeauthoryear{{Cervantes} \& {Vazdekis}}{{Cervantes} \&
  {Vazdekis}}{2009}]{cervantes09}
{Cervantes} J.~L.,  {Vazdekis} A.,  2009, \mnras, 392, 691

\bibitem[\protect\citeauthoryear{{Chung} \& {Bureau}}{{Chung} \&
  {Bureau}}{2004}]{chun2004}
{Chung} A.,  {Bureau} M.,  2004, \aj, 127, 3192

\bibitem[\protect\citeauthoryear{{Clarkson}, {Sahu}, {Anderson}, {Smith},
  {Brown}, {Rich}, {Casertano}, {Bond}, {Livio}, {Minniti}, {Panagia},
  {Renzini}, {Valenti} \& {Zoccali}}{{Clarkson} et~al.}{2008}]{clar2008}
{Clarkson} W.,  {Sahu} K.,  {Anderson} J.,  {Smith} T.~E.,  {Brown} T.~M.,
  {Rich} R.~M.,  {Casertano} S.,  {Bond} H.~E.,  {Livio} M.,  {Minniti} D.,
  {Panagia} N.,  {Renzini} A.,  {Valenti} J.,    {Zoccali} M.,  2008, \apj,
  684, 1110

\bibitem[\protect\citeauthoryear{{Clarkson}, {Sahu}, {Anderson}, {Rich},
  {Smith}, {Brown}, {Bond}, {Livio}, {Minniti}, {Renzini} \&
  {Zoccali}}{{Clarkson} et~al.}{2011}]{clar2011}
{Clarkson} W.~I.,  {Sahu} K.~C.,  {Anderson} J.,  {Rich} R.~M.,  {Smith} T.~E.,
   {Brown} T.~M.,  {Bond} H.~E.,  {Livio} M.,  {Minniti} D.,  {Renzini} A.,
  {Zoccali} M.,  2011, \apj, 735, 37

\bibitem[\protect\citeauthoryear{{Coccato}, {Morelli}, {Pizzella}, {Corsini},
  {Buson} \& {Dalla Bont{\`a}}}{{Coccato} et~al.}{2013}]{coccato13}
{Coccato} L.,  {Morelli} L.,  {Pizzella} A.,  {Corsini} E.~M.,  {Buson} L.~M.,
    {Dalla Bont{\`a}} E.,  2013, \aap, 549, A3

\bibitem[\protect\citeauthoryear{{Coelho}, {Barbuy}, {Mel{\'e}ndez}, {Schiavon}
  \& {Castilho}}{{Coelho} et~al.}{2005}]{coel2005}
{Coelho} P.,  {Barbuy} B.,  {Mel{\'e}ndez} J.,  {Schiavon} R.~P.,    {Castilho}
  B.~V.,  2005, \aap, 443, 735

\bibitem[\protect\citeauthoryear{{Coelho}, {Bruzual}, {Charlot}, {Weiss},
  {Barbuy} \& {Ferguson}}{{Coelho} et~al.}{2007}]{coel2007}
{Coelho} P.,  {Bruzual} G.,  {Charlot} S.,  {Weiss} A.,  {Barbuy} B.,
  {Ferguson} J.~W.,  2007, \mnras, 382, 498

\bibitem[\protect\citeauthoryear{{Coelho} \& {Gadotti}}{{Coelho} \&
  {Gadotti}}{2011}]{coel2011}
{Coelho} P.,  {Gadotti} D.~A.,  2011, \apjl, 743, L13

\bibitem[\protect\citeauthoryear{{Cunha} \& {Smith}}{{Cunha} \&
  {Smith}}{2006}]{cunh2006}
{Cunha} K.,  {Smith} V.~V.,  2006, \apj, 651, 491

\bibitem[\protect\citeauthoryear{{Debattista}, {Mayer}, {Carollo}, {Moore},
  {Wadsley} \& {Quinn}}{{Debattista} et~al.}{2006}]{deba2006}
{Debattista} V.~P.,  {Mayer} L.,  {Carollo} C.~M.,  {Moore} B.,  {Wadsley} J.,
    {Quinn} T.,  2006, \apj, 645, 209

\bibitem[\protect\citeauthoryear{{Di Matteo}, {Haywood}, {G{\'o}mez}, {van
  Damme}, {Combes}, {Hall{\'e}}, {Semelin}, {Lehnert} \& {Katz}}{{Di Matteo}
  et~al.}{2014}]{dima2014}
{Di Matteo} P.,  {Haywood} M.,  {G{\'o}mez} A.,  {van Damme} L.,  {Combes} F.,
  {Hall{\'e}} A.,  {Semelin} B.,  {Lehnert} M.~D.,    {Katz} D.,  2014, \aap,
  567, A122

\bibitem[\protect\citeauthoryear{{Eggen}, {Lynden-Bell} \& {Sandage}}{{Eggen}
  et~al.}{1962}]{egge1962}
{Eggen} O.~J.,  {Lynden-Bell} D.,    {Sandage} A.~R.,  1962, \apj, 136, 748

\bibitem[\protect\citeauthoryear{{Eskridge}, {Frogel}, {Pogge}, {Quillen},
  {Davies}, {DePoy}, {Houdashelt}, {Kuchinski}, {Ram{\'{\i}}rez}, {Sellgren},
  {Terndrup} \& {Tiede}}{{Eskridge} et~al.}{2000}]{eskr2000}
{Eskridge} P.~B.,  {Frogel} J.~A.,  {Pogge} R.~W.,  {Quillen} A.~C.,  {Davies}
  R.~L.,  {DePoy} D.~L.,  {Houdashelt} M.~L.,  {Kuchinski} L.~E.,
  {Ram{\'{\i}}rez} S.~V.,  {Sellgren} K.,  {Terndrup} D.~M.,    {Tiede} G.~P.,
  2000, \aj, 119, 536

\bibitem[\protect\citeauthoryear{{Falc{\'o}n-Barroso}, {Bacon}, {Bureau},
  {Cappellari}, {Davies}, {de Zeeuw}, {Emsellem}, {Fathi}, {Krajnovi{\'c}},
  {Kuntschner}, {McDermid}, {Peletier} \& {Sarzi}}{{Falc{\'o}n-Barroso}
  et~al.}{2006}]{falc2006}
{Falc{\'o}n-Barroso} J.,  {Bacon} R.,  {Bureau} M.,  {Cappellari} M.,  {Davies}
  R.~L.,  {de Zeeuw} P.~T.,  {Emsellem} E.,  {Fathi} K.,  {Krajnovi{\'c}} D.,
  {Kuntschner} H.,  {McDermid} R.~M.,  {Peletier} R.~F.,    {Sarzi} M.,  2006,
  \mnras, 369, 529

\bibitem[\protect\citeauthoryear{{Falc{\'o}n-Barroso}, {Peletier} \&
  {Balcells}}{{Falc{\'o}n-Barroso} et~al.}{2002}]{falc2002}
{Falc{\'o}n-Barroso} J.,  {Peletier} R.~F.,    {Balcells} M.,  2002, \mnras,
  335, 741

\bibitem[\protect\citeauthoryear{{Freeman}}{{Freeman}}{2008}]{free2008}
{Freeman} K.~C.,  2008, in {Bureau} M.,  {Athanassoula} E.,   {Barbuy} B.,
  eds, Formation and Evolution of Galaxy Bulges Vol.~245 of IAU Symposium,
  {Galactic bulges: overview}.
pp 3--10

\bibitem[\protect\citeauthoryear{{Gonzalez} \& {Gadotti}}{{Gonzalez} \&
  {Gadotti}}{2016}]{gonz2016}
{Gonzalez} O.~A.,  {Gadotti} D.,  2016, Galactic Bulges, 418, 199

\bibitem[\protect\citeauthoryear{{Gonzalez}, {Rejkuba}, {Zoccali}, {Valent},
  {Minniti} \& {Tobar}}{{Gonzalez} et~al.}{2013}]{gonz2013}
{Gonzalez} O.~A.,  {Rejkuba} M.,  {Zoccali} M.,  {Valent} E.,  {Minniti} D.,
  {Tobar} R.,  2013, \aap, 552, A110

\bibitem[\protect\citeauthoryear{{Grosb{\o}l}, {Patsis} \&
  {Pompei}}{{Grosb{\o}l} et~al.}{2004}]{grosb2004}
{Grosb{\o}l} P.,  {Patsis} P.~A.,    {Pompei} E.,  2004, \aap, 423, 849

\bibitem[\protect\citeauthoryear{{Iannuzzi} \& {Athanassoula}}{{Iannuzzi} \&
  {Athanassoula}}{2015}]{lann2015}
{Iannuzzi} F.,  {Athanassoula} E.,  2015, \mnras, 450, 2514

\bibitem[\protect\citeauthoryear{{Ibata} \& {Gilmore}}{{Ibata} \&
  {Gilmore}}{1995}]{lbat1995}
{Ibata} R.~A.,  {Gilmore} G.~F.,  1995, \mnras, 275, 605

\bibitem[\protect\citeauthoryear{{Jablonka}, {Gorgas} \&
  {Goudfrooij}}{{Jablonka} et~al.}{2007}]{jabl20072}
{Jablonka} P.,  {Gorgas} J.,    {Goudfrooij} P.,  2007, \aap, 474, 763

\bibitem[\protect\citeauthoryear{{Johnston}, {Merrifield},
  {Arag{\'o}n-Salamanca} \& {Cappellari}}{{Johnston} et~al.}{2013}]{johnston13}
{Johnston} E.~J.,  {Merrifield} M.~R.,  {Arag{\'o}n-Salamanca} A.,
  {Cappellari} M.,  2013, \mnras, 428, 1296

\bibitem[\protect\citeauthoryear{{Knapen}, {Shlosman} \& {Peletier}}{{Knapen}
  et~al.}{2000}]{knapen2000}
{Knapen} J.~H.,  {Shlosman} I.,    {Peletier} R.~F.,  2000, \apj, 529, 93

\bibitem[\protect\citeauthoryear{{Kormendy}}{{Kormendy}}{2013}]{korm2013}
{Kormendy} J.,  2013, {Secular Evolution in Disk Galaxies}.
p.~1

\bibitem[\protect\citeauthoryear{{Kormendy} \& {Illingworth}}{{Kormendy} \&
  {Illingworth}}{1982}]{korm1982}
{Kormendy} J.,  {Illingworth} G.,  1982, \apj, 256, 460

\bibitem[\protect\citeauthoryear{{Kormendy} \& {Kennicutt} Jr.}{{Kormendy} \&
  {Kennicutt}}{2004}]{korm2004}
{Kormendy} J.,  {Kennicutt} Jr. R.~C.,  2004, \araa, 42, 603

\bibitem[\protect\citeauthoryear{{Krajnovi{\'c}}, {Emsellem}, {Cappellari},
  {Alatalo}, {Blitz}, {Bois}, {Bournaud}, {Bureau}, {Davies} \&
  {Davis}}{{Krajnovi{\'c}} et~al.}{2011}]{kraj2011}
{Krajnovi{\'c}} D.,  {Emsellem} E.,  {Cappellari} M.,  {Alatalo} K.,  {Blitz}
  L.,  {Bois} M.,  {Bournaud} F.,  {Bureau} M.,  {Davies} R.~L.,    {Davis}
  T.~A.,  2011, \mnras, 414, 2923

\bibitem[\protect\citeauthoryear{{Kuijken} \& {Merrifield}}{{Kuijken} \&
  {Merrifield}}{1995}]{kuij1995}
{Kuijken} K.,  {Merrifield} M.~R.,  1995, \apjl, 443, L13

\bibitem[\protect\citeauthoryear{{Larson}}{{Larson}}{1974}]{lars1974}
{Larson} R.~B.,  1974, \mnras, 166, 585

\bibitem[\protect\citeauthoryear{{Lecureur}, {Hill}, {Zoccali}, {Barbuy},
  {G{\'o}mez}, {Minniti}, {Ortolani} \& {Renzini}}{{Lecureur}
  et~al.}{2007}]{lecu2007}
{Lecureur} A.,  {Hill} V.,  {Zoccali} M.,  {Barbuy} B.,  {G{\'o}mez} A.,
  {Minniti} D.,  {Ortolani} S.,    {Renzini} A.,  2007, \aap, 465, 799

\bibitem[\protect\citeauthoryear{{Marinova} \& {Jogee}}{{Marinova} \&
  {Jogee}}{2007}]{marin2007}
{Marinova} I.,  {Jogee} S.,  2007, \apj, 659, 1176

\bibitem[\protect\citeauthoryear{{Mart{\'{\i}}n-Navarro}}{{Mart{\'{\i}}n-Navarro}}{2016}]{mart2016}
{Mart{\'{\i}}n-Navarro} I.,  2016, \mnras, 456, L104

\bibitem[\protect\citeauthoryear{{Martinez-Valpuesta} \&
  {Gerhard}}{{Martinez-Valpuesta} \& {Gerhard}}{2013}]{mart2013}
{Martinez-Valpuesta} I.,  {Gerhard} O.,  2013, \apjl, 766, L3

\bibitem[\protect\citeauthoryear{{Martinez-Valpuesta} \&
  {Shlosman}}{{Martinez-Valpuesta} \& {Shlosman}}{2004}]{IMV+Shlosman04}
{Martinez-Valpuesta} I.,  {Shlosman} I.,  2004, \apjl, 613, L29

\bibitem[\protect\citeauthoryear{{Martinez-Valpuesta}, {Shlosman} \&
  {Heller}}{{Martinez-Valpuesta} et~al.}{2006}]{MV+06}
{Martinez-Valpuesta} I.,  {Shlosman} I.,    {Heller} C.,  2006, \apj, 637, 214

\bibitem[\protect\citeauthoryear{{Mehlert}, {Thomas}, {Saglia}, {Bender} \&
  {Wegner}}{{Mehlert} et~al.}{2003}]{mehl2003}
{Mehlert} D.,  {Thomas} D.,  {Saglia} R.~P.,  {Bender} R.,    {Wegner} G.,
  2003, \aap, 407, 423

\bibitem[\protect\citeauthoryear{{Melvin}, {Masters}, {Lintott}, {Nichol},
  {Simmons}, {Bamford}, {Casteels}, {Cheung}, {Edmondson}, {Fortson},
  {Schawinski}, {Skibba}, {Smith} \& {Willett}}{{Melvin}
  et~al.}{2014}]{melv2014}
{Melvin} T.,  {Masters} K.,  {Lintott} C.,  {Nichol} R.~C.,  {Simmons} B.,
  {Bamford} S.~P.,  {Casteels} K.~R.~V.,  {Cheung} E.,  {Edmondson} E.~M.,
  {Fortson} L.,  {Schawinski} K.,  {Skibba} R.~A.,  {Smith} A.~M.,    {Willett}
  K.~W.,  2014, \mnras, 438, 2882

\bibitem[\protect\citeauthoryear{{Merrifield} \& {Kuijken}}{{Merrifield} \&
  {Kuijken}}{1999}]{merr1999}
{Merrifield} M.~R.,  {Kuijken} K.,  1999, \aap, 345, L47

\bibitem[\protect\citeauthoryear{{Molaeinezhad}, {Falc{\'o}n-Barroso},
  {Mart{\'{\i}}nez-Valpuesta}, {Khosroshahi}, {Balcells} \&
  {Peletier}}{{Molaeinezhad} et~al.}{2016}]{mola2016}
{Molaeinezhad} A.,  {Falc{\'o}n-Barroso} J.,  {Mart{\'{\i}}nez-Valpuesta} I.,
  {Khosroshahi} H.~G.,  {Balcells} M.,    {Peletier} R.~F.,  2016, \mnras, 456,
  692

\bibitem[\protect\citeauthoryear{{Moorthy} \& {Holtzman}}{{Moorthy} \&
  {Holtzman}}{2006}]{moor2006}
{Moorthy} B.~K.,  {Holtzman} J.~A.,  2006, \mnras, 371, 583

\bibitem[\protect\citeauthoryear{{Morelli}, {Pompei}, {Pizzella},
  {M{\'e}ndez-Abreu}, {Corsini}, {Coccato}, {Saglia}, {Sarzi} \&
  {Bertola}}{{Morelli} et~al.}{2008}]{more2008}
{Morelli} L.,  {Pompei} E.,  {Pizzella} A.,  {M{\'e}ndez-Abreu} J.,  {Corsini}
  E.~M.,  {Coccato} L.,  {Saglia} R.~P.,  {Sarzi} M.,    {Bertola} F.,  2008,
  \mnras, 389, 341

\bibitem[\protect\citeauthoryear{{Ness}, {Freeman}, {Athanassoula},
  {Wylie-de-Boer}, {Bland-Hawthorn}, {Asplund}, {Lewis}, {Yong}, {Lane} \&
  {Kiss}}{{Ness} et~al.}{2013}]{ness2013}
{Ness} M.,  {Freeman} K.,  {Athanassoula} E.,  {Wylie-de-Boer} E.,
  {Bland-Hawthorn} J.,  {Asplund} M.,  {Lewis} G.~F.,  {Yong} D.,  {Lane}
  R.~R.,    {Kiss} L.~L.,  2013, \mnras, 430, 836

\bibitem[\protect\citeauthoryear{{Paturel}, {Petit}, {Prugniel}, {Theureau},
  {Rousseau}, {Brouty}, {Dubois} \& {Cambr{\'e}sy}}{{Paturel}
  et~al.}{2003}]{patu2003}
{Paturel} G.,  {Petit} C.,  {Prugniel} P.,  {Theureau} G.,  {Rousseau} J.,
  {Brouty} M.,  {Dubois} P.,    {Cambr{\'e}sy} L.,  2003, \aap, 412, 45

\bibitem[\protect\citeauthoryear{{Peletier} \& {Balcells}}{{Peletier} \&
  {Balcells}}{1997}]{pele1997}
{Peletier} R.~F.,  {Balcells} M.,  1997, \na, 1, 349

\bibitem[\protect\citeauthoryear{{Peletier}, {Falc{\'o}n-Barroso}, {Bacon},
  {Cappellari}, {Davies}, {de Zeeuw}, {Emsellem}, {Ganda}, {Krajnovi{\'c}},
  {Kuntschner}, {McDermid}, {Sarzi} \& {van de Ven}}{{Peletier}
  et~al.}{2007}]{pele2007}
{Peletier} R.~F.,  {Falc{\'o}n-Barroso} J.,  {Bacon} R.,  {Cappellari} M.,
  {Davies} R.~L.,  {de Zeeuw} P.~T.,  {Emsellem} E.,  {Ganda} K.,
  {Krajnovi{\'c}} D.,  {Kuntschner} H.,  {McDermid} R.~M.,  {Sarzi} M.,    {van
  de Ven} G.,  2007, \mnras, 379, 445

\bibitem[\protect\citeauthoryear{{Peng}, {Ho}, {Impey} \& {Rix}}{{Peng}
  et~al.}{2002}]{peng2002}
{Peng} C.~Y.,  {Ho} L.~C.,  {Impey} C.~D.,    {Rix} H.-W.,  2002, \aj, 124, 266

\bibitem[\protect\citeauthoryear{{P{\'e}rez} \&
  {S{\'a}nchez-Bl{\'a}zquez}}{{P{\'e}rez} \&
  {S{\'a}nchez-Bl{\'a}zquez}}{2011}]{perez2011}
{P{\'e}rez} I.,  {S{\'a}nchez-Bl{\'a}zquez} P.,  2011, \aap, 529, A64

\bibitem[\protect\citeauthoryear{{Pietrinferni}, {Cassisi}, {Salaris} \&
  {Castelli}}{{Pietrinferni} et~al.}{2004}]{piet2004}
{Pietrinferni} A.,  {Cassisi} S.,  {Salaris} M.,    {Castelli} F.,  2004, \apj,
  612, 168

\bibitem[\protect\citeauthoryear{{Pietrinferni}, {Cassisi}, {Salaris} \&
  {Castelli}}{{Pietrinferni} et~al.}{2006}]{piet2006}
{Pietrinferni} A.,  {Cassisi} S.,  {Salaris} M.,    {Castelli} F.,  2006, \apj,
  642, 797

\bibitem[\protect\citeauthoryear{{Rich}}{{Rich}}{1990}]{rich1990}
{Rich} R.~M.,  1990, \apj, 362, 604

\bibitem[\protect\citeauthoryear{{Rich}, {Reitzel}, {Howard} \& {Zhao}}{{Rich}
  et~al.}{2007}]{rich2007}
{Rich} R.~M.,  {Reitzel} D.~B.,  {Howard} C.~D.,    {Zhao} H.,  2007, \apjl,
  658, L29

\bibitem[\protect\citeauthoryear{{Saha}, {Martinez-Valpuesta} \&
  {Gerhard}}{{Saha} et~al.}{2012}]{saha2012}
{Saha} K.,  {Martinez-Valpuesta} I.,    {Gerhard} O.,  2012, \mnras, 421, 333

\bibitem[\protect\citeauthoryear{{S{\'a}nchez-Bl{\'a}zquez}}{{S{\'a}nchez-Bl{\'a}zquez}}{2016}]{sanc2016}
{S{\'a}nchez-Bl{\'a}zquez} P.,  2016, Galactic Bulges, 418, 127

\bibitem[\protect\citeauthoryear{{S{\'a}nchez-Bl{\'a}zquez}, {Gorgas},
  {Cardiel} \& {Gonz{\'a}lez}}{{S{\'a}nchez-Bl{\'a}zquez}
  et~al.}{2006}]{sanc2006}
{S{\'a}nchez-Bl{\'a}zquez} P.,  {Gorgas} J.,  {Cardiel} N.,    {Gonz{\'a}lez}
  J.~J.,  2006, \aap, 457, 809

\bibitem[\protect\citeauthoryear{{S{\'a}nchez-Bl{\'a}zquez}, {Ocvirk},
  {Gibson}, {P{\'e}rez} \& {Peletier}}{{S{\'a}nchez-Bl{\'a}zquez}
  et~al.}{2011}]{sanc2011}
{S{\'a}nchez-Bl{\'a}zquez} P.,  {Ocvirk} P.,  {Gibson} B.~K.,  {P{\'e}rez} I.,
    {Peletier} R.~F.,  2011, \mnras, 415, 709

\bibitem[\protect\citeauthoryear{{Seidel}, {Cacho}, {Ruiz-Lara},
  {Falc{\'o}n-Barroso}, {P{\'e}rez}, {S{\'a}nchez-Bl{\'a}zquez}, {Vogt},
  {Ness}, {Freeman} \& {Aniyan}}{{Seidel} et~al.}{2015}]{seidel15}
{Seidel} M.~K.,  {Cacho} R.,  {Ruiz-Lara} T.,  {Falc{\'o}n-Barroso} J.,
  {P{\'e}rez} I.,  {S{\'a}nchez-Bl{\'a}zquez} P.,  {Vogt} F.~P.~A.,  {Ness} M.,
   {Freeman} K.,    {Aniyan} S.,  2015, \mnras, 446, 2837

\bibitem[\protect\citeauthoryear{{Seidel}, {Falc{\'o}n-Barroso},
  {Mart{\'{\i}}nez-Valpuesta}, {S{\'a}nchez-Bl{\'a}zquez}, {P{\'e}rez},
  {Peletier} \& {Vazdekis}}{{Seidel} et~al.}{2016}]{seidel16}
{Seidel} M.~K.,  {Falc{\'o}n-Barroso} J.,  {Mart{\'{\i}}nez-Valpuesta} I.,
  {S{\'a}nchez-Bl{\'a}zquez} P.,  {P{\'e}rez} I.,  {Peletier} R.,    {Vazdekis}
  A.,  2016, ArXiv e-prints

\bibitem[\protect\citeauthoryear{{Sheth}, {Elmegreen}, {Elmegreen}, {Capak},
  {Abraham}, {Athanassoula}, {Ellis}, {Mobasher}, {Salvato}, {Schinnerer},
  {Scoville}, {Spalsbury}, {Strubbe}, {Carollo}, {Rich} \& {West}}{{Sheth}
  et~al.}{2008}]{shet2008}
{Sheth} K.,  {Elmegreen} D.~M.,  {Elmegreen} B.~G.,  {Capak} P.,  {Abraham}
  R.~G.,  {Athanassoula} E.,  {Ellis} R.~S.,  {Mobasher} B.,  {Salvato} M.,
  {Schinnerer} E.,  {Scoville} N.~Z.,  {Spalsbury} L.,  {Strubbe} L.,
  {Carollo} M.,  {Rich} M.,    {West} A.~A.,  2008, \apj, 675, 1141

\bibitem[\protect\citeauthoryear{Taylor}{Taylor}{1982}]{tayl1982}
Taylor J.~R.,  1982, Univ. Sci. Books, Mill Valley, CA, 270p

\bibitem[\protect\citeauthoryear{{Thomas}, {Saglia}, {Bender}, {Thomas},
  {Gebhardt}, {Magorrian}, {Corsini}, {Wegner} \& {Seitz}}{{Thomas}
  et~al.}{2011}]{thom2011}
{Thomas} J.,  {Saglia} R.~P.,  {Bender} R.,  {Thomas} D.,  {Gebhardt} K.,
  {Magorrian} J.,  {Corsini} E.~M.,  {Wegner} G.,    {Seitz} S.,  2011, \mnras,
  415, 545

\bibitem[\protect\citeauthoryear{{Trager}, {Worthey}, {Faber}, {Burstein} \&
  {Gonz{\'a}lez}}{{Trager} et~al.}{1998}]{trag1998}
{Trager} S.~C.,  {Worthey} G.,  {Faber} S.~M.,  {Burstein} D.,
  {Gonz{\'a}lez} J.~J.,  1998, \apjs, 116, 1

\bibitem[\protect\citeauthoryear{{Vazdekis}, {Coelho}, {Cassisi},
  {Ricciardelli}, {Falc{\'o}n-Barroso}, {S{\'a}nchez-Bl{\'a}zquez}, {Barbera},
  {Beasley} \& {Pietrinferni}}{{Vazdekis} et~al.}{2015}]{vazd2015}
{Vazdekis} A.,  {Coelho} P.,  {Cassisi} S.,  {Ricciardelli} E.,
  {Falc{\'o}n-Barroso} J.,  {S{\'a}nchez-Bl{\'a}zquez} P.,  {Barbera} F.~L.,
  {Beasley} M.~A.,    {Pietrinferni} A.,  2015, \mnras, 449, 1177

\bibitem[\protect\citeauthoryear{{Vazdekis}, {S{\'a}nchez-Bl{\'a}zquez},
  {Falc{\'o}n-Barroso}, {Cenarro}, {Beasley}, {Cardiel}, {Gorgas} \&
  {Peletier}}{{Vazdekis} et~al.}{2010}]{vazd2010}
{Vazdekis} A.,  {S{\'a}nchez-Bl{\'a}zquez} P.,  {Falc{\'o}n-Barroso} J.,
  {Cenarro} A.~J.,  {Beasley} M.~A.,  {Cardiel} N.,  {Gorgas} J.,    {Peletier}
  R.~F.,  2010, \mnras, 404, 1639

\bibitem[\protect\citeauthoryear{{Whyte}, {Abraham}, {Merrifield}, {Eskridge},
  {Frogel} \& {Pogge}}{{Whyte} et~al.}{2002}]{whyt2002}
{Whyte} L.~F.,  {Abraham} R.~G.,  {Merrifield} M.~R.,  {Eskridge} P.~B.,
  {Frogel} J.~A.,    {Pogge} R.~W.,  2002, \mnras, 336, 1281

\bibitem[\protect\citeauthoryear{{Williams}, {Bureau} \&
  {Kuntschner}}{{Williams} et~al.}{2012}]{will2012}
{Williams} M.~J.,  {Bureau} M.,    {Kuntschner} H.,  2012, \mnras, 427, L99

\bibitem[\protect\citeauthoryear{{Williams}, {Zamojski}, {Bureau},
  {Kuntschner}, {Merrifield}, {de Zeeuw} \& {Kuijken}}{{Williams}
  et~al.}{2011}]{will2011}
{Williams} M.~J.,  {Zamojski} M.~A.,  {Bureau} M.,  {Kuntschner} H.,
  {Merrifield} M.~R.,  {de Zeeuw} P.~T.,    {Kuijken} K.,  2011, \mnras, 414,
  2163

\bibitem[\protect\citeauthoryear{{Zoccali}, {Hill}, {Lecureur}, {Barbuy},
  {Renzini}, {Minniti}, {G{\'o}mez} \& {Ortolani}}{{Zoccali}
  et~al.}{2008}]{zocc2008}
{Zoccali} M.,  {Hill} V.,  {Lecureur} A.,  {Barbuy} B.,  {Renzini} A.,
  {Minniti} D.,  {G{\'o}mez} A.,    {Ortolani} S.,  2008, \aap, 486, 177

\bibitem[\protect\citeauthoryear{{Zoccali}, {Renzini}, {Ortolani}, {Greggio},
  {Saviane}, {Cassisi}, {Rejkuba}, {Barbuy}, {Rich} \& {Bica}}{{Zoccali}
  et~al.}{2003}]{zocc2003}
{Zoccali} M.,  {Renzini} A.,  {Ortolani} S.,  {Greggio} L.,  {Saviane} I.,
  {Cassisi} S.,  {Rejkuba} M.,  {Barbuy} B.,  {Rich} R.~M.,    {Bica} E.,
  2003, \aap, 399, 931

\end{thebibliography}

\label{lastpage}

\appendix

\section{Unsharp images for individual galaxies}
\label{app:unsharp}

For each galaxy in our sample, we present here the SDSS colour image and an unsharp masked image in the $i$-band, obtained from SDSS DR10 \citep{ahn2014}. The bulge region, obtained from our Galfit analysis is shown as a blue rectangle on top of the colour image. We also present the $V$--h$_3$ correlation profiles for all galaxies in our sample. These profiles show the level of correlation of V and h$_3$ within the bulge zone in pseudo-slits parallel to the major axis of the galaxies at different heights ($z$) from the disc plane. These measurements have been performed on the dust-free side of the galaxies. Negative values indicate negative correlation, while values closer to +1 mean stronger positive correlation. Different shaded regions mark the disc plane, the bulge/bar dominated parts and the regions beyond $z_{\rm{B}}$.

 \begin{landscape}
 	\begin{figure}
 		\captionsetup[subfigure]{labelformat=empty}
 		\centering
 		\begin{subfigure}{0.44\textwidth}
 			\centering
 			\includegraphics[width=0.95\textwidth]{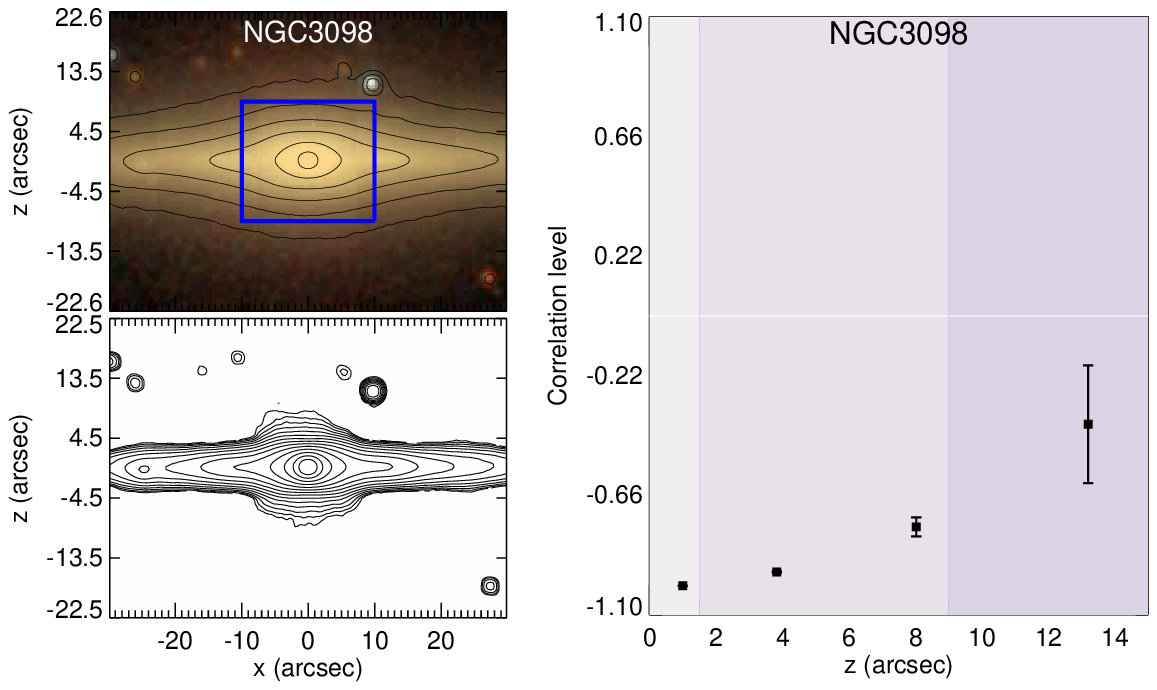}
 			\label{fig:suba1}
 		\end{subfigure}%
 		\hspace*{-0.9em}
 		\begin{subfigure}{0.44\textwidth}
 			\centering
 			\includegraphics[width=0.95\textwidth]{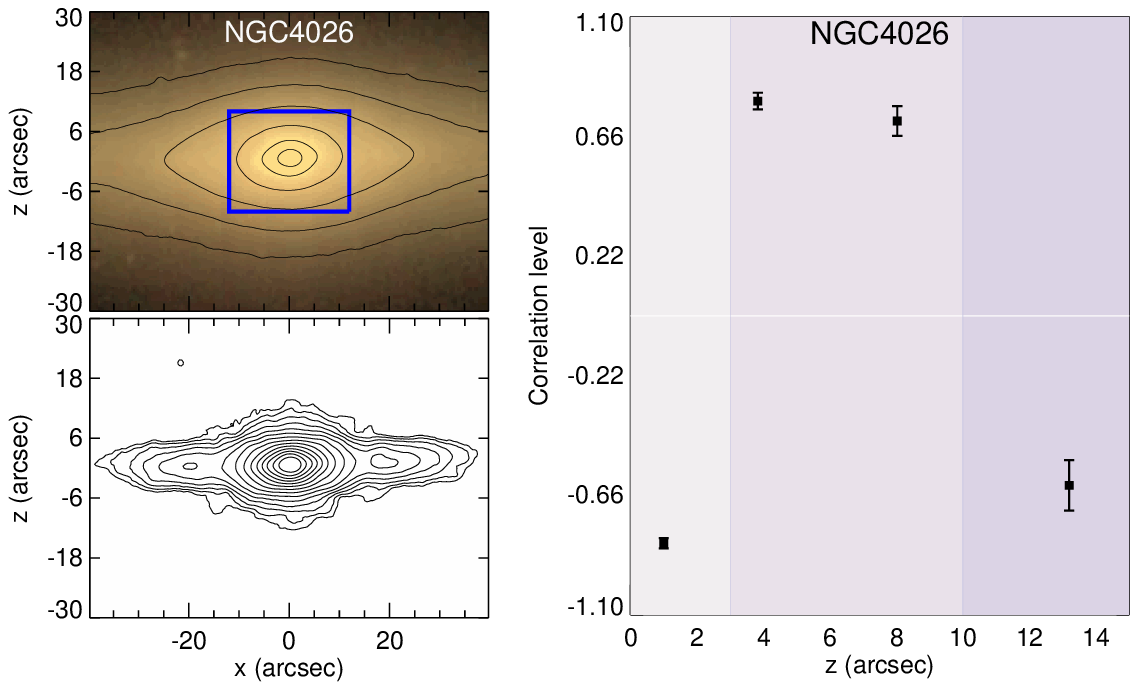}
 			\label{fig:suba2}
 		\end{subfigure}
 		\begin{subfigure}{0.44\textwidth}
 			\centering
 			\includegraphics[width=0.95\textwidth]{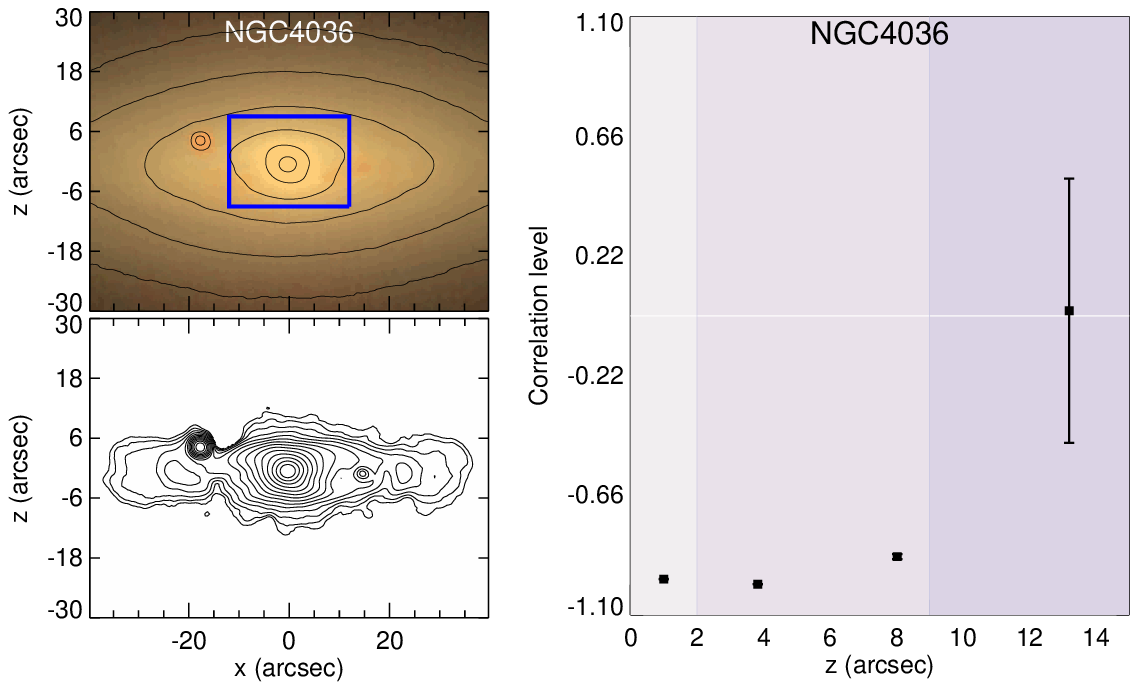}
 			\label{fig:suba3}
 		\end{subfigure}
 		\begin{subfigure}{0.44\textwidth}
 			\centering
 			\includegraphics[width=0.95\textwidth]{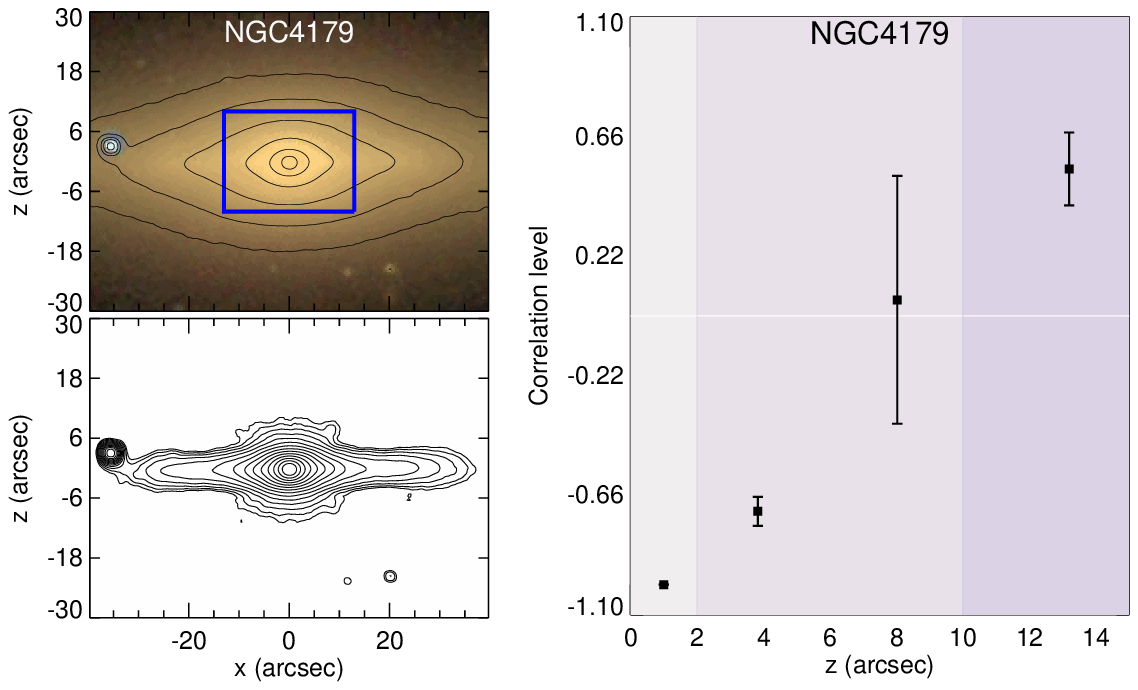}
 			\label{fig:suba4}
 		\end{subfigure}
 		\begin{subfigure}{0.44\textwidth}
 			\centering
 			\includegraphics[width=0.95\textwidth]{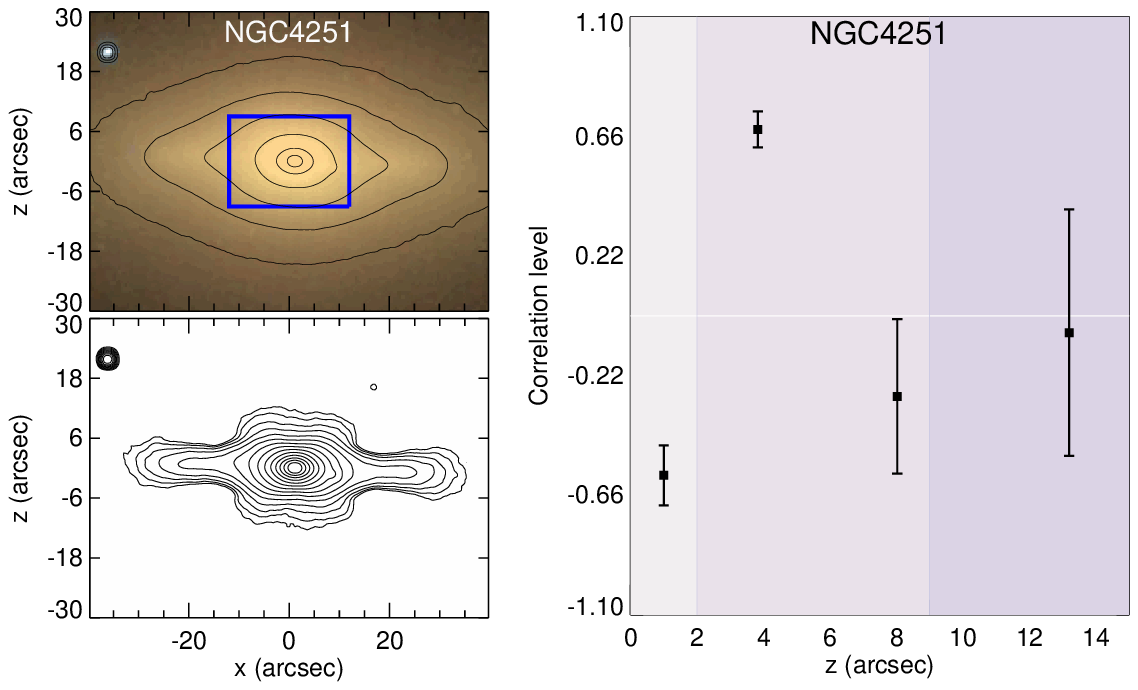}
 			\label{fig:suba5}
 		\end{subfigure}%
 		\begin{subfigure}{0.44\textwidth}
 			\centering
 			\includegraphics[width=0.95\textwidth]{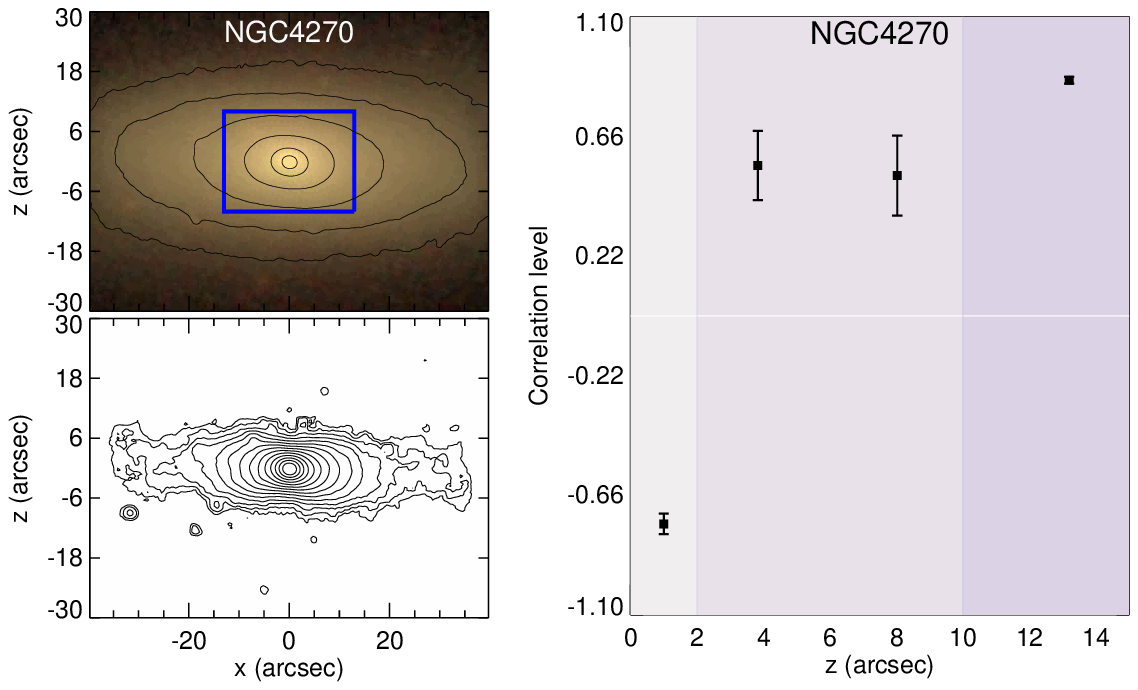}
 			\label{fig:suba6}
 		\end{subfigure}
 		\begin{subfigure}{0.44\textwidth}
 			\centering
 			\includegraphics[width=0.95\textwidth]{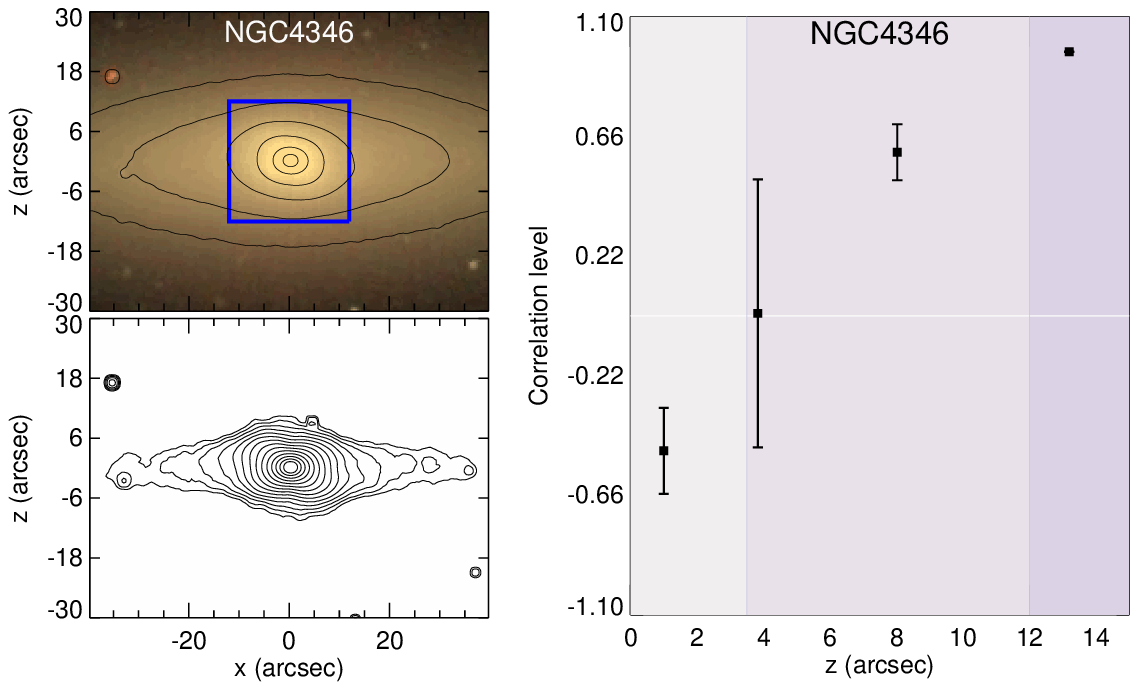}
 			\label{fig:suba7}
 		\end{subfigure}
 		\begin{subfigure}{0.44\textwidth}
 			\centering
 			\includegraphics[width=0.95\textwidth]{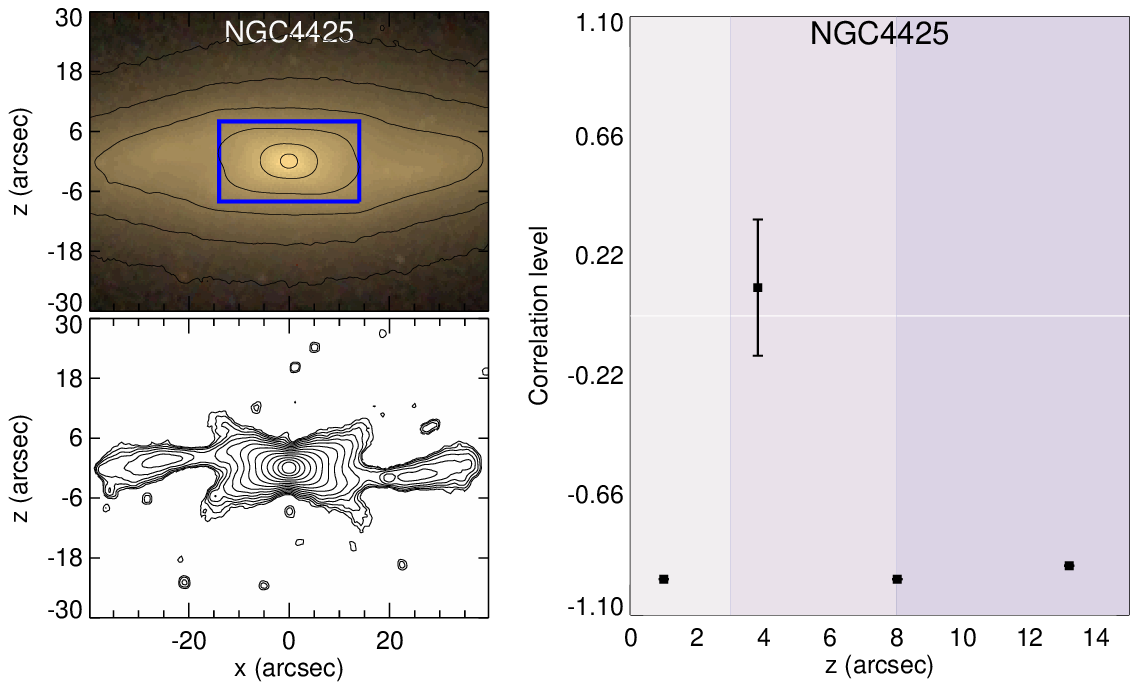}
 			\label{fig:suba8}
 		\end{subfigure}
 		\begin{subfigure}{0.44\textwidth}
 			\centering
 			\includegraphics[width=0.95\textwidth]{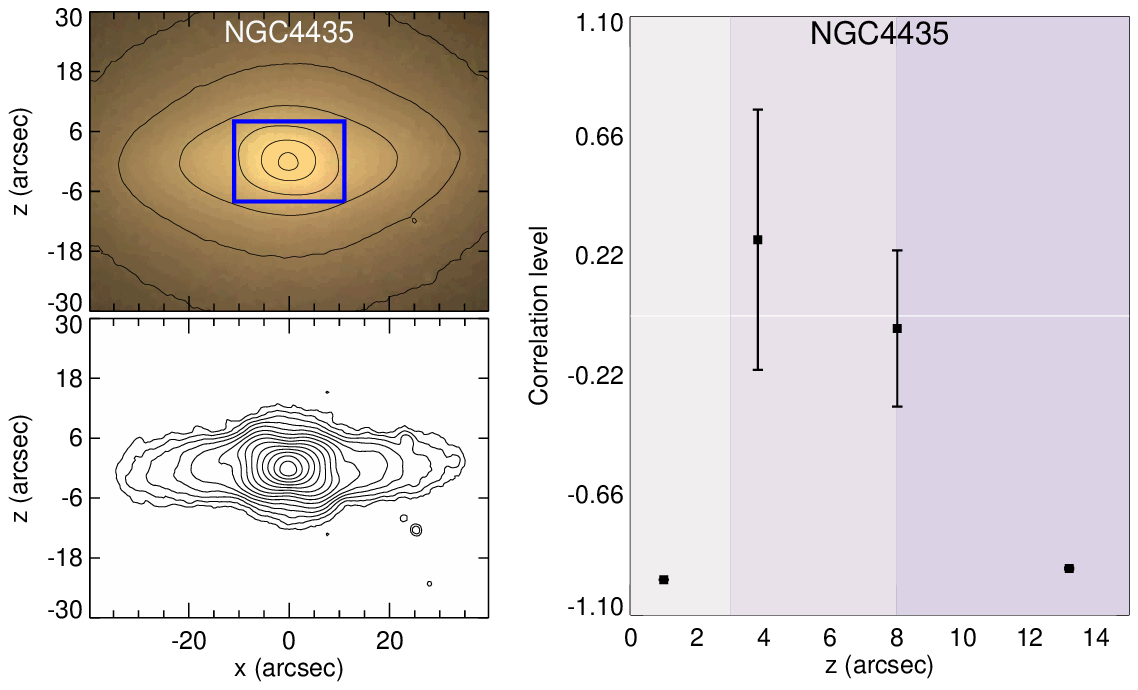}
 			\label{fig:suba9}
 		\end{subfigure}%
 		\setcounter{figure}{1}
 		\centering
 		\centering
\caption{On the top left we demonstrate a colour SDSS image, while the blue rectangle shows the bulge region, obtained from our Galfit analysis. Below, we show the unsharp masked $i-$band image. Right panel show the $V$--h$_3$ correlation profile. Different shaded regions mark the disc plane, the bulge/bar dominated parts and the regions beyond $z_{\rm{B}}$. For more details please refer to the text.}
 	\end{figure}
 \end{landscape} 
 
 \begin{landscape} 	 
 	\begin{figure}
 		\captionsetup[subfigure]{labelformat=empty}
 		\begin{subfigure}{0.44\textwidth}
 			 		 	 		\centering
 			\includegraphics[width=0.95\textwidth]{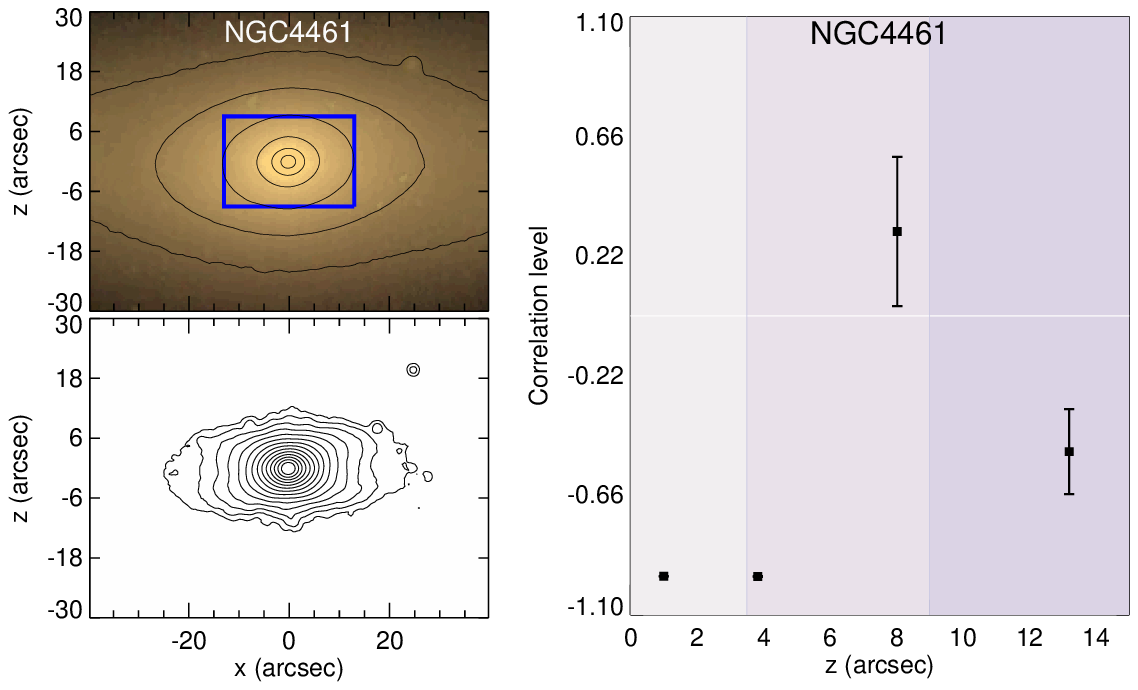}
 			\label{fig:suba10}
 		\end{subfigure}
 		\begin{subfigure}{0.44\textwidth}
 			 		 	 		\centering
 			\includegraphics[width=0.95\textwidth]{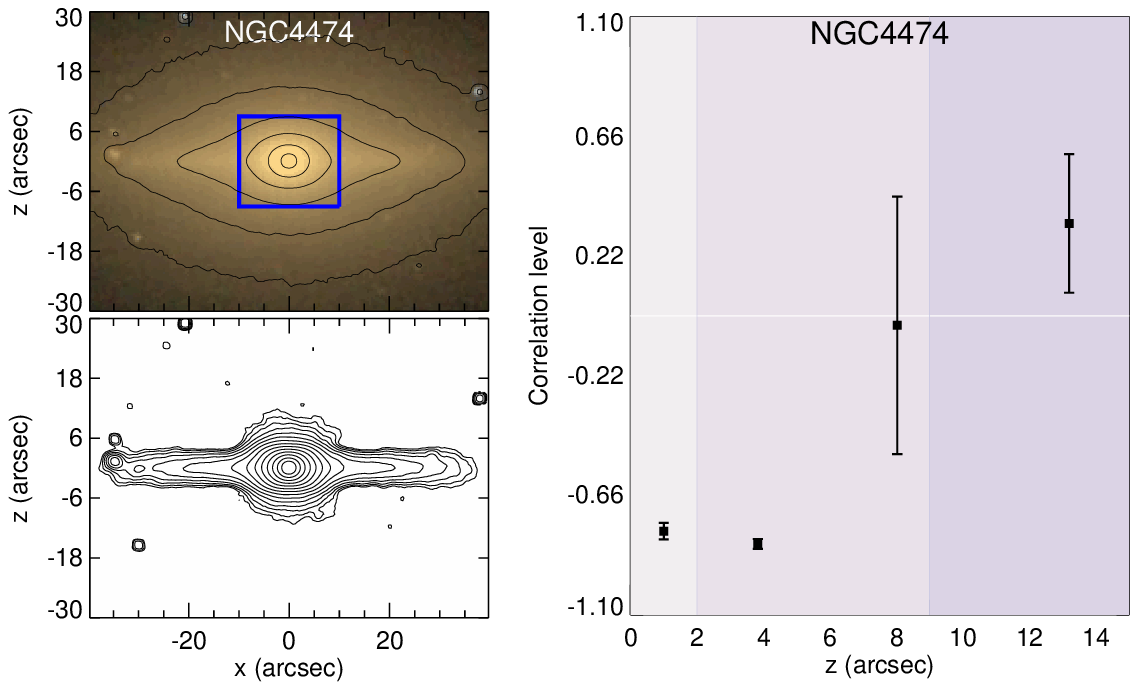}
 			\label{fig:suba11}
 		\end{subfigure}
 		\begin{subfigure}{0.44\textwidth}
 			 		 	 		\centering
 			\includegraphics[width=0.95\textwidth]{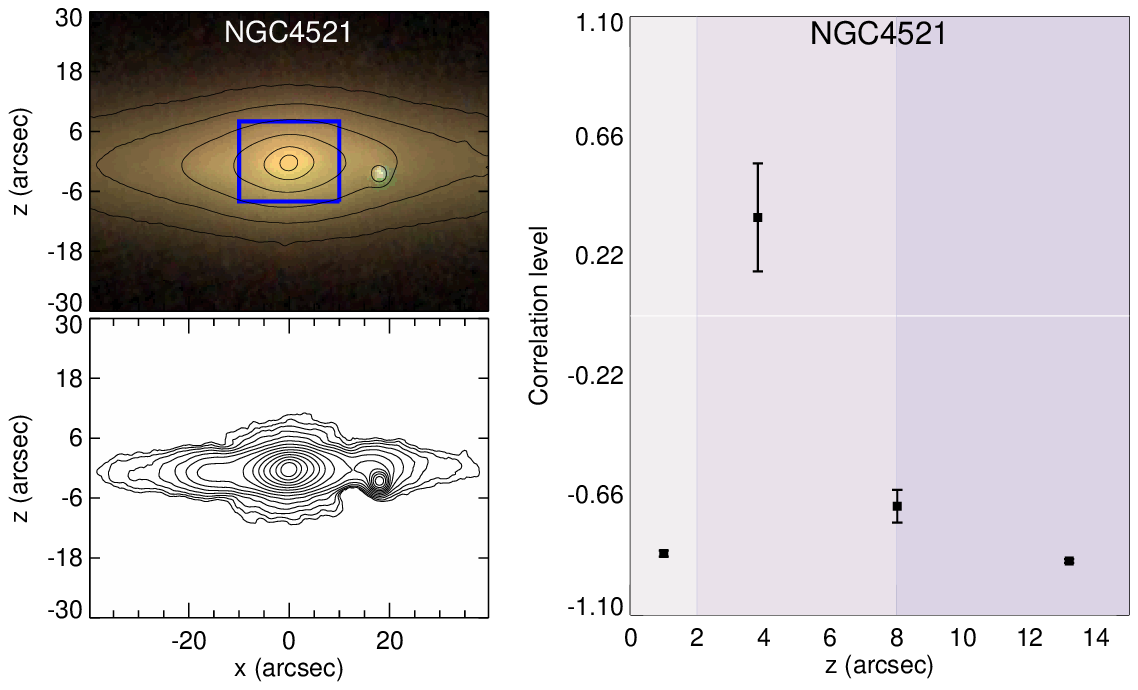}
 			\label{fig:suba12}
 		\end{subfigure}
 		 		\begin{subfigure}{0.44\textwidth}
 		 					 	 		\centering
 		 			\includegraphics[width=0.95\textwidth]{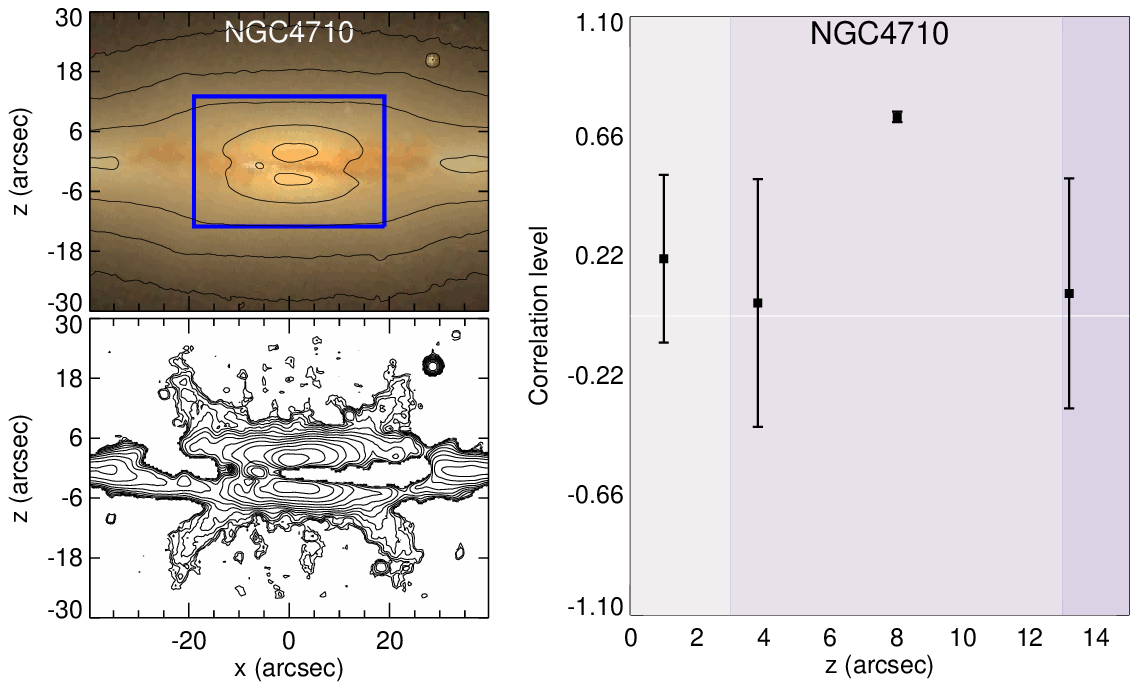}
 		 			\label{fig:suba13}
 		 		\end{subfigure}%
 		 		\begin{subfigure}{0.44\textwidth}
 		 				 	 		\centering
 		 			\includegraphics[width=0.95\textwidth]{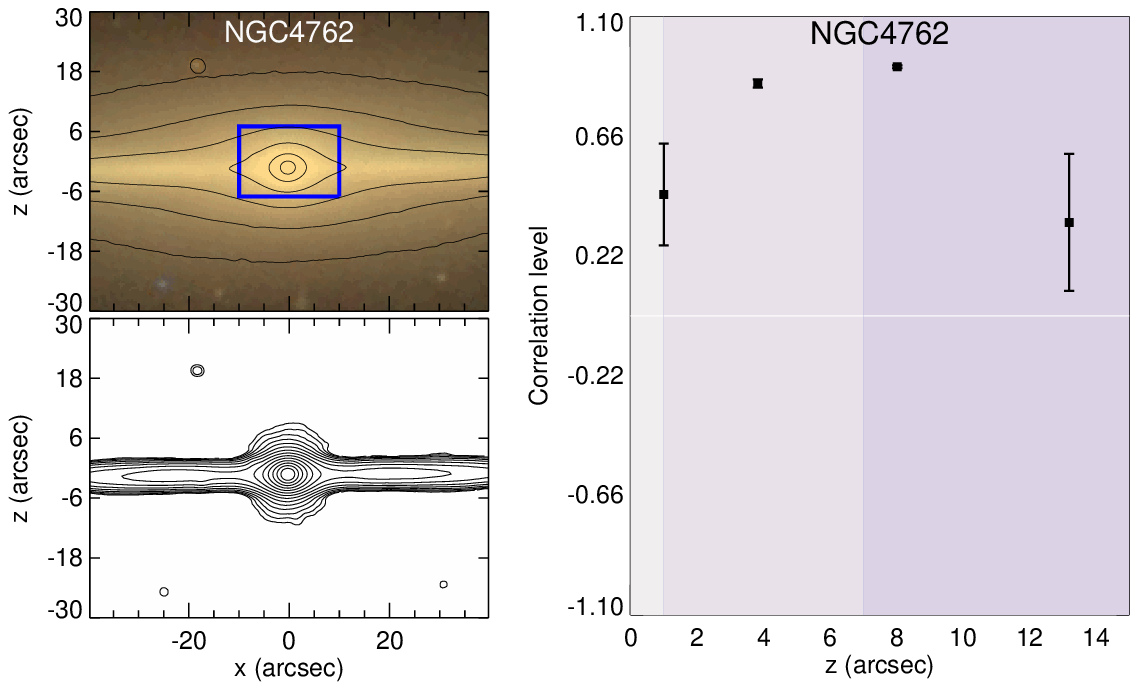}
 		 			\label{fig:suba14}
 		 		\end{subfigure}
 		 		\begin{subfigure}{0.44\textwidth}
 		 				 	 		\centering
 		 			\includegraphics[width=0.95\textwidth]{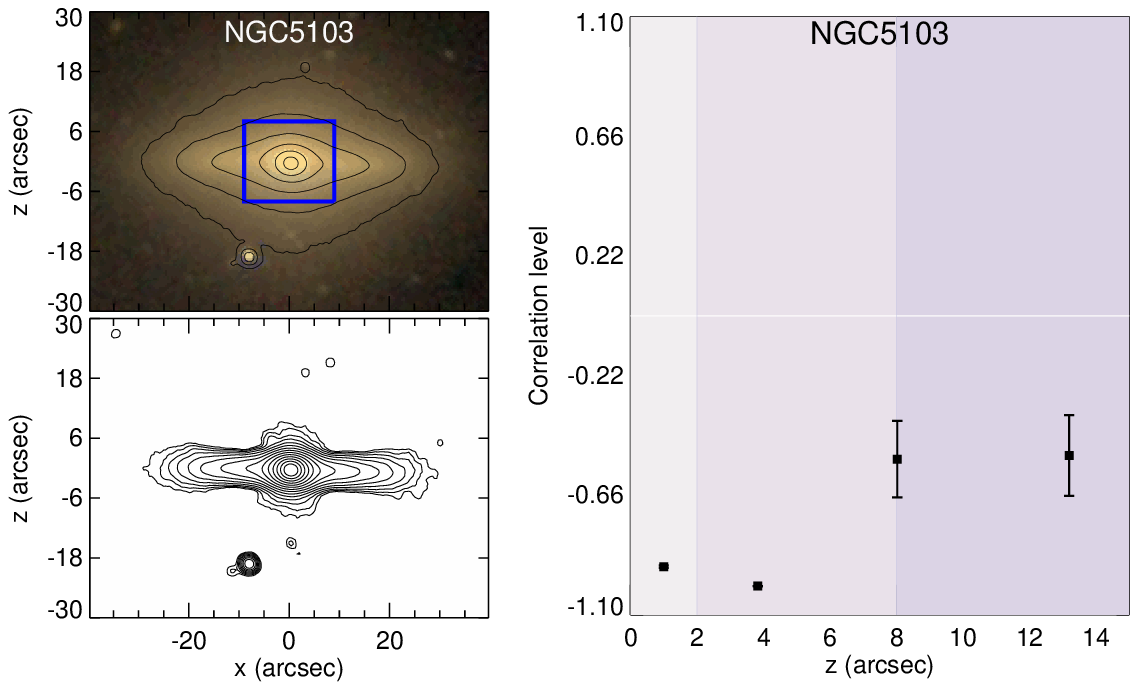}
 		 			\label{fig:suba15}
 		 		\end{subfigure}
 		 		\begin{subfigure}{0.44\textwidth}
 		 				 	 		\centering
 		 			\includegraphics[width=0.95\textwidth]{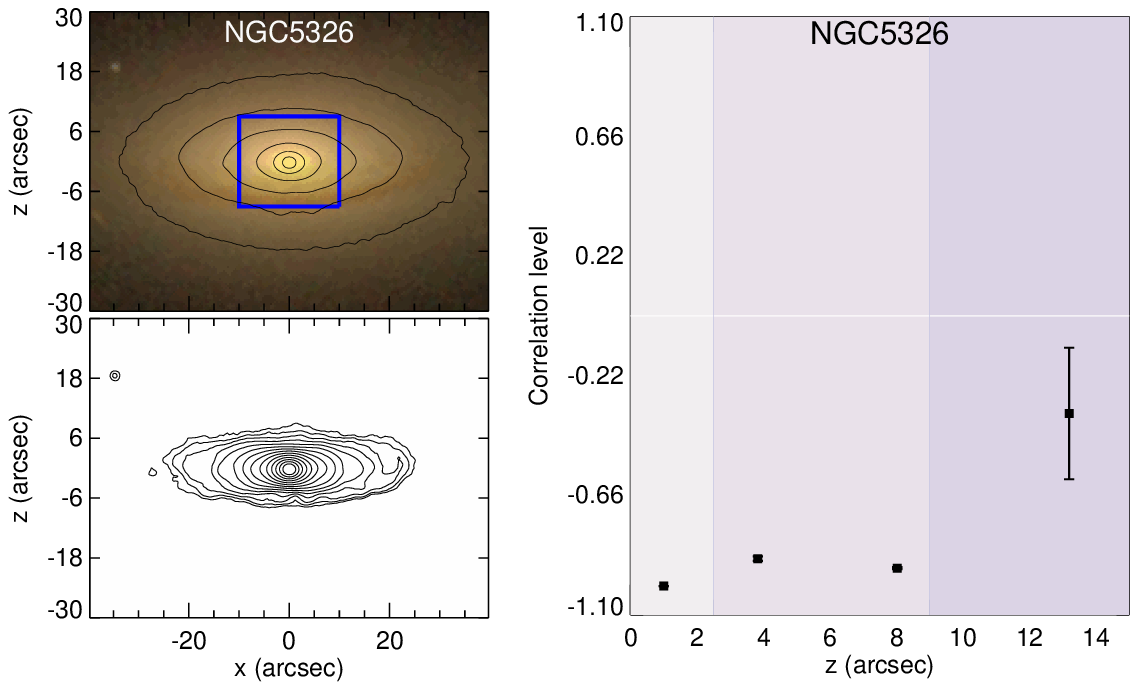}
 		 			\label{fig:suba16}
 		 		\end{subfigure}
 		 		\begin{subfigure}{0.44\textwidth}
 		 					 	 		\centering
 		 			\includegraphics[width=0.95\textwidth]{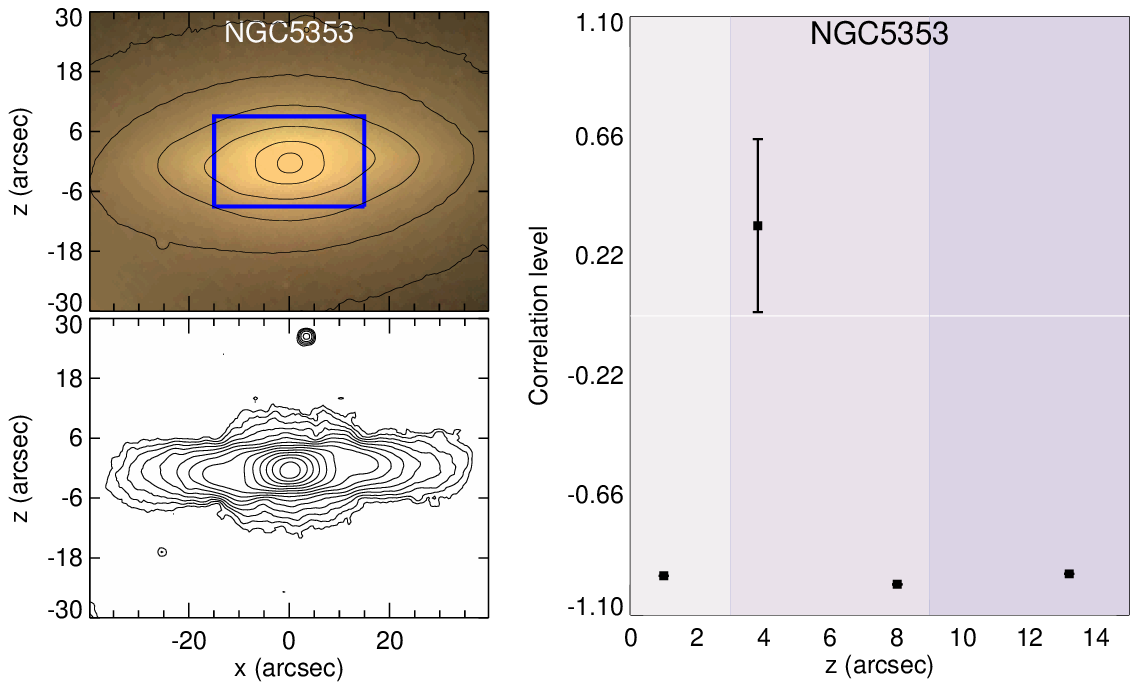}
 		 			\label{fig:suba17}
 		 		\end{subfigure}%
 		 		\begin{subfigure}{0.44\textwidth}
 		 					 	 		\centering
 		 			\includegraphics[width=0.95\textwidth]{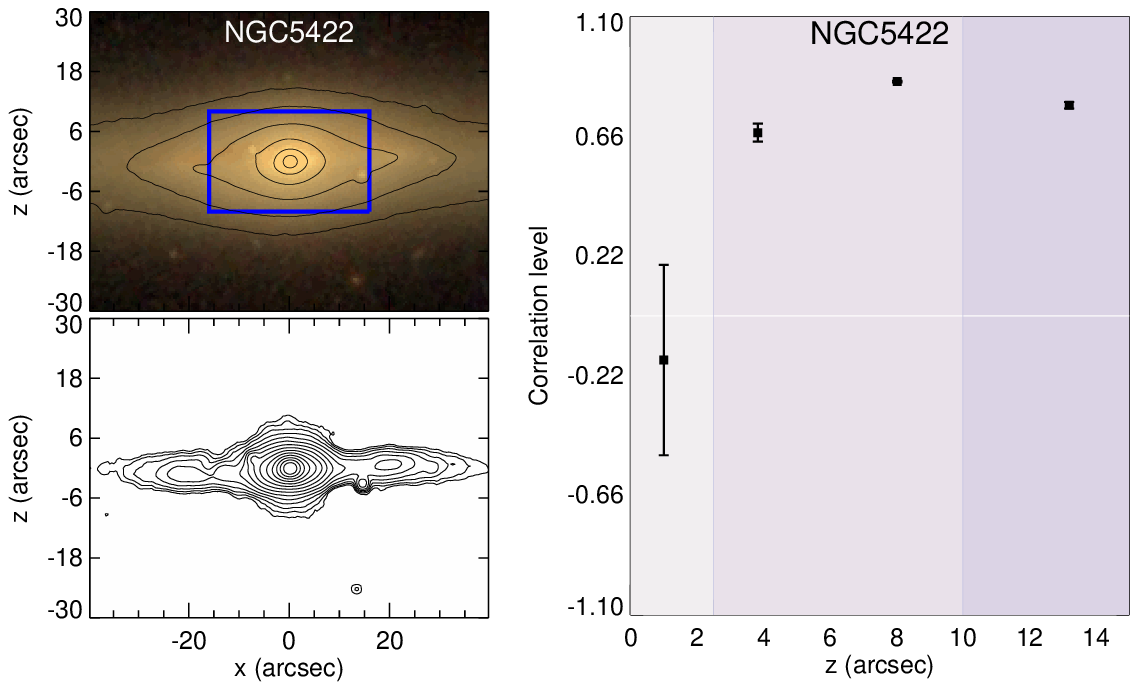}
 		 			\label{fig:suba18}
 		 		\end{subfigure}
 		\setcounter{figure}{1}
 		\centering
 		\centering
 		\caption{Continued.}
 	\end{figure}
 \end{landscape} 

 \begin{landscape}  
 	\begin{figure}
 		\captionsetup[subfigure]{labelformat=empty}
 			 		 	 	\centering
 		\begin{subfigure}{0.44\textwidth}
 			\includegraphics[width=0.95\textwidth]{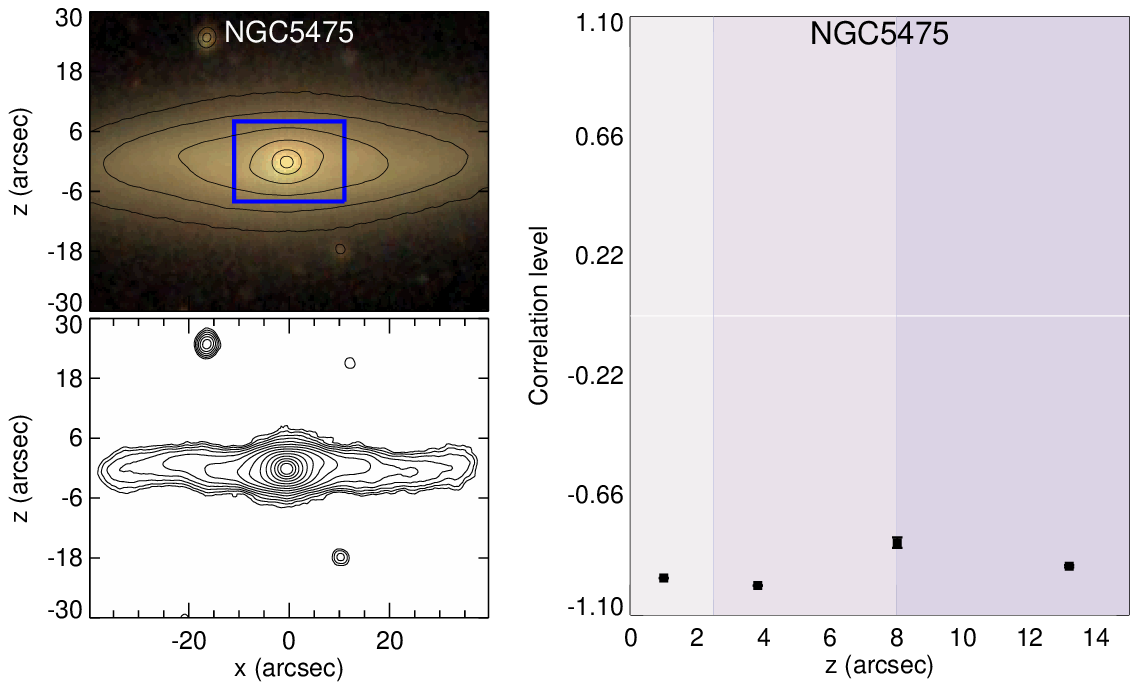}
 			\label{fig:subb19}
 		\end{subfigure}
 		\begin{subfigure}{0.44\textwidth}
 			\includegraphics[width=0.95\textwidth]{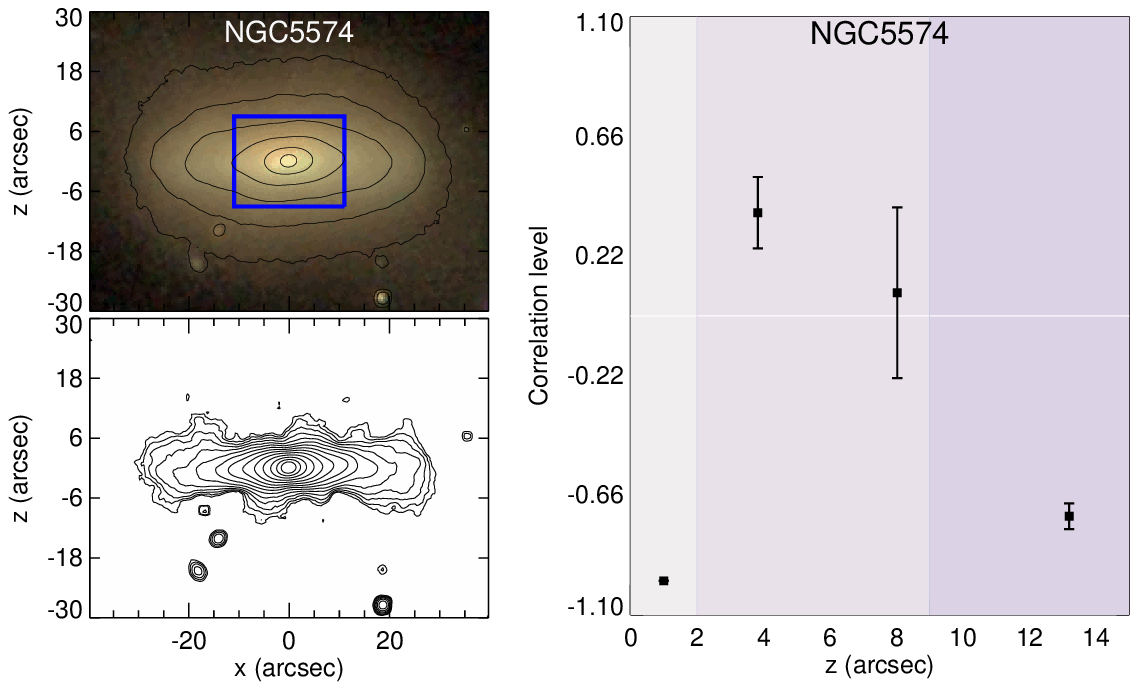}
 			\label{fig:suba20}
 		\end{subfigure}
 		\begin{subfigure}{0.44\textwidth}
 			\includegraphics[width=0.95\textwidth]{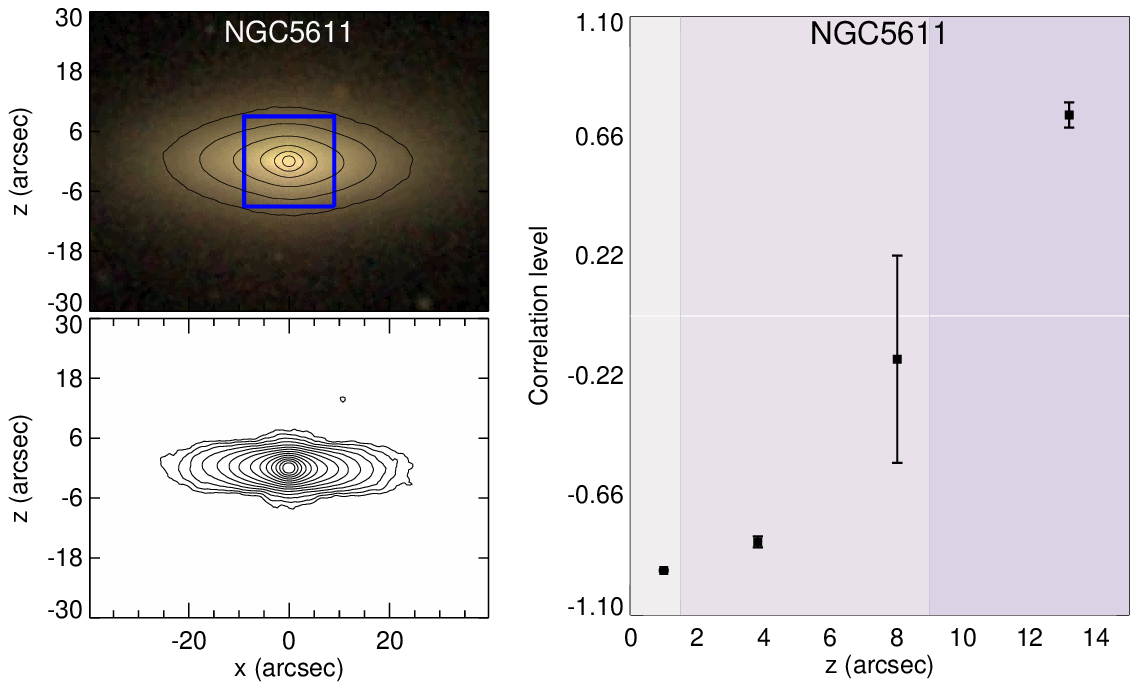}
 			\label{fig:suba21}
 		\end{subfigure}%
 		\\
 		\begin{subfigure}{0.44\textwidth}
 			\includegraphics[width=0.95\textwidth]{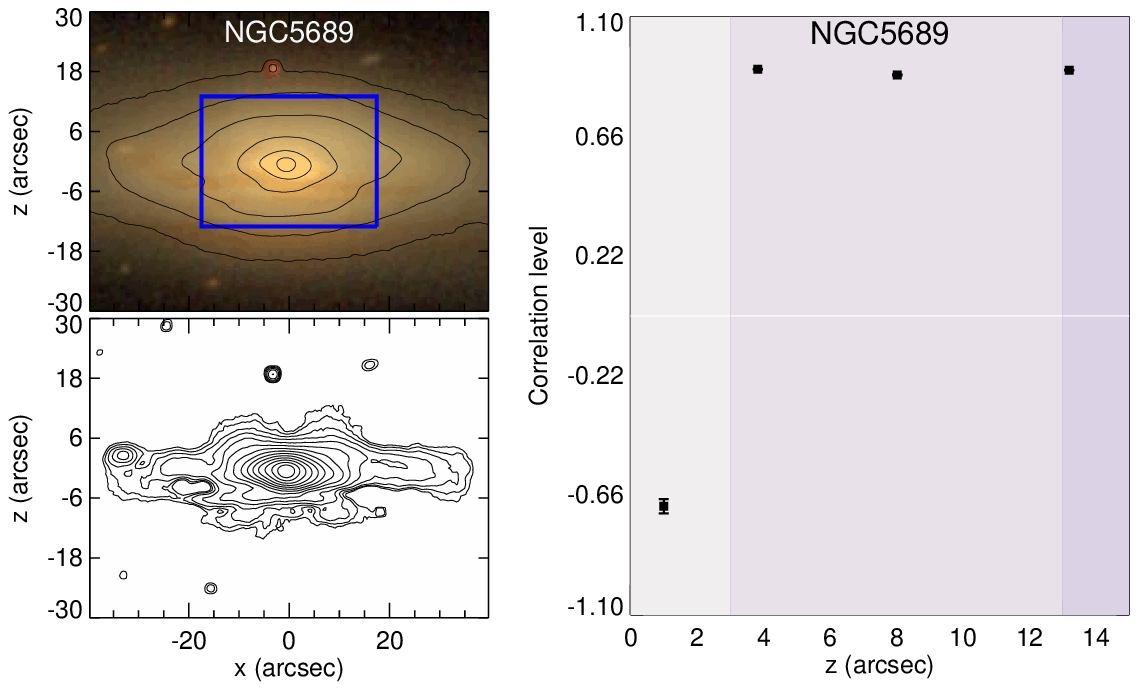}
 			\label{fig:suba22}
 		\end{subfigure}
 		\begin{subfigure}{0.44\textwidth}
 			\includegraphics[width=0.95\textwidth]{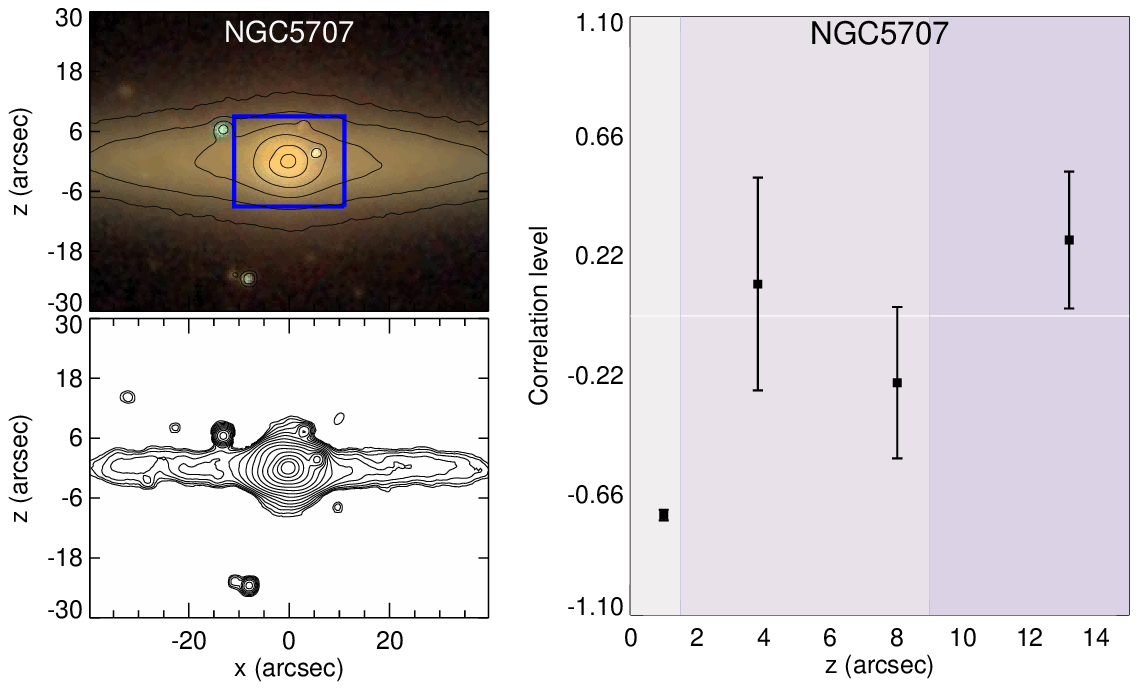}
 			\label{fig:suba23}
 		\end{subfigure}
 		\begin{subfigure}{0.44\textwidth}
 			\includegraphics[width=0.95\textwidth]{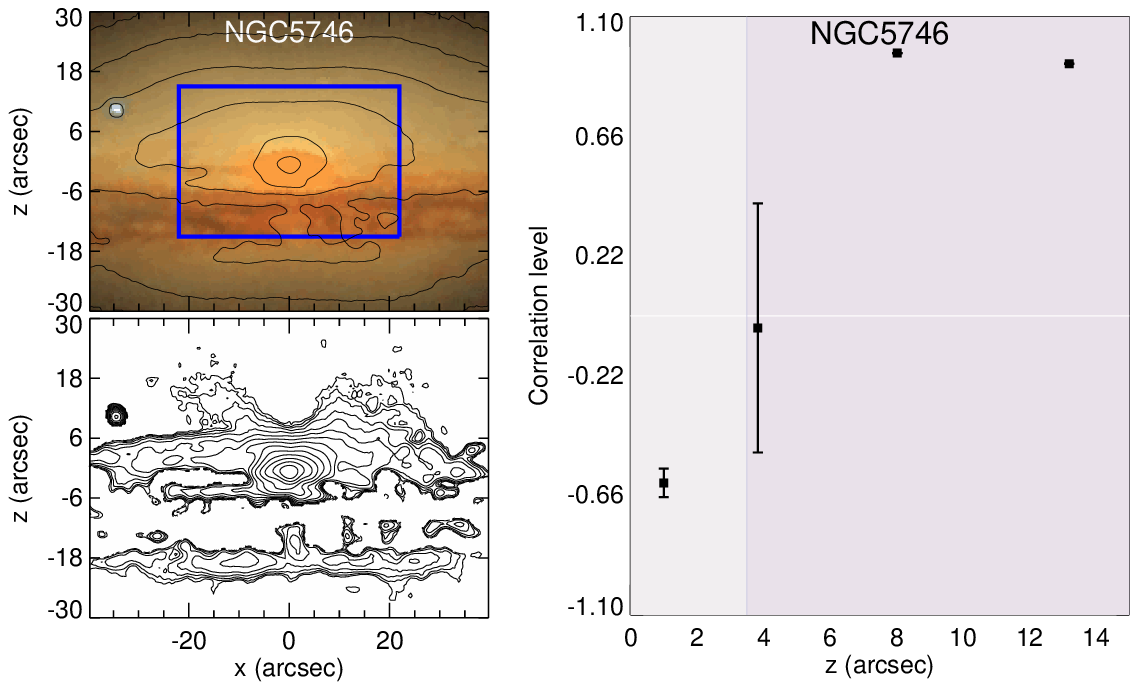}
 			\label{fig:suba24}
 		\end{subfigure}
 			\begin{subfigure}{0.44\textwidth}
 				\includegraphics[width=0.95\textwidth]{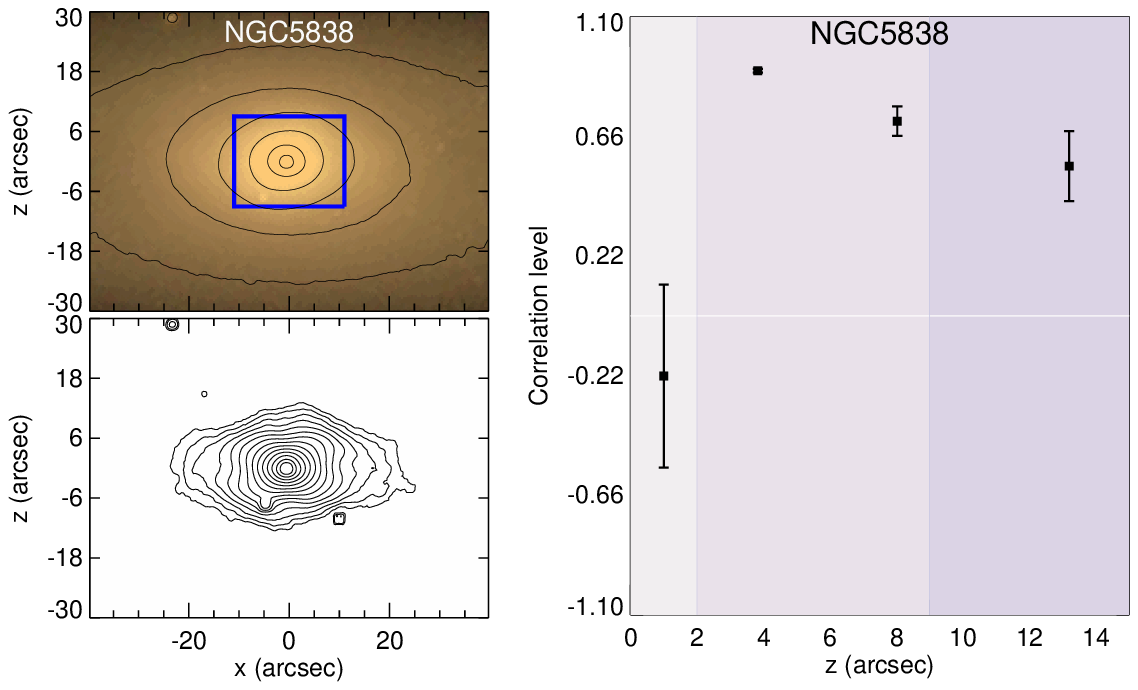}
 				\label{fig:suba25}
 			\end{subfigure}%
 			\begin{subfigure}{0.44\textwidth}
 				\includegraphics[width=0.95\textwidth]{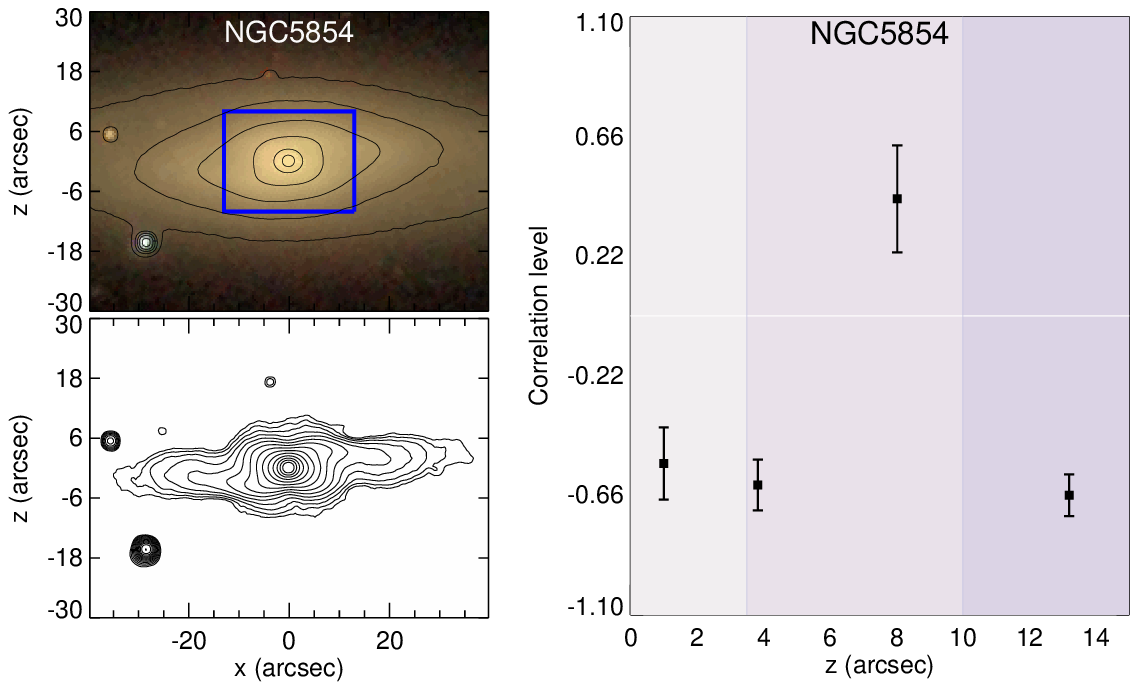}
 				\label{fig:suba26}
 			\end{subfigure}
 			\begin{subfigure}{0.44\textwidth}
 				\includegraphics[width=0.95\textwidth]{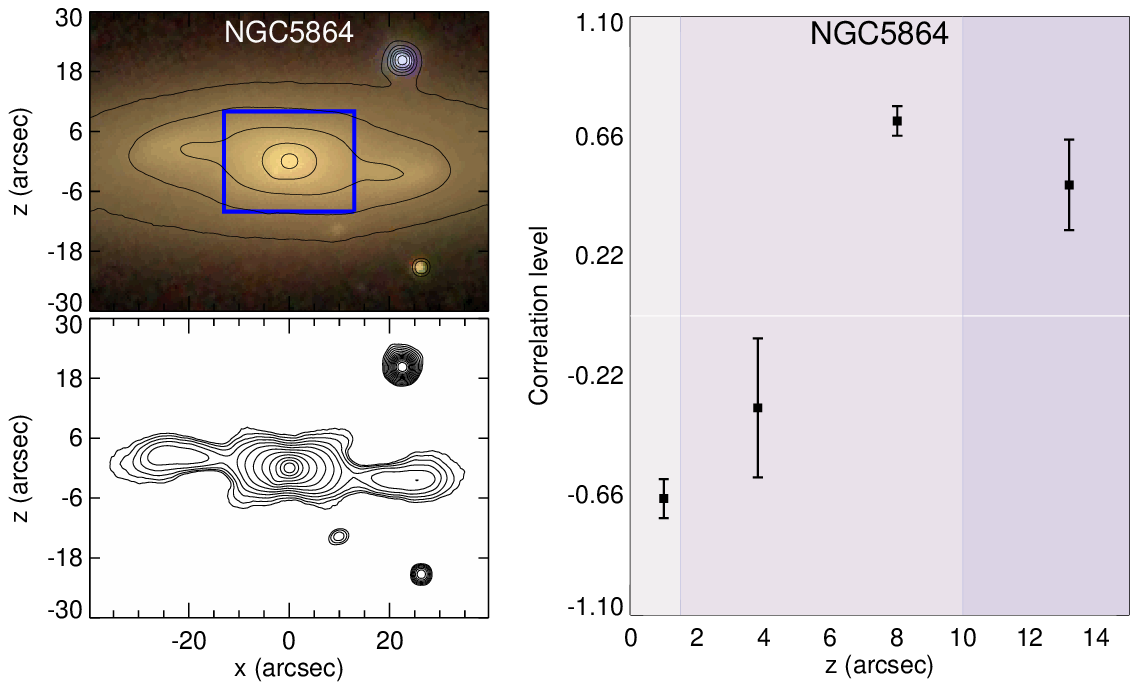}
 				\label{fig:suba27}
 			\end{subfigure}
 			 		\begin{subfigure}{0.44\textwidth}
 			 			\includegraphics[width=0.95\textwidth]{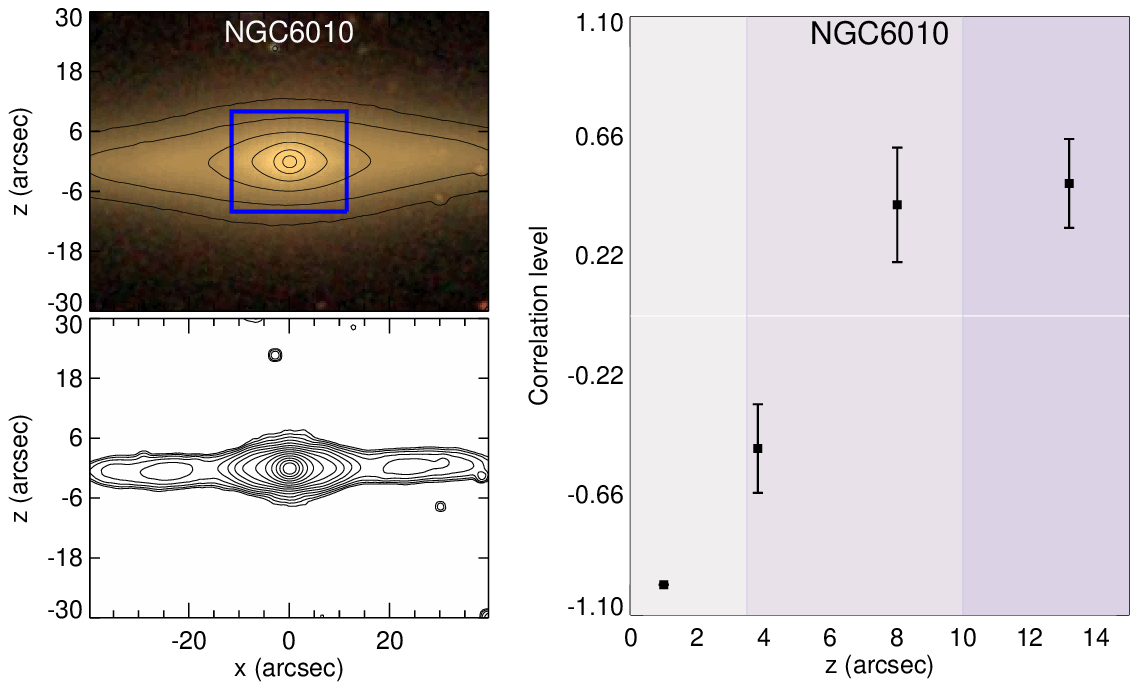}
 			 			\label{fig:suba28}
 			 		\end{subfigure}
 		\setcounter{figure}{1}
 		\centering
 		\centering
 		\caption{Continued.}
 	\end{figure} 
 \end{landscape}   
 
%

\section{SSP profiles  for individual galaxies}
\label{app:single_profiles}
Here, we present the individual SSP-equivalent stellar population profiles, along the minor axis for all bulges in our sample.
The top axis is in physical units, showing the physical extent of the bulge in kpc, while the bottom axis is normalised to $z_{\rm{B}}$, in arcsec. The gradients have been calculated in dex/$z_{\rm{B}}$.
 We also present the index-index diagram of H$\beta_{o}$ versus [MgFe50]' (and the Mg$b$ versus Fe5015) of the average Voronoi bin values at various heights from the disc plane. Overlaid are the MILES SSP models for different ages and metallicities. The green and red grids correspond to the scaled-solar and $\alpha$--enhanced SSP models, respectively. The points are colour-coded according to their distances from the disc plane.
	 \begin{landscape}
	 \begin{figure}
	 	\captionsetup[subfigure]{labelformat=empty}
	 		\centering
	 	\begin{subfigure}{0.65\textwidth}
	 		\centering
	 		\includegraphics[width=0.95\textwidth]{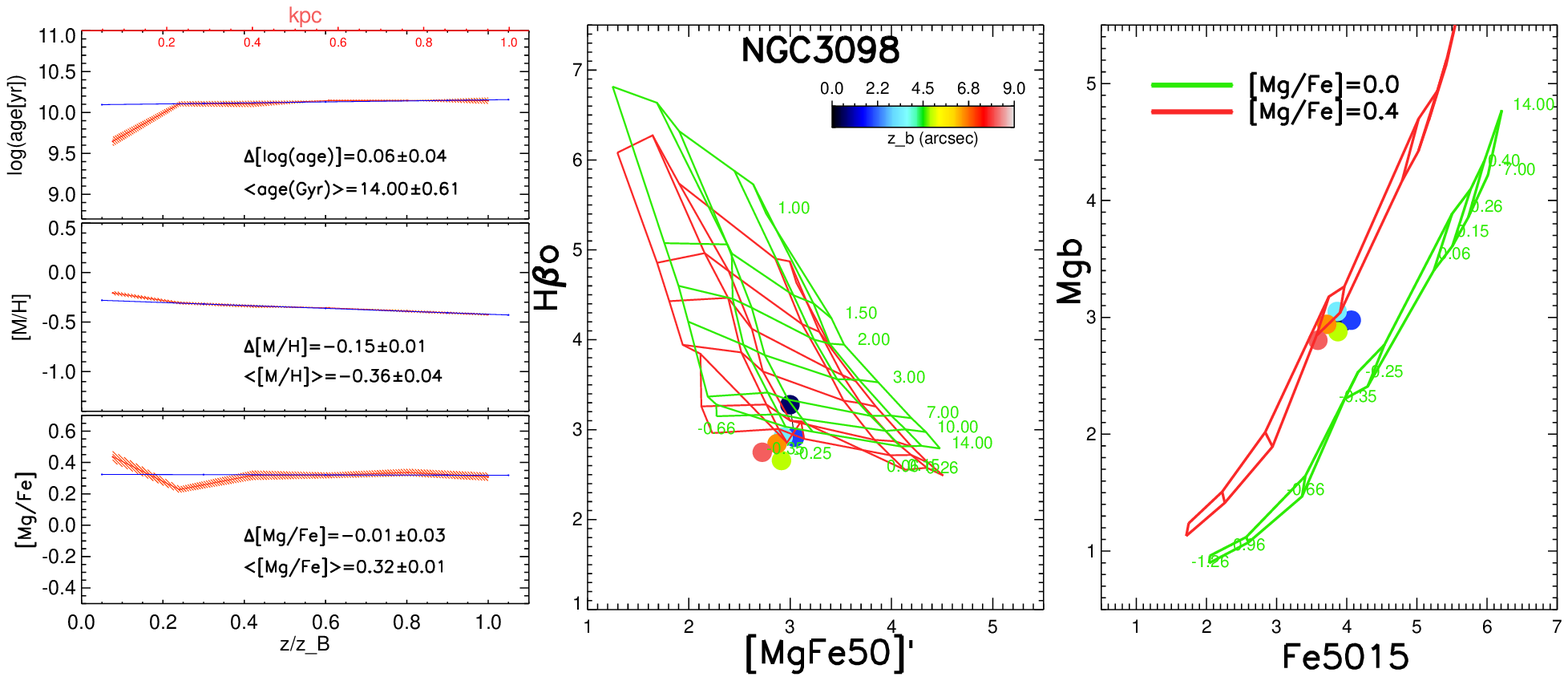}
	 		\label{fig:subb1}
	 	\end{subfigure}%
	 	\hspace*{-0.9em}
	 	\begin{subfigure}{0.65\textwidth}
	 		\centering
	 		\includegraphics[width=0.95\textwidth]{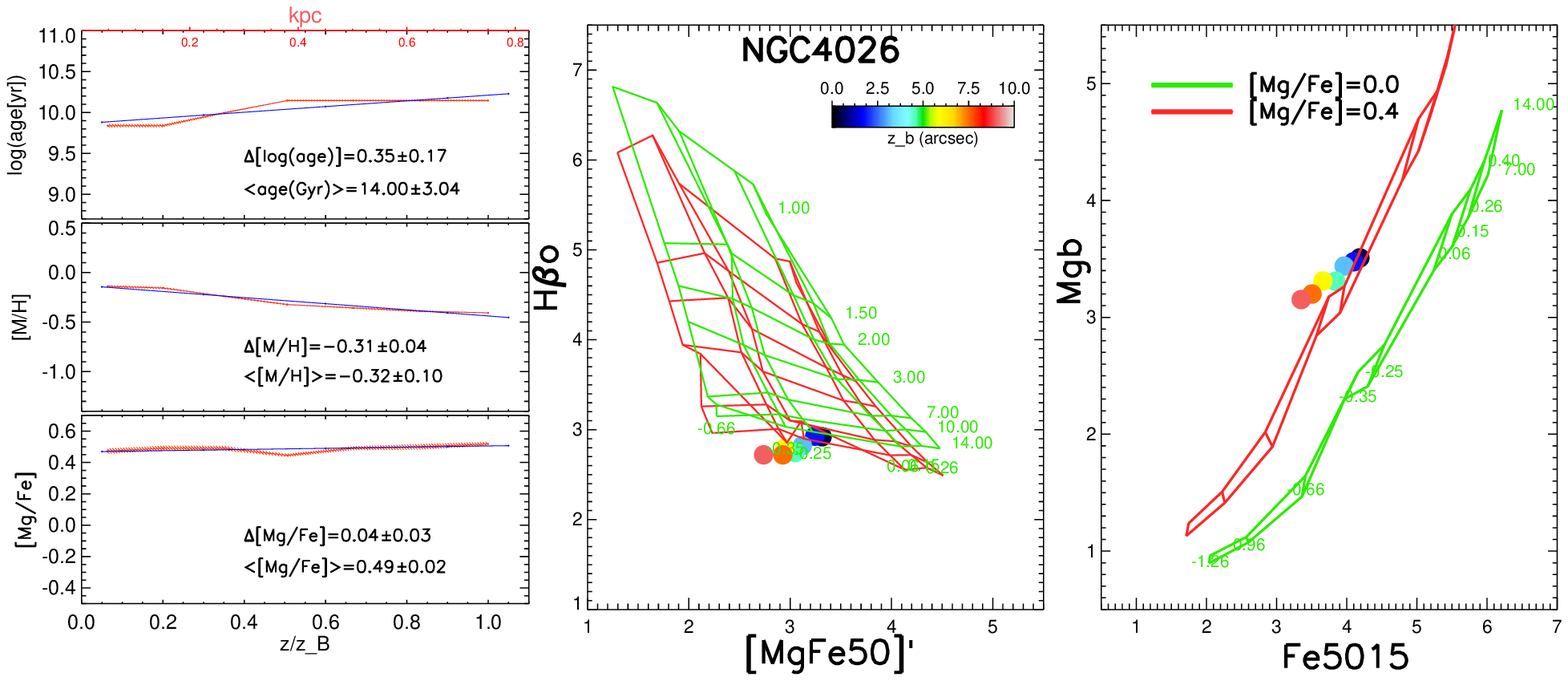}
	 		\label{fig:subb2}
	 	\end{subfigure}
	 	\begin{subfigure}{0.65\textwidth}
	 		\centering
	 		\includegraphics[width=0.95\textwidth]{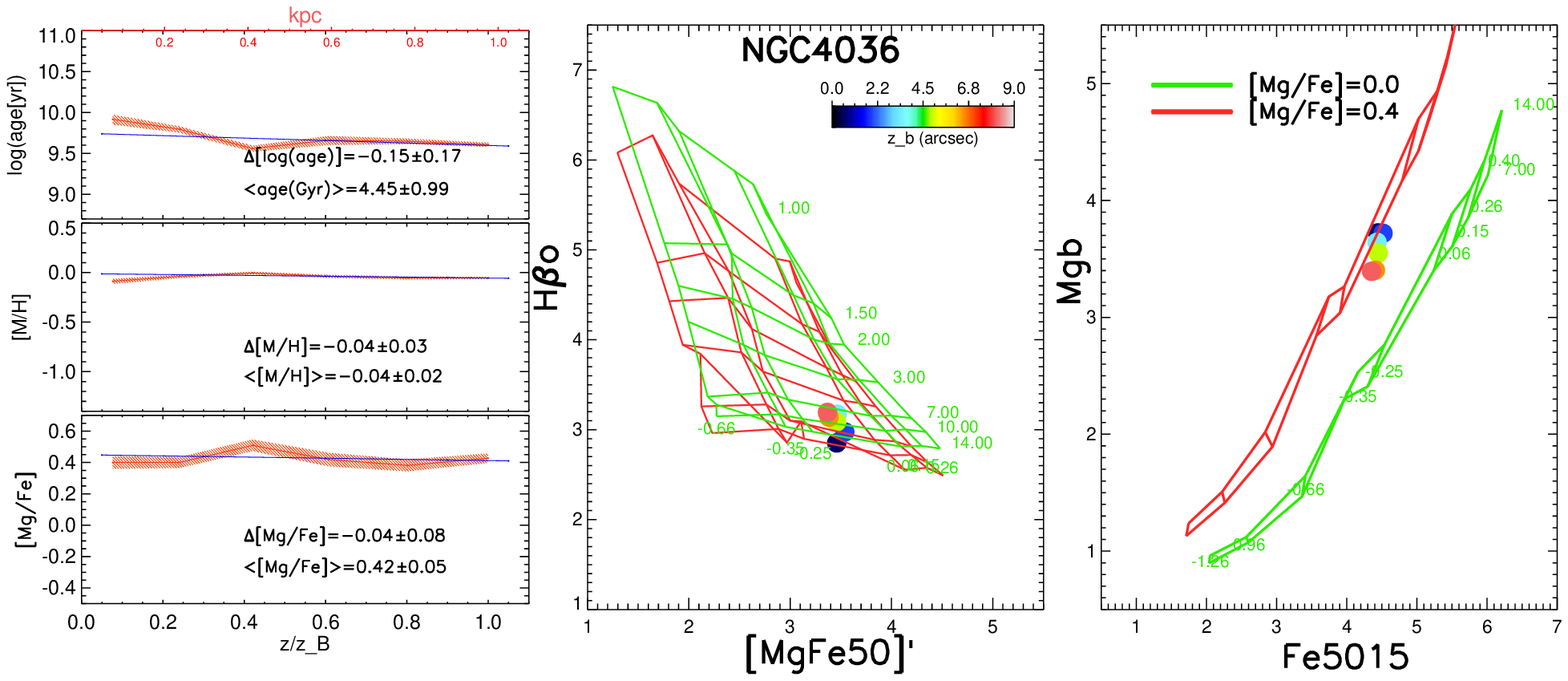}
	 		\label{fig:subb3}
	 	\end{subfigure}
	 	\begin{subfigure}{0.65\textwidth}
	 		\centering
	 		\includegraphics[width=0.95\textwidth]{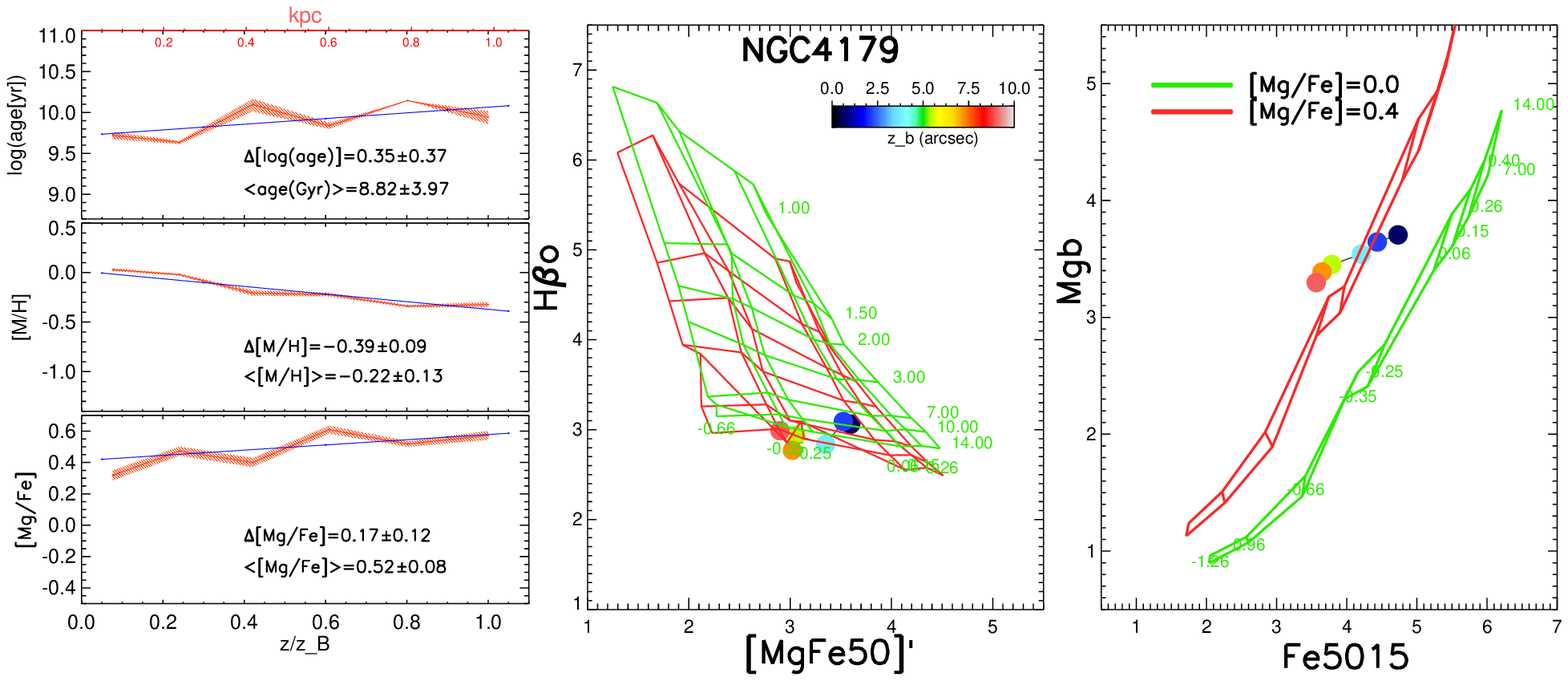}
	 		\label{fig:subb4}
	 	\end{subfigure}
	 	\begin{subfigure}{0.65\textwidth}
	 		\centering
	 		\includegraphics[width=0.95\textwidth]{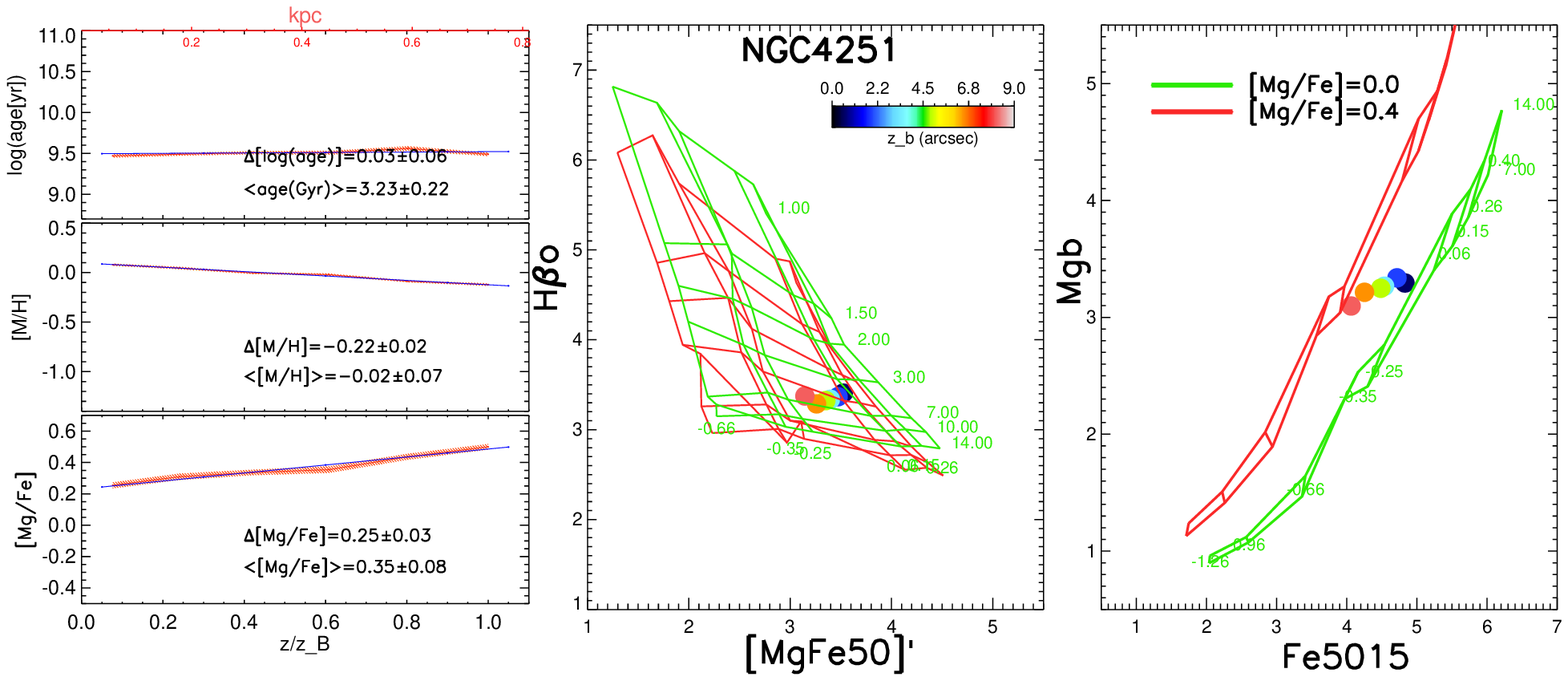}
	 		\label{fig:subb5}
	 	\end{subfigure}%
	 	\begin{subfigure}{0.65\textwidth}
	 		\centering
	 		\includegraphics[width=0.95\textwidth]{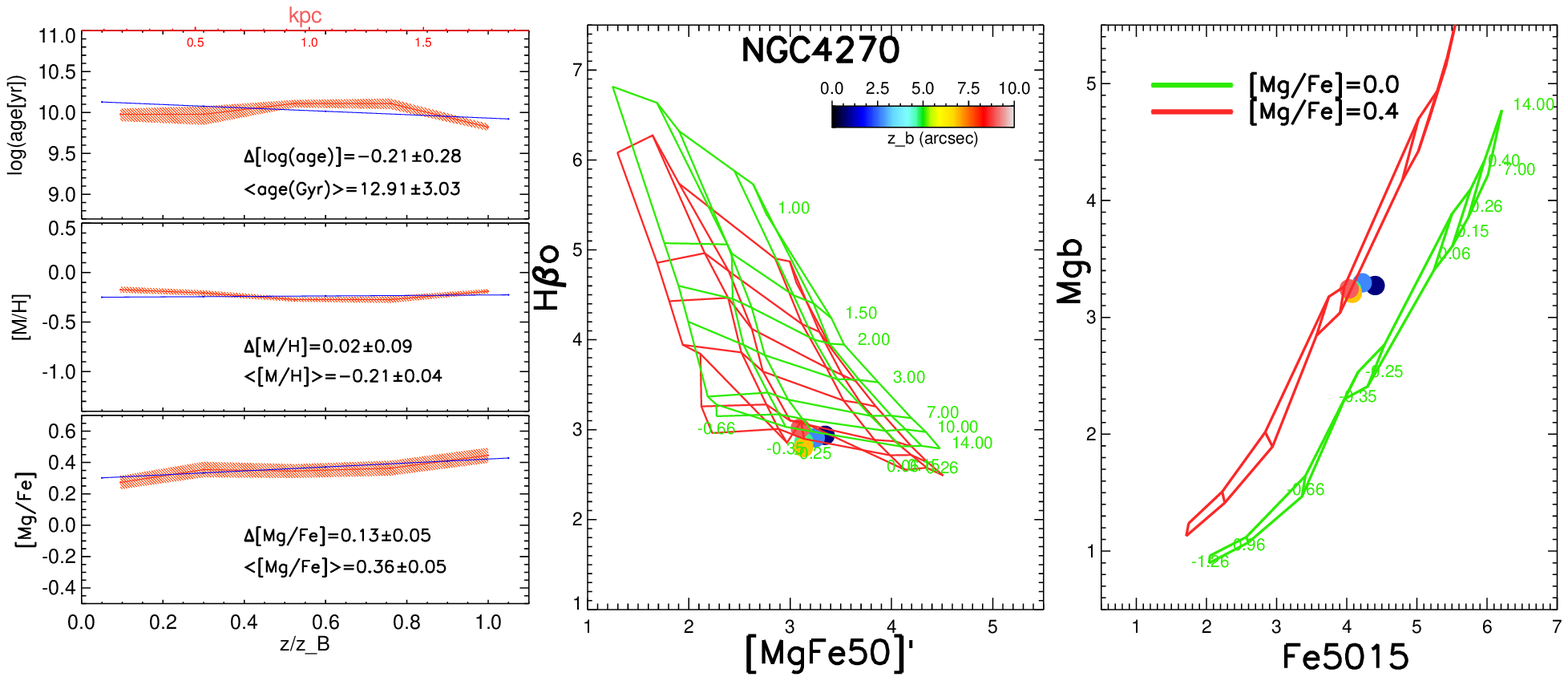}
	 		\label{fig:subb6}
	 	\end{subfigure}
	 	\setcounter{figure}{1}
	 	\centering
	 	\centering
	 	\caption{Left panel: The SSP-equivalent age, metallicity and [Mg/Fe] profiles, along the minor axis. The top axis is showing the physical extent of the bulge in kpc and the bottom axis is normalised to $z_{\rm{B}}$, in arcsec.
        In each panel, we indicate the mean SSP-equivalent age, metallicity and [Mg/Fe] within the bulge analysis window, as well as the vertical gradients (measured in dex/$z_{\rm{B}}$). Middle (and right) panel: the index-index diagram of H$\beta_{o}$ versus [MgFe50]' (and the Mg$b$ versus Fe5015) of the average Voronoi bin values at various heights from the disc plane. Overlaid are the MILES SSP models for different ages and metallicities, correspond to the scaled-solar (green) and $\alpha$--enhanced SSP models (red), respectively.
	 		}
	 \end{figure}
		 \end{landscape} 

		 \begin{landscape} 	 
	 \begin{figure}
	 	\captionsetup[subfigure]{labelformat=empty}
	 	\begin{subfigure}{0.65\textwidth}
	 		\includegraphics[width=0.95\textwidth]{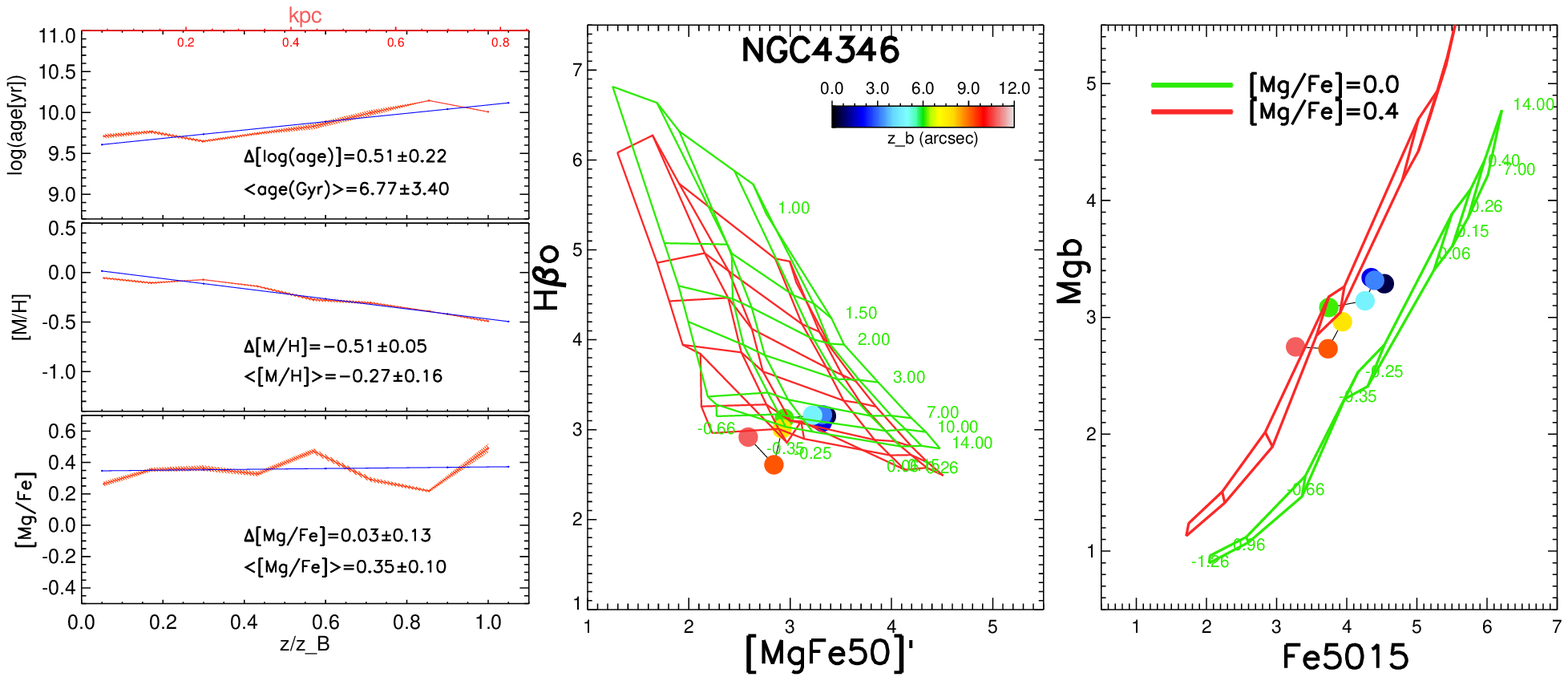}
	 		\label{fig:subb7}
	 	\end{subfigure}
	 	\begin{subfigure}{0.65\textwidth}
	 		\includegraphics[width=0.95\textwidth]{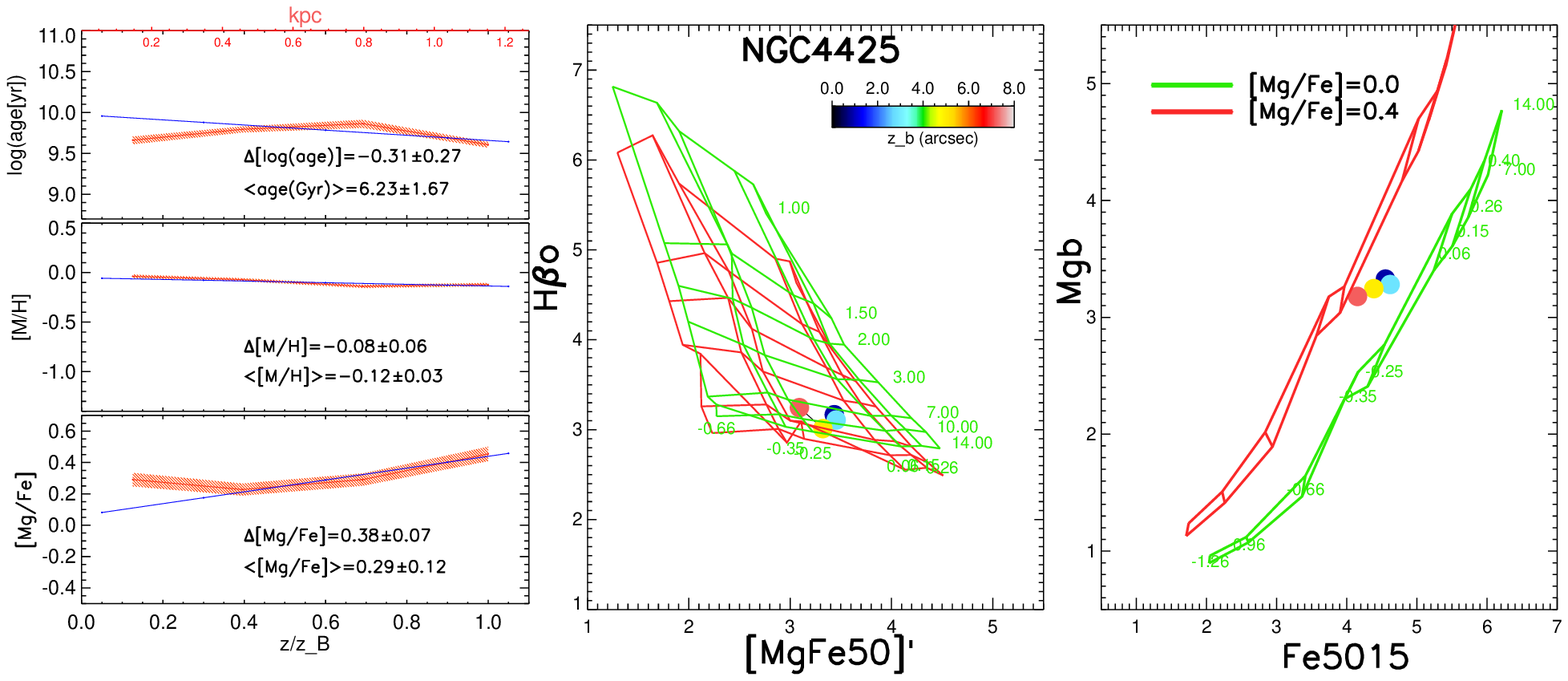}
	 		\label{fig:subb8}
	 	\end{subfigure}
	 		 	 	\begin{subfigure}{0.65\textwidth}
	 		 	 		\includegraphics[width=0.95\textwidth]{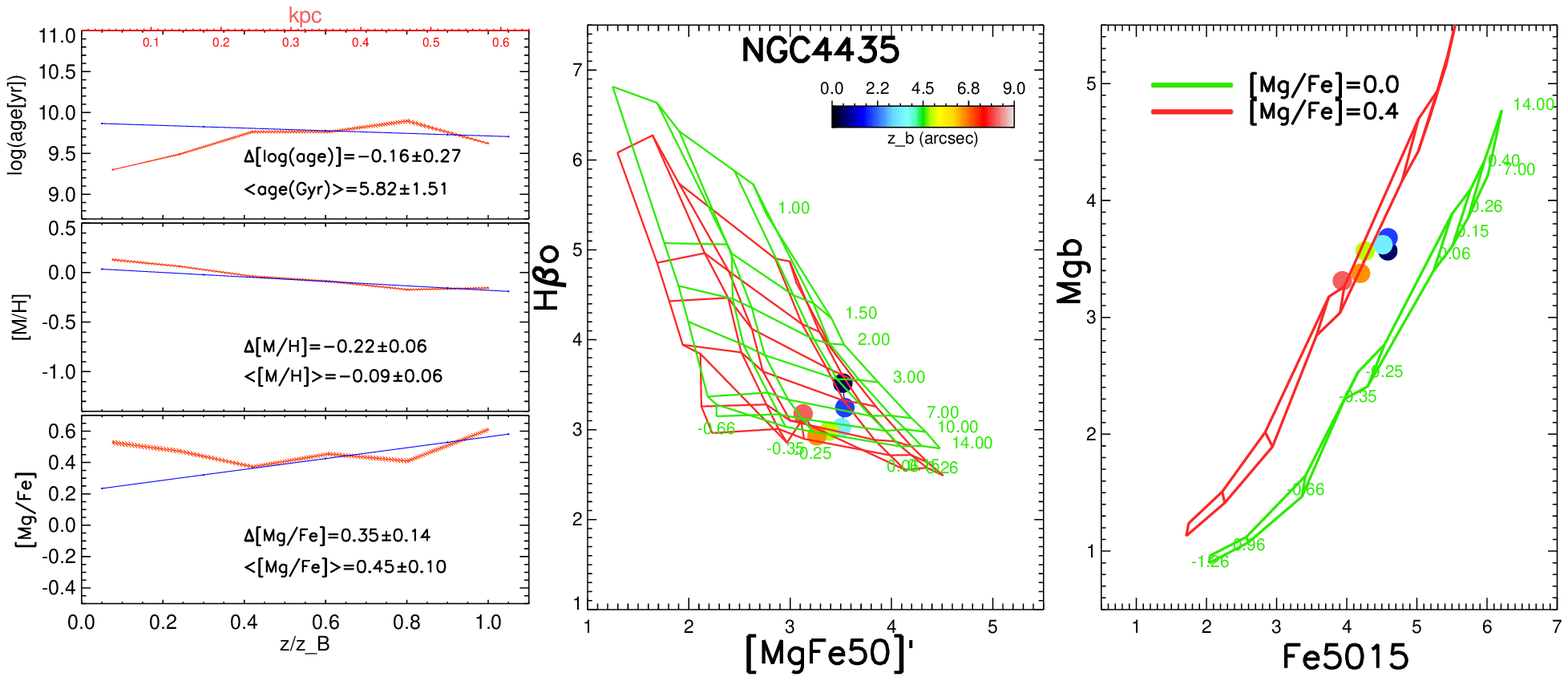}
	 		 	 		\label{fig:subb9}
	 		 	 	\end{subfigure}%
	 		 	 	\begin{subfigure}{0.65\textwidth}
	 		 	 		\includegraphics[width=0.95\textwidth]{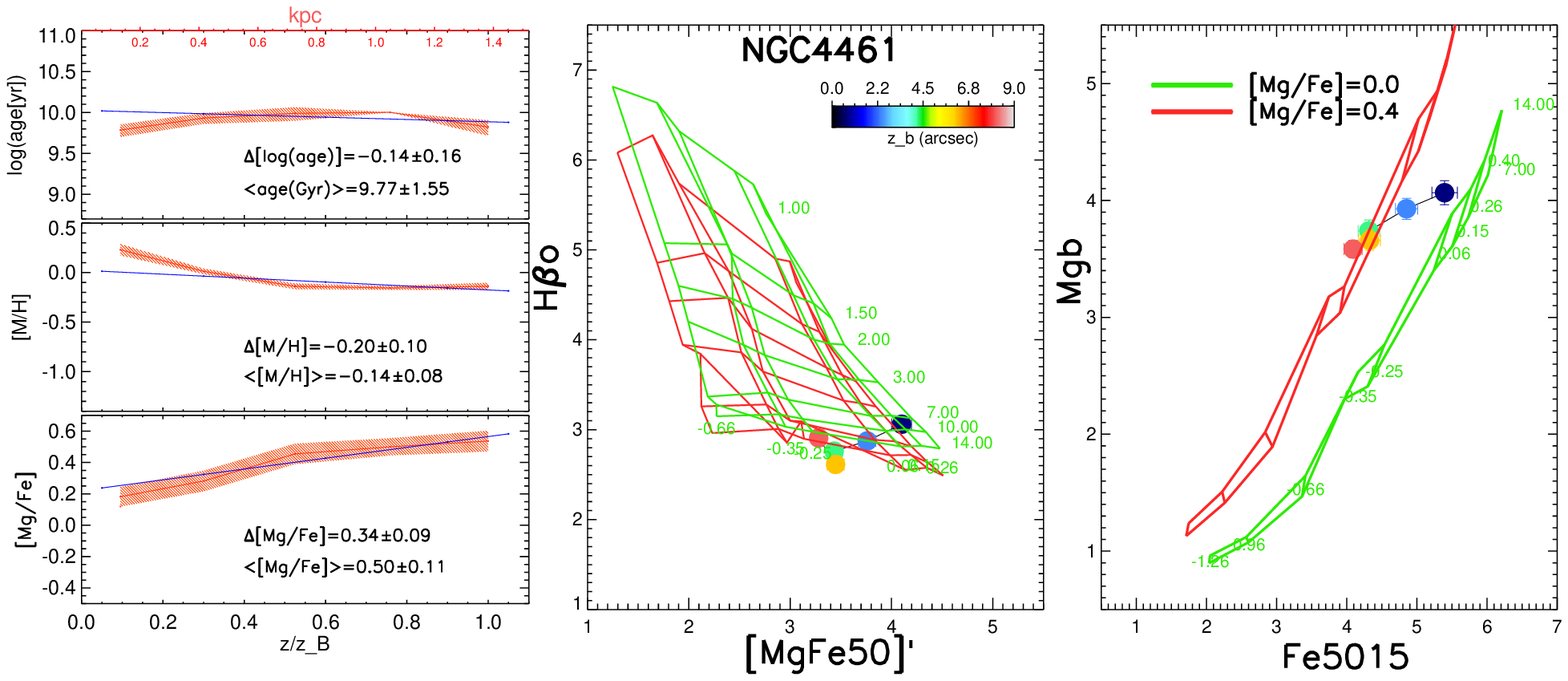}
	 		 	 		\label{fig:subb10}
	 		 	 	\end{subfigure}
	 		 	 	\begin{subfigure}{0.65\textwidth}
	 		 	 		\includegraphics[width=0.95\textwidth]{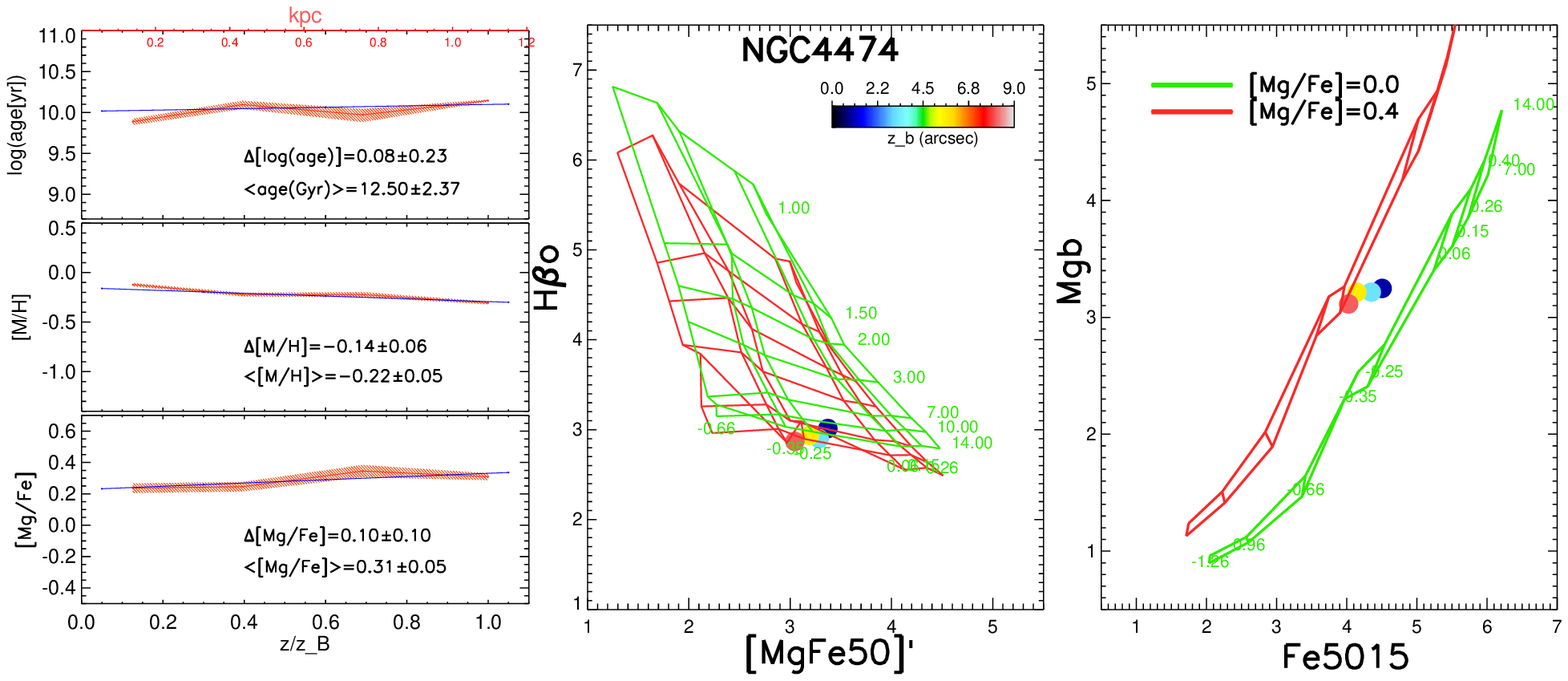}
	 		 	 		\label{fig:subb11}
	 		 	 	\end{subfigure}
	 		 	 	\begin{subfigure}{0.65\textwidth}
	 		 	 		\includegraphics[width=0.95\textwidth]{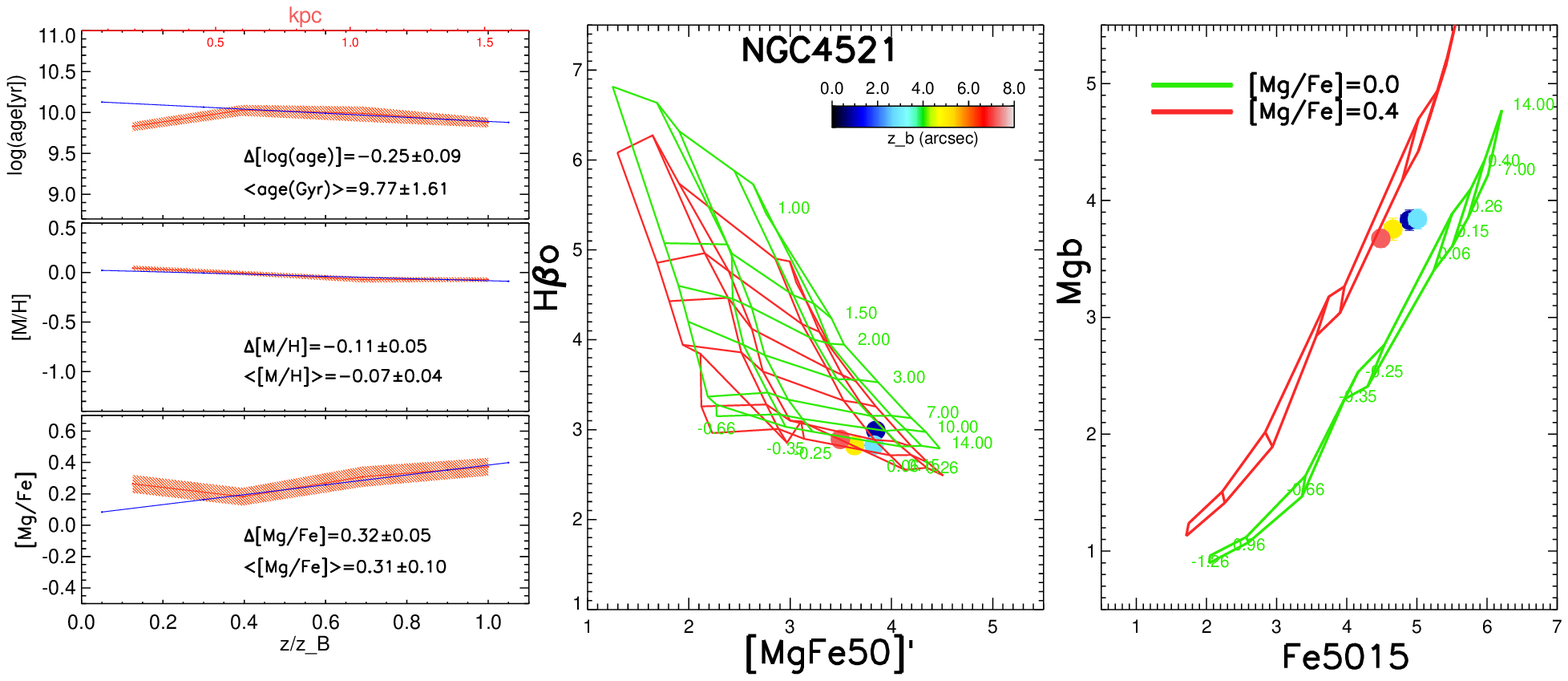}
	 		 	 		\label{fig:subb12}
	 		 	 	\end{subfigure}
	 	\setcounter{figure}{1}
	 	\centering
	 	\centering
	 	\caption{Continued.}
	 \end{figure}
		 \end{landscape} 

		 \begin{landscape} 	 
		 	 \begin{figure}
		 	 	\captionsetup[subfigure]{labelformat=empty}
		 	 	\begin{subfigure}{0.65\textwidth}
		 	 		\includegraphics[width=0.95\textwidth]{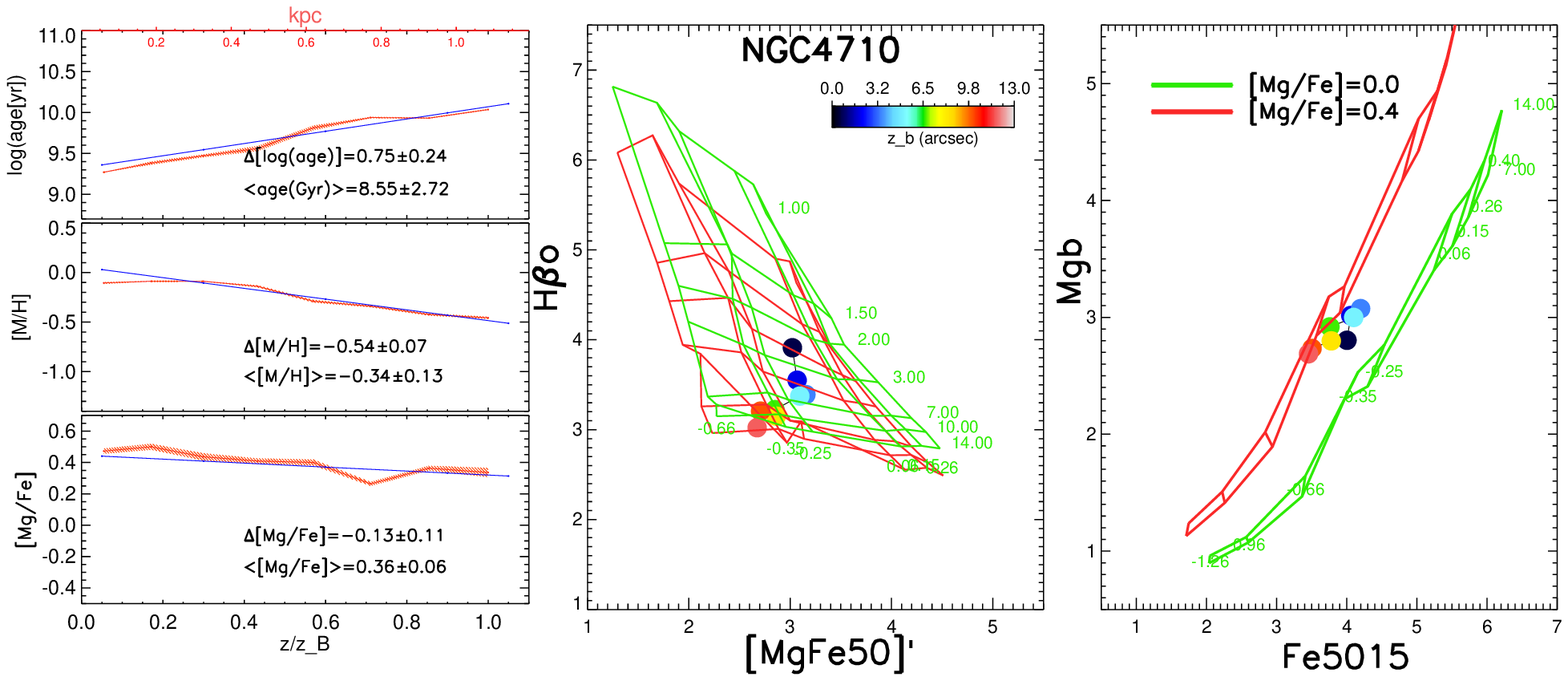}
		 	 		\label{fig:subb13}
		 	 	\end{subfigure}%
		 	 	\begin{subfigure}{0.65\textwidth}
		 	 		\includegraphics[width=0.95\textwidth]{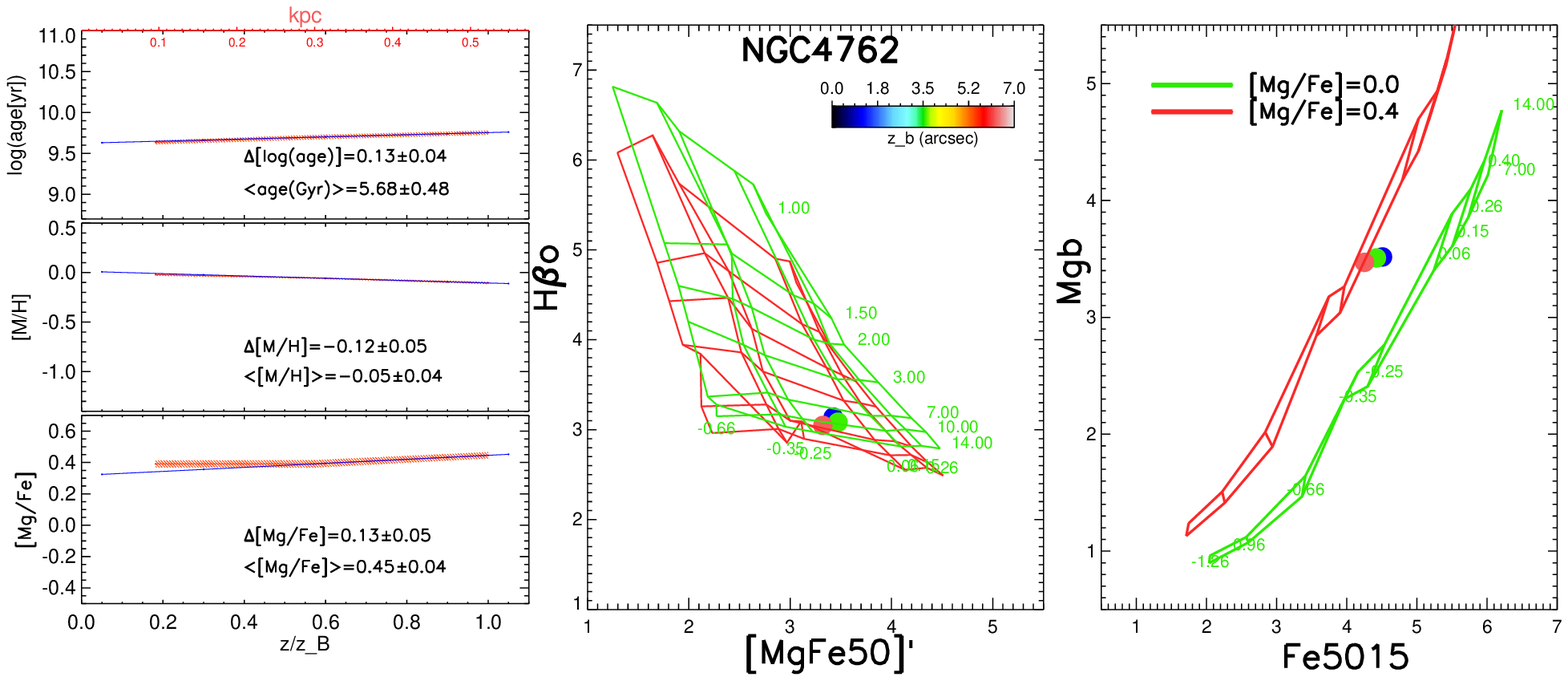}
		 	 		\label{fig:subb14}
		 	 	\end{subfigure}
		 	 	\begin{subfigure}{0.65\textwidth}
		 	 		\includegraphics[width=0.95\textwidth]{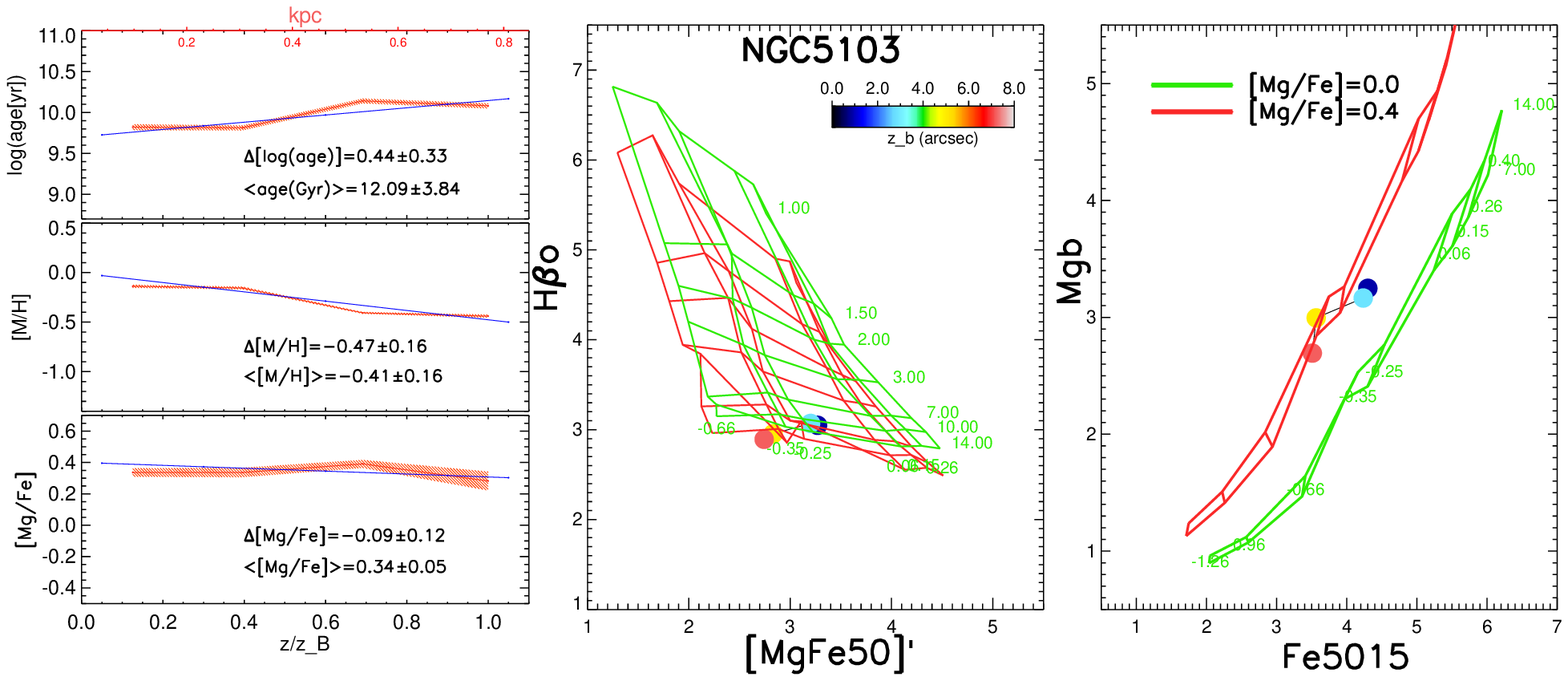}
		 	 		\label{fig:subb15}
		 	 	\end{subfigure}
		 	 	\begin{subfigure}{0.65\textwidth}
		 	 		\includegraphics[width=0.95\textwidth]{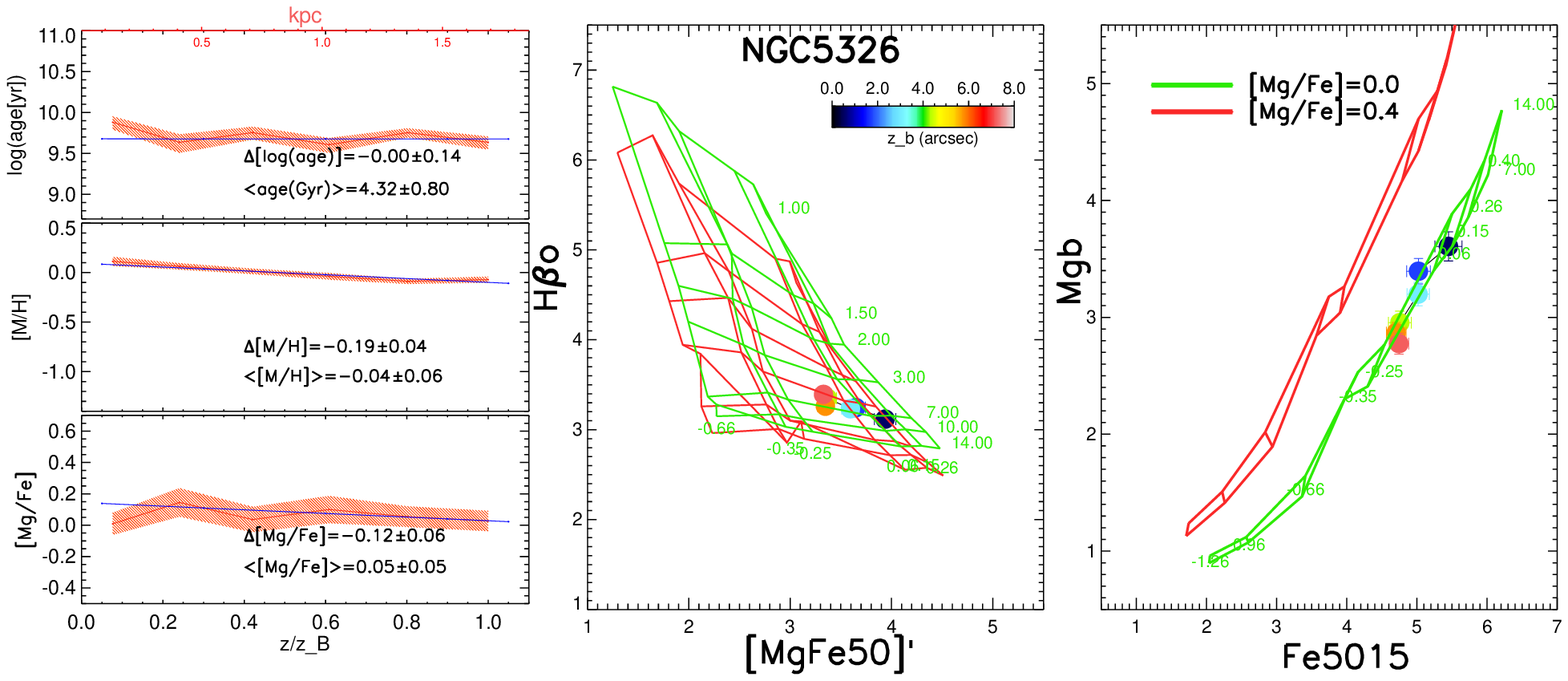}
		 	 		\label{fig:subb16}
		 	 	\end{subfigure}
		 	 	\begin{subfigure}{0.65\textwidth}
		 	 		\includegraphics[width=0.95\textwidth]{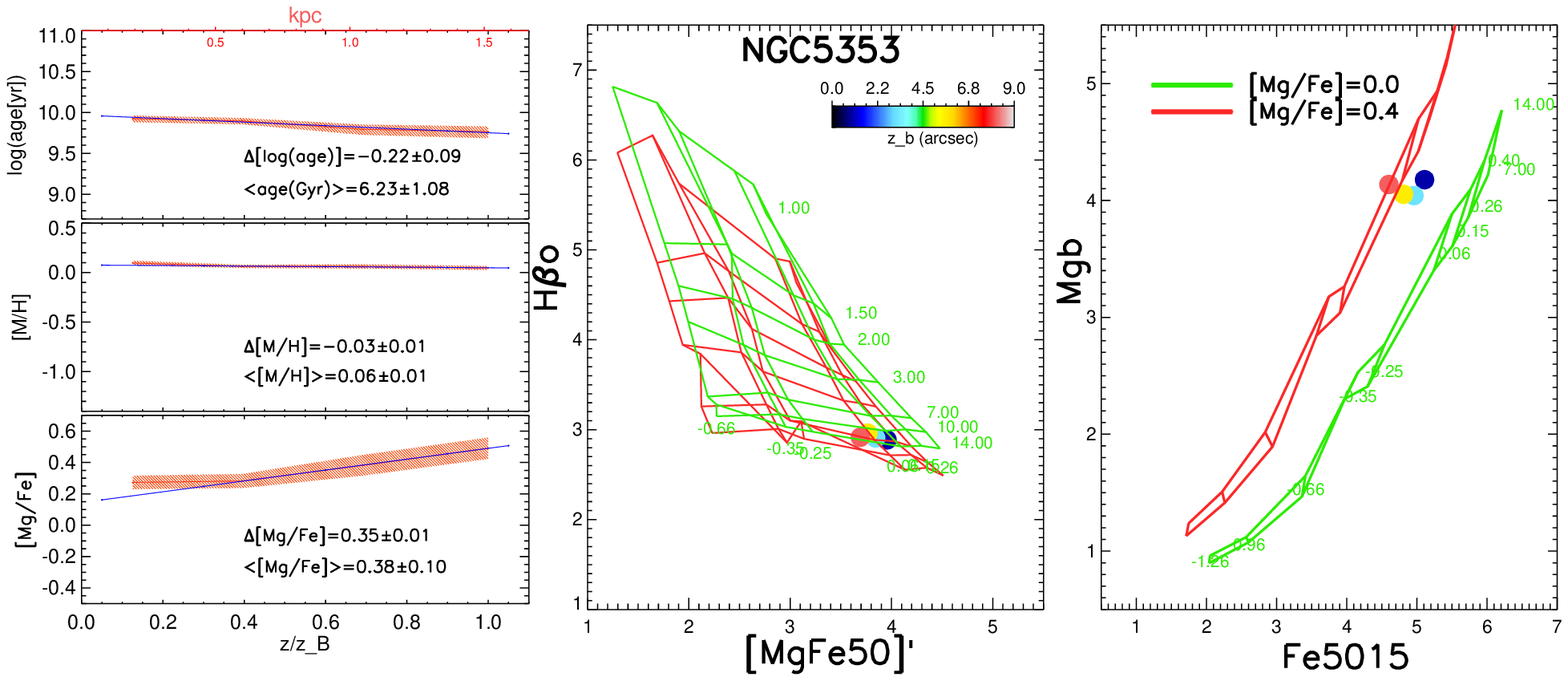}
		 	 		\label{fig:subb17}
		 	 	\end{subfigure}%
		 	 	\begin{subfigure}{0.65\textwidth}
		 	 		\includegraphics[width=0.95\textwidth]{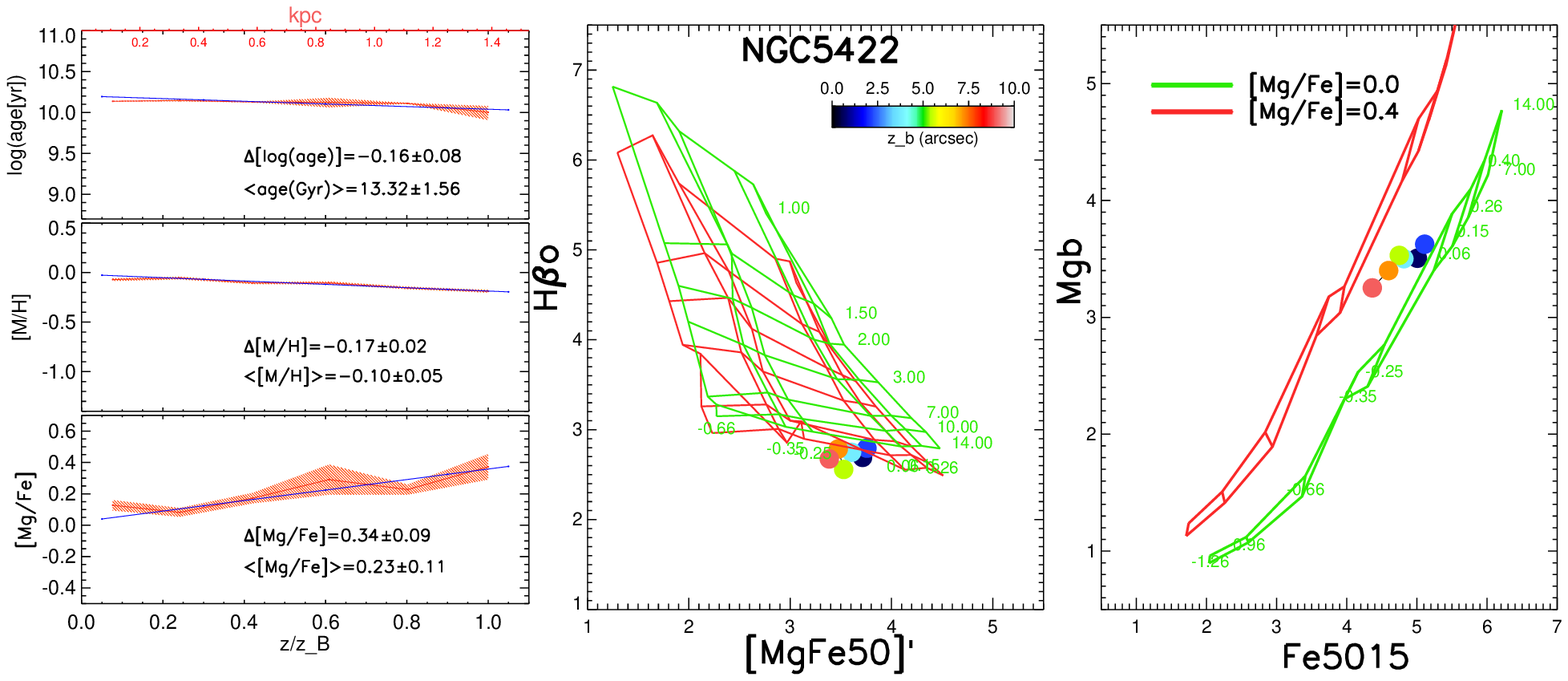}
		 	 		\label{fig:subb18}
		 	 	\end{subfigure}
		 	 	\setcounter{figure}{1}
		 	 	\centering
		 	 	\centering
		 	 	\caption{Continued.}
		 	 \end{figure}
 \end{landscape}  
 
	  \begin{landscape}  
	 		 	 \begin{figure}
	 		 	 	\captionsetup[subfigure]{labelformat=empty}
	 		 	 	\begin{subfigure}{0.65\textwidth}
	 		 	 		\includegraphics[width=0.95\textwidth]{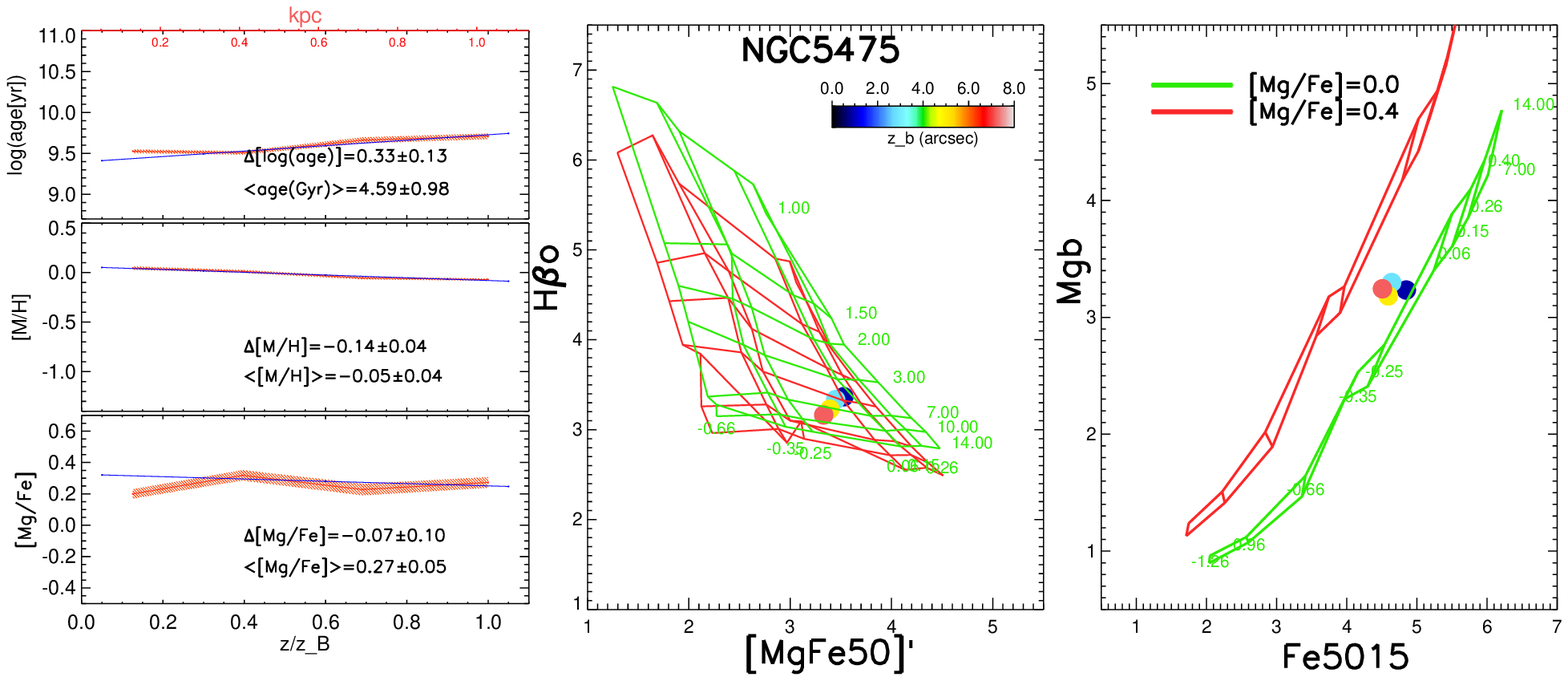}
	 		 	 		\label{fig:subb19}
	 		 	 	\end{subfigure}
	 		 	 	\begin{subfigure}{0.65\textwidth}
	 		 	 		\includegraphics[width=0.95\textwidth]{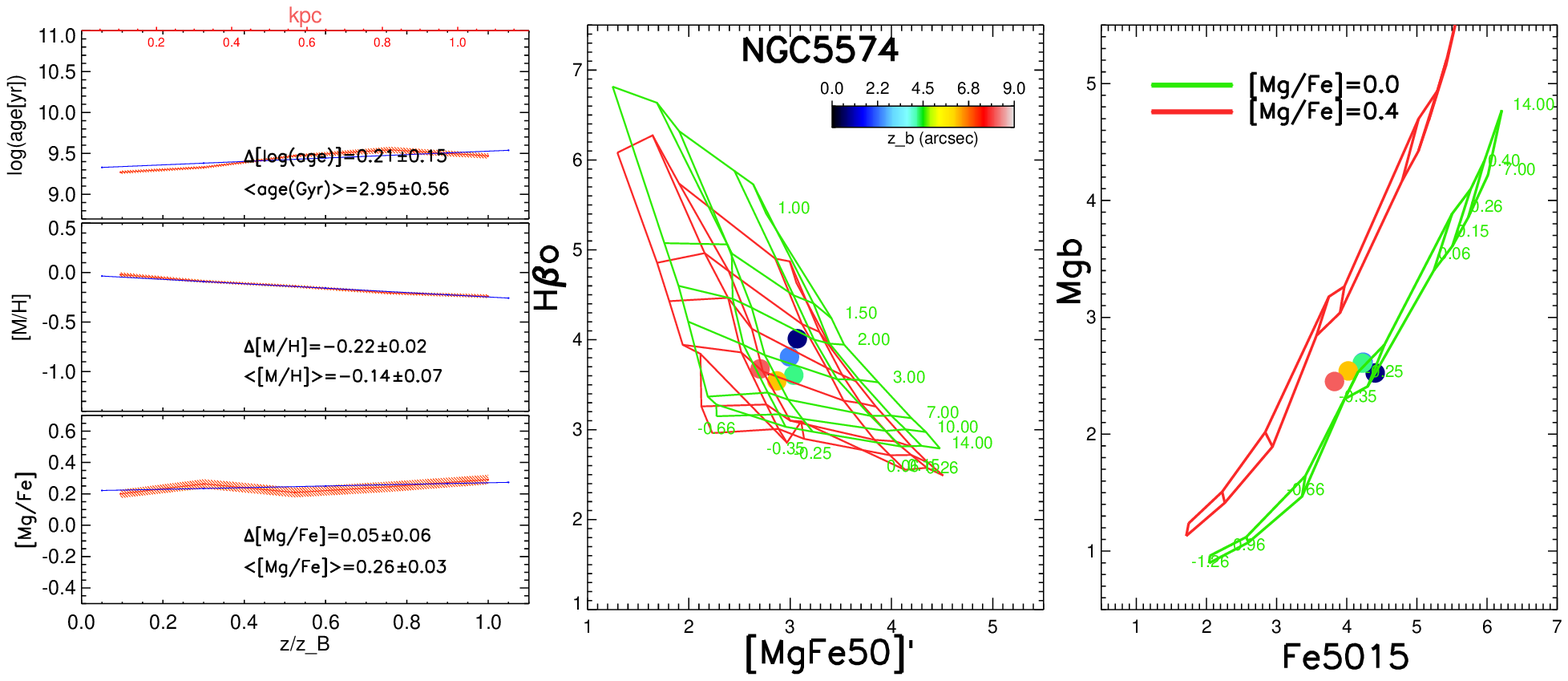}
	 		 	 		\label{fig:subb20}
	 		 	 	\end{subfigure}
	 		 	 		\begin{subfigure}{0.65\textwidth}
	 		 	 			\includegraphics[width=0.95\textwidth]{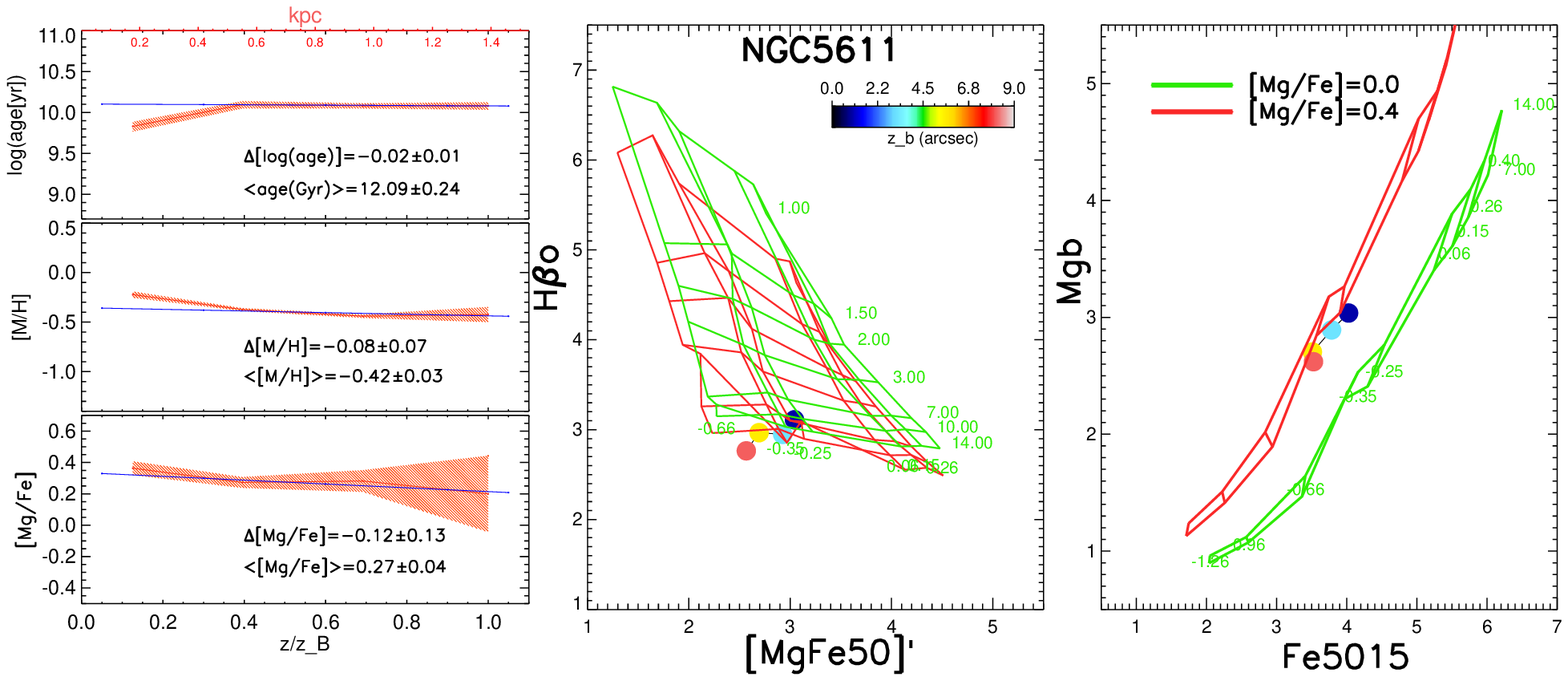}
	 		 	 			\label{fig:subb21}
	 		 	 		\end{subfigure}%
	 		 	 		\begin{subfigure}{0.65\textwidth}
	 		 	 			\includegraphics[width=0.95\textwidth]{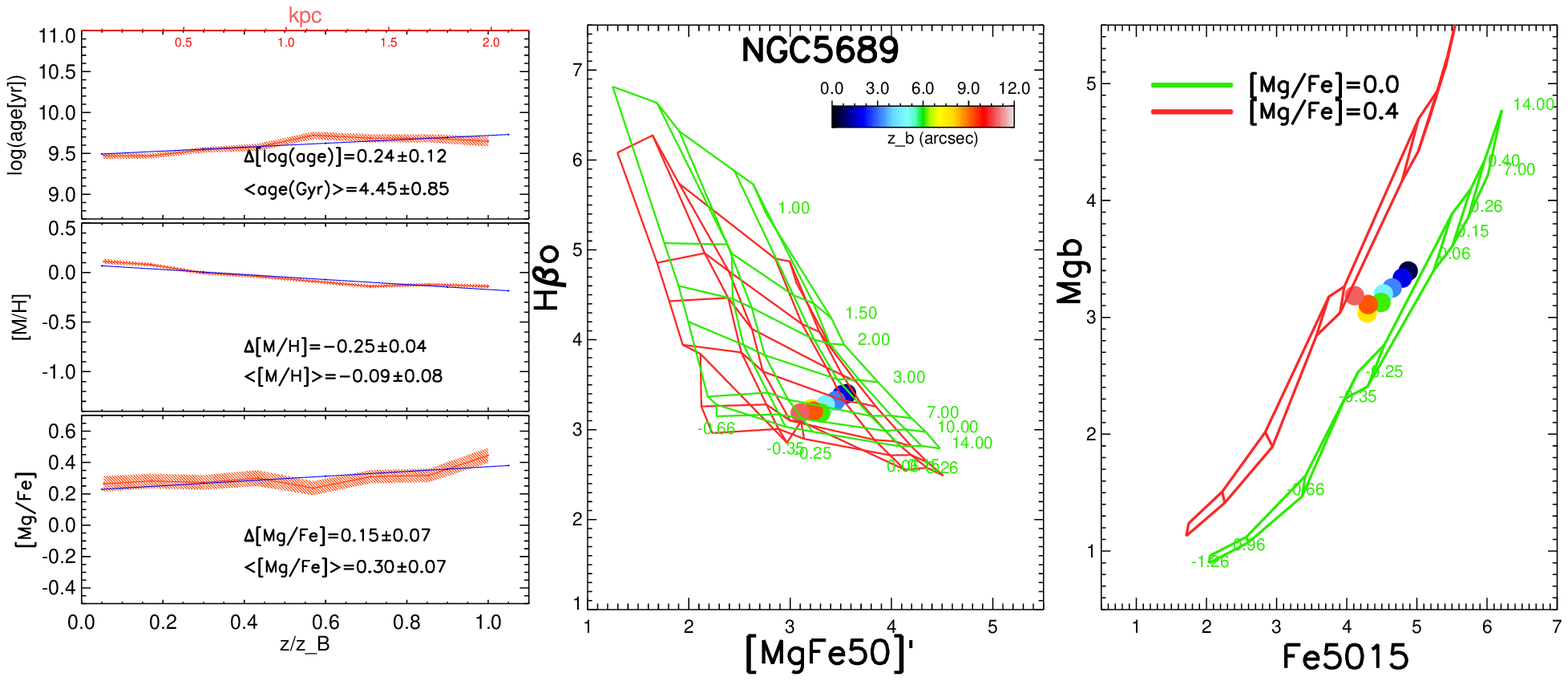}
	 		 	 			\label{fig:subb22}
	 		 	 		\end{subfigure}
	 		 	 		\begin{subfigure}{0.65\textwidth}
	 		 	 			\includegraphics[width=0.95\textwidth]{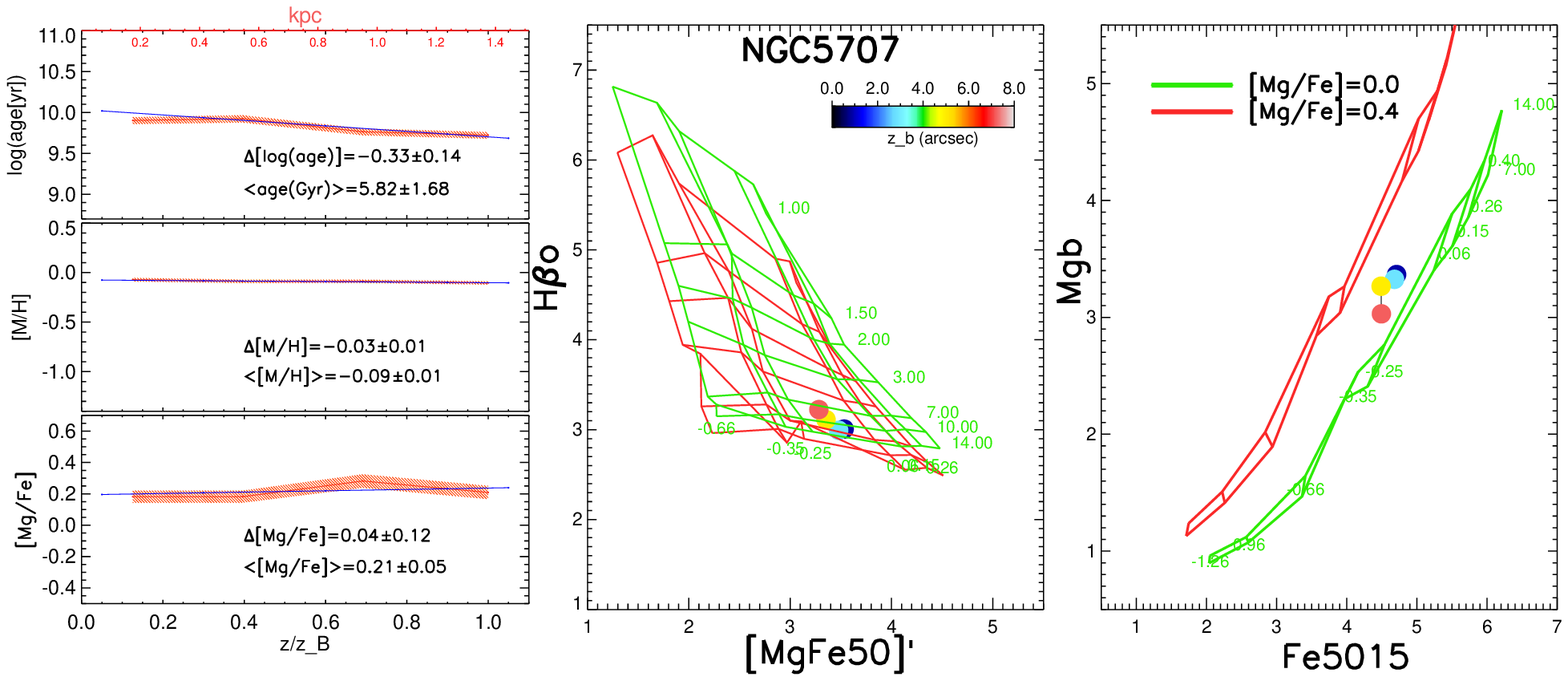}
	 		 	 			\label{fig:subb23}
	 		 	 		\end{subfigure}
	 		 	 		\begin{subfigure}{0.65\textwidth}
	 		 	 			\includegraphics[width=0.95\textwidth]{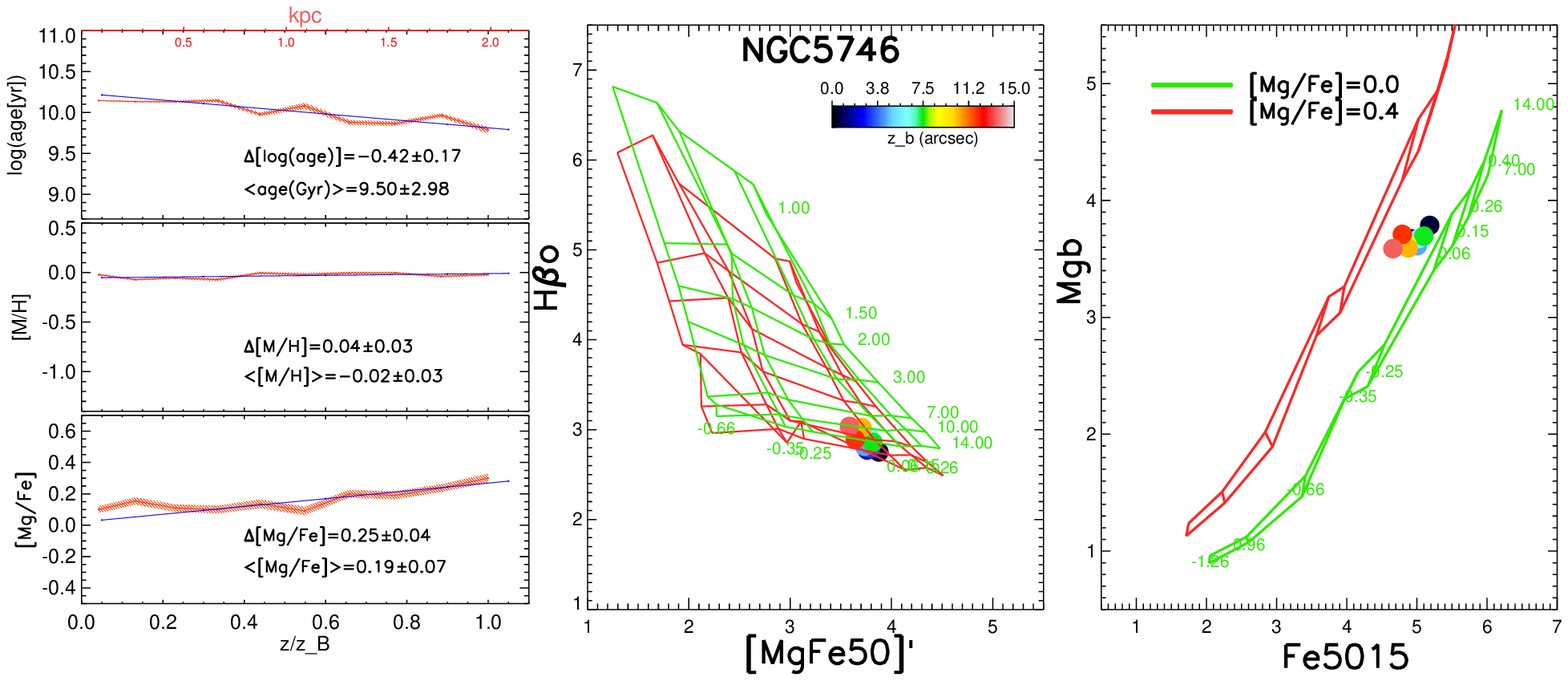}
	 		 	 			\label{fig:subb24}
	 		 	 		\end{subfigure}
	 		 	 	\setcounter{figure}{1}
	 		 	 	\centering
	 		 	 	\centering
	 		 	 	\caption{Continued.}
	 		 	 \end{figure} 
	 \end{landscape}   

	 \begin{landscape}   
			 		 	 \begin{figure}
			 		 	 	\captionsetup[subfigure]{labelformat=empty}
			 		 	 	\begin{subfigure}{0.65\textwidth}
			 		 	 		\includegraphics[width=0.95\textwidth]{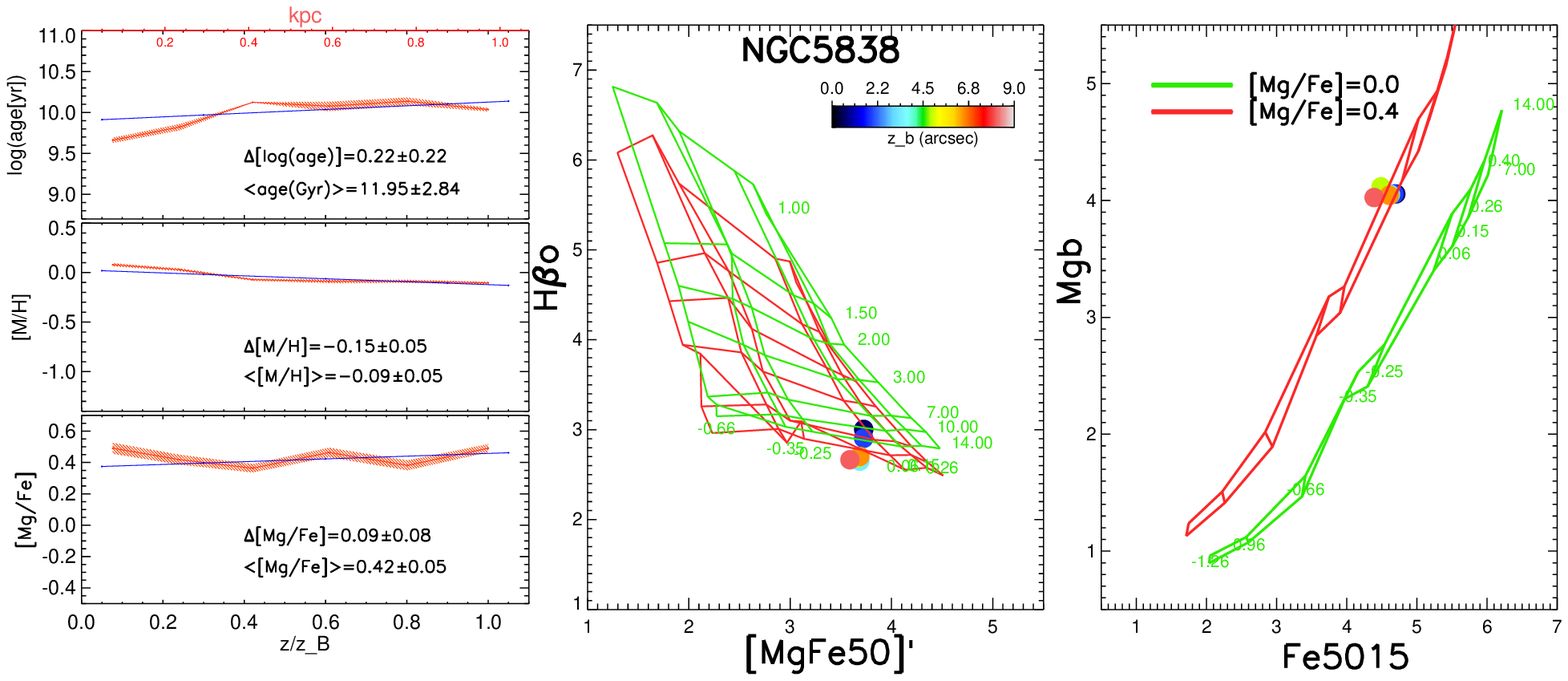}
			 		 	 		\label{fig:subb25}
			 		 	 	\end{subfigure}%
			 		 	 	\begin{subfigure}{0.65\textwidth}
			 		 	 		\includegraphics[width=0.95\textwidth]{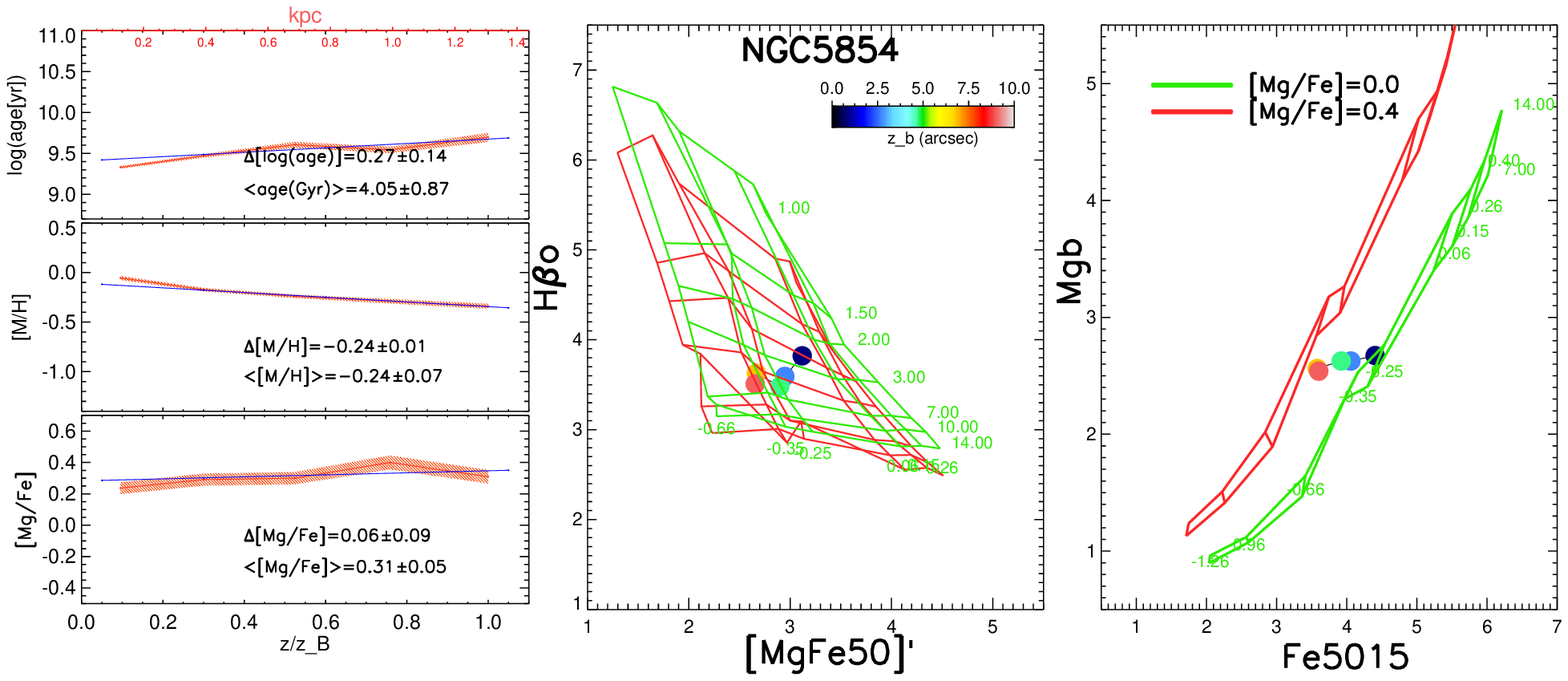}
			 		 	 		\label{fig:subb26}
			 		 	 	\end{subfigure}
			 		 	 	\begin{subfigure}{0.65\textwidth}
			 		 	 		\includegraphics[width=0.95\textwidth]{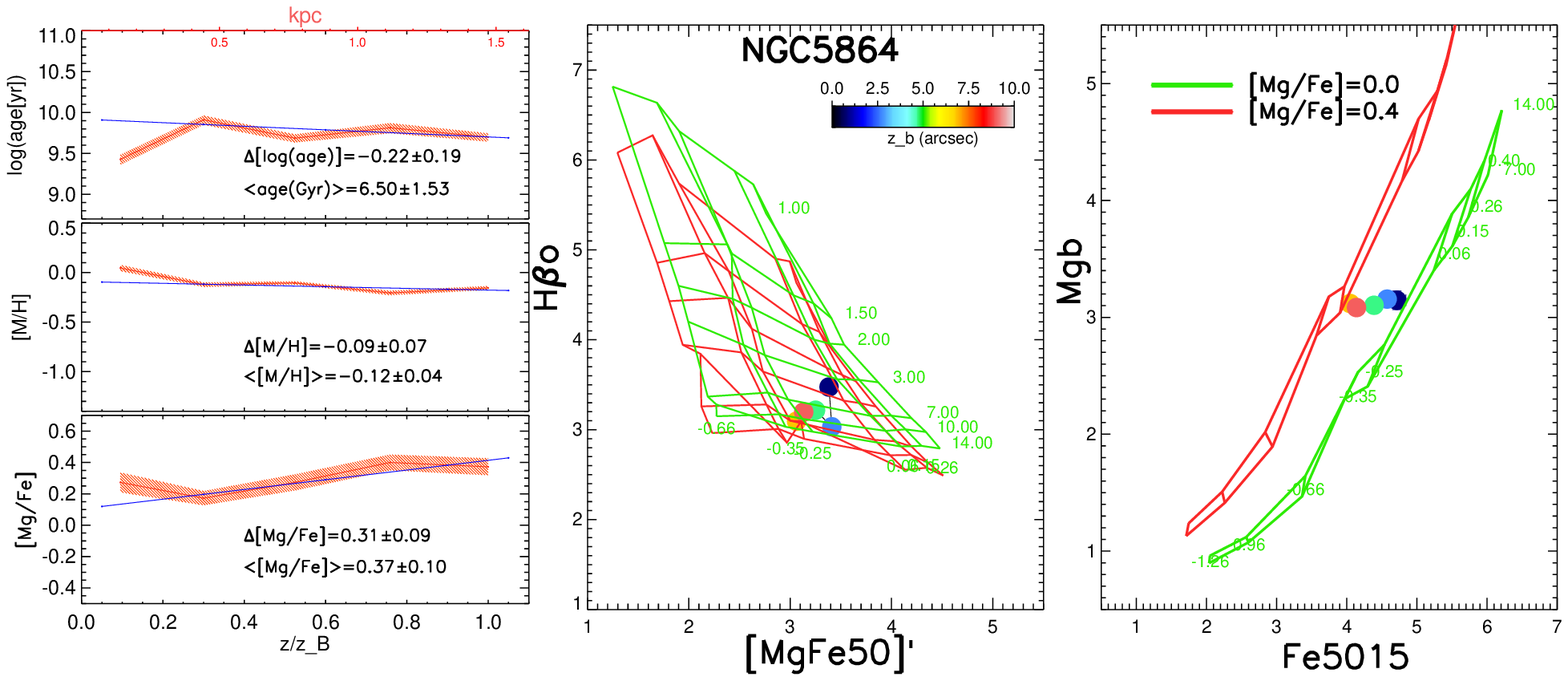}
			 		 	 		\label{fig:subb27}
			 		 	 	\end{subfigure}
			 		 	 	\begin{subfigure}{0.65\textwidth}
			 		 	 		\includegraphics[width=0.95\textwidth]{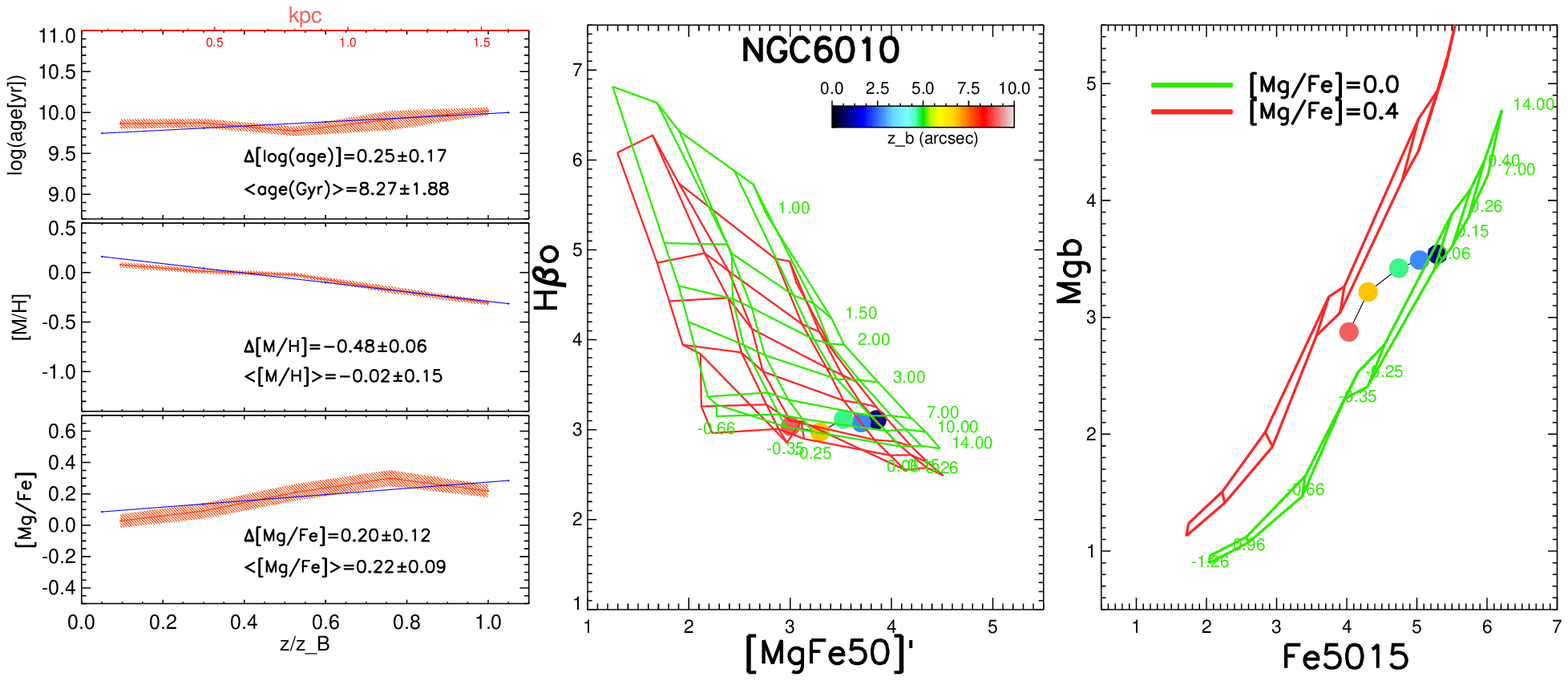}
			 		 	 		\label{fig:subb28}
			 		 	 	\end{subfigure}
			 		 	 	\setcounter{figure}{1}
			 		 	 	\centering
			 		 	 	\centering
			 		 	 	\caption{Continued.}
			 		 	 \end{figure} 
	 	 \end{landscape}   		 		 	 

\end{document}